\documentclass[fleqn,usenatbib]{mnras}

\usepackage[T1]{fontenc}

\DeclareRobustCommand{\VAN}[3]{#2}
\let\VANthebibliography\thebibliography
\def\thebibliography{\DeclareRobustCommand{\VAN}[3]{##3}\VANthebibliography}

\usepackage{graphicx}	
\usepackage{amsmath}	
\usepackage{amssymb}	
\usepackage{adjustbox}
\usepackage{verbatim}
\usepackage{multirow}
\usepackage{multicol}
\usepackage{pdflscape}
\usepackage[table,xcdraw]{xcolor}
\usepackage{placeins}
\usepackage{orcidlink}
\usepackage{afterpage}
\usepackage{caption}
\usepackage{subcaption}
\usepackage[normalem]{ulem}
\useunder{\uline}{\ul}{}

\usepackage{newtxtext,newtxmath}

\DeclareMathAlphabet{\mathsc}{OT1}{cmr}{m}{sc}
\def\testbx{bx}%
\DeclareRobustCommand{\ion}[2]{%
\relax\ifmmode
\ifx\testbx\f@series
{\mathbf{#1\,\mathsc{#2}}}\else
{\mathrm{#1\,\mathsc{#2}}}\fi
\else\textup{#1\,{\mdseries\textsc{#2}}}%
\fi}

\newcommand{\fspectralline}[3]{[\ion{#1}{#2}]~$\uplambda#3$~\AA}

\newcommand{\Hei}{\ion{He}{i}}
\newcommand{\Heii}{\ion{He}{ii}}
\newcommand{\Oi}{[\ion{O}{i}]}
\newcommand{\Oii}{[\ion{O}{ii}]}
\newcommand{\Oiii}{[\ion{O}{iii}]}

\newcommand{\CaII}{\ion{Ca}{ii}}

\newcommand{\Nii}{[\ion{N}{ii}]}

\newcommand{\Sii}{[\ion{S}{ii}]}

\newcommand{\Fevii}{[\ion{Fe}{vii}]}

\newcommand{\Fex}{[\ion{Fe}{x}]}
\newcommand{\Fexi}{[\ion{Fe}{xi}]}

\newcommand{\Fexiv}{[\ion{Fe}{xiv}]}

\newcommand{\Neiii}{[\ion{Ne}{iii}]}
\newcommand{\Nev}{[\ion{Ne}{v}]}

\newcommand{\msol}{\mbox{M$_{\odot}$}}

\newcommand{\kms}{\mbox{$\rm{km}~s^{-1}$}}

\newcommand{\dyk}{AT~2018dyk}
\newcommand{\dykhost}{SDSS~J1533+4432}

\newcommand{\marvin}{\textsc{Marvin}}
\newcommand{\tdemass}{\textsc{TDEmass}}

\title[AT~2018dyk: TDE or AGN?]{AT~2018dyk: tidal disruption event or active galactic nucleus?\\ Follow-up observations of an extreme coronal line emitter with the Dark Energy Spectroscopic Instrument}

\author[P.~Clark~et~al.]{Peter~Clark$^{\orcidlink{0000-0002-6576-7400}}$,$^{1,2}$\thanks{E-mail:~P.S.J.Clark@soton.ac.uk}
Joseph~Callow$^{\orcidlink{0000-0002-0804-9533}}$,$^{2}$
Or~Graur$^{\orcidlink{0000-0002-4391-6137}}$,$^{2,3}$
Claire~Greenwell$^{\orcidlink{0000-0002-7719-5809}}$,$^{4}$
Lei~Hu$^{\orcidlink{0000-0001-7201-1938}}$,$^{5}$
Jessica~Aguilar$^{\orcidlink{0000-0003-0822-452X}}$,$^{6}$
\newauthor
Steven~Ahlen$^{\orcidlink{0000-0001-6098-7247}}$,$^{7}$
Davide~Bianchi$^{\orcidlink{0000-0001-9712-0006}}$,$^{8}$
David~Brooks$^{\orcidlink{0000-0002-8458-5047}}$,$^{9}$
Todd~Claybaugh$^{\orcidlink{0000-0002-5024-6987}}$,$^{6}$
Kyle~Dawson$^{\orcidlink{0000-0002-0553-3805}}$,$^{10}$
\newauthor
Axel~de~la~Macorra$^{\orcidlink{0000-0002-1769-1640}}$,$^{11}$
Peter~Doel$^{\orcidlink{0000-0002-6397-4457}}$,$^{9}$
Satya~Gontcho~A~Gontcho$^{\orcidlink{0000-0003-3142-233X}}$,$^{6}$
Gaston~Gutierrez$^{\orcidlink{0000-0003-0825-0517}}$,$^{12}$
\newauthor
Klaus~Honscheid$^{\orcidlink{0000-0002-6550-2023}}$,$^{13,14,15}$
Stephanie~Juneau$^{\orcidlink{0000-0002-0000-2394}}$,$^{16}$
Robert~Kehoe,$^{17}$
Theodore~Kisner$^{\orcidlink{0000-0003-3510-7134}}$,$^{6}$
Anthony~Kremin$^{\orcidlink{0000-0001-6356-7424}}$,$^{6}$
\newauthor
Martin~Landriau$^{\orcidlink{0000-0003-1838-8528}}$,$^{6}$
Laurent~Le~Guillou$^{\orcidlink{0000-0001-7178-8868}}$,$^{18}$
Aaron~Meisner$^{\orcidlink{0000-0002-1125-7384}}$,$^{16}$
Ramon~Miquel$^{\orcidlink{0000-0002-6610-4836}}$,$^{19,20}$
John~Moustakas$^{\orcidlink{0000-0002-2733-4559}}$,$^{21}$
\newauthor
Ignasi~P\'erez-R\`afols$^{\orcidlink{0000-0001-6979-0125}}$,$^{22}$
Eusebio~Sanchez$^{\orcidlink{0000-0002-9646-8198}}$,$^{23}$
Michael~Schubnell$^{\orcidlink{0000-0001-9504-2059}}$,$^{24,25}$
David~Sprayberry,$^{16}$
\newauthor
Gregory~Tarl\'{e}$^{\orcidlink{0000-0003-1704-0781}}$,$^{25}$
Benjamin~A.~Weaver,$^{16}$
and Hu~Zou$^{\orcidlink{0000-0002-6684-3997}}$$^{26}$
\medskip
\\
$^{1}$ School of Physics and Astronomy, University of Southampton, Southampton, SO17 1BJ, UK\\
$^{2}$ Institute of Cosmology and Gravitation, University of Portsmouth, Portsmouth, PO1 3FX, UK \\
$^{3}$ Department of Astrophysics, American Museum of Natural History, New York, NY 10024, USA \\
$^{4}$ Centre for Extragalactic Astronomy, Department of Physics, Durham University, South Road, Durham DH1 3LE, UK\\
$^{5}$ McWilliams Center for Cosmology and Astrophysics, Department of Physics,
Carnegie Mellon University, 5000 Forbes Ave, Pittsburgh, 15213, PA, USA\\
$^{6}$ Lawrence Berkeley National Laboratory, 1 Cyclotron Road, Berkeley, CA 94720, USA\\
$^{7}$ Physics Dept., Boston University, 590 Commonwealth Avenue, Boston, MA 02215, USA\\
$^{8}$ Dipartimento di Fisica ``Aldo Pontremoli'', Universit\`a degli Studi di Milano, Via Celoria 16, I-20133 Milano, Italy\\
$^{9}$ Department of Physics \& Astronomy, University College London, Gower Street, London, WC1E 6BT, UK\\
$^{10}$ Department of Physics and Astronomy, The University of Utah, 115 South 1400 East, Salt Lake City, UT 84112, USA\\
$^{11}$ Instituto de F\'{\i}sica, Universidad Nacional Aut\'{o}noma de M\'{e}xico, Circuito de la Investigaci\'{o}n Cient\'{\i}fica, Ciudad Universitaria, Cd. de M\'{e}xico C.~P.~04510, M\'{e}xico\\
$^{12}$ Fermi National Accelerator Laboratory, PO Box 500, Batavia, IL 60510, USA \\
$^{13}$ Center for Cosmology and AstroParticle Physics, The Ohio State University, 191 West Woodruff Avenue, Columbus, OH 43210, USA\\
$^{14}$ Department of Physics, The Ohio State University, 191 West Woodruff Avenue, Columbus, OH 43210, USA\\
$^{15}$ The Ohio State University, Columbus, 43210 OH, USA\\
$^{16}$ NSF NOIRLab, 950 N. Cherry Ave., Tucson, AZ 85719, USA \\
$^{17}$ Department of Physics, Southern Methodist University, 3215 Daniel Avenue, Dallas, TX 75275, USA\\
$^{18}$ Sorbonne Universit\'{e}, CNRS/IN2P3, Laboratoire de Physique Nucl\'{e}aire et de Hautes Energies (LPNHE), FR-75005 Paris, France\\
$^{19}$ Instituci\'{o} Catalana de Recerca i Estudis Avan\c{c}ats, Passeig de Llu\'{\i}s Companys, 23, 08010 Barcelona, Spain \\
$^{20}$ Institut de F\'{i}sica d’Altes Energies (IFAE), The Barcelona Institute of Science and Technology, Edifici Cn, Campus UAB, 08193, Bellaterra (Barcelona), Spain\\
$^{21}$ Department of Physics and Astronomy, Siena College, 515 Loudon Road, Loudonville, NY 12211, USA\\
$^{22}$ Departament de F\'isica, EEBE, Universitat Polit\`ecnica de Catalunya, c/Eduard Maristany 10, 08930 Barcelona, Spain\\
$^{23}$ CIEMAT, Avenida Complutense 40, E-28040 Madrid, Spain\\
$^{24}$ Department of Physics, University of Michigan, 450 Church Street, Ann Arbor, MI 48109, USA\\
$^{25}$ University of Michigan, 500 S. State Street, Ann Arbor, MI 48109, USA\\
$^{26}$ National Astronomical Observatories, Chinese Academy of Sciences, A20 Datun Rd., Chaoyang District, Beijing, 100012, P.R. China\\
}

\date{Accepted 2025 April 29. Received 2025 April 24; in original form 2025 February 6}

\pubyear{2025}

\begin{document}
\label{firstpage}
\pagerange{\pageref{firstpage}--\pageref{lastpage}}
\maketitle

\begin{abstract}
We present fresh insights into the nature of the tidal disruption event (TDE) candidate AT~2018dyk. AT~2018dyk has sparked a debate in the literature around its classification as either a bona-fide TDE or as an active galactic nucleus (AGN) turn-on state change. A new follow-up spectrum taken with the Dark Energy Spectroscopic Instrument, in combination with host-galaxy analysis using archival SDSS-MaNGA data, supports the identification of AT~2018dyk as a TDE. Specifically, we classify this object as a TDE that occurred within a gas-rich environment, which was responsible for both its mid-infrared (MIR) outburst and development of Fe coronal emission lines. Comparison with the known sample of TDE-linked extreme coronal line emitters (TDE-ECLEs) and other TDEs displaying coronal emission lines (CrL-TDEs) reveals similar characteristics and shared properties. For example, the MIR properties of both groups appear to form a continuum with links to the content and density of the material in their local environments. This includes evidence for a MIR colour-luminosity relationship in TDEs occurring within such gas-rich environments, with those with larger MIR outbursts also exhibiting redder peaks.
\end{abstract}




\section{Introduction}
\label{Sec:Introduction}

The supermassive black holes (SMBHs) at the centres of galaxies can produce numerous astrophysical phenomena, including tidal disruption events (TDEs). These are luminous flaring transients produced by the gravitational shredding of a star that passes too close to its galaxy's SMBH and result in a portion of the star's mass being accreted onto the SMBH with the remaining being ejected from the system \citep{ulmer_1999_FlaresTidalDisruption}. Whilst the specific mechanisms responsible for the associated UV/optical emission remain debated, a combination of the circularisation of the disrupted material to form an accretion disk around the SMBH and collisions within the infalling material streams are likely the primary processes involved (e.g., \citealt{lacy_1982_NatureCentralParsec, rees_1988_TidalDisruptionStars, phinney_1989_CosmicMergerMania, evans_1989_TidalDisruptionStar}). TDEs were first identified in the 1990s within X-ray surveys, the energy regime where the overall peak of TDE emission occurs \citep{bade_1996_DetectionExtremelySoft}. TDEs are now routinely detected by wide-field optical surveys, with subsequent follow-up observations also having detected TDEs at radio and infrared wavelengths --- for example, \cite{alexander_2017_RadioObservationsTidal} and \cite{dou_2017_DiscoveryMidinfraredEcho}, respectively.

TDEs cause rapid increases in the accretion rates of material onto SMBHs and can occur around SMBHs regardless of previous levels of activity. Already `active' SMBHs (i.e., those with ongoing accretion rates sufficient to form accretion disks) are located within active galactic nuclei (AGNs). The presence of an AGN can have a significant effect on the resulting spectrum of its host galaxy, with the energy output of an active SMBH potentially exceeding that of the galaxy's stellar population. The spectra of AGN can display narrow or broad emission lines overlaid on a power law continuum of emission, with the AGN unification model positing the observer's viewing angle of the accretion disk explains the significant observed diversity despite the identical physical processes involved across AGN classifications \citep{antonucci_1993_UnifiedModelsActive,netzer_2015_RevisitingUnifiedModelb}. AGNs are known to display variability on a range of timescales and across the electromagnetic spectrum as a result of instabilities within the feeder accretion disks. For example, a drop in the SMBH's accretion rate will lead to a reduction of an AGN's output, while a temporary increase in the accretion rate can lead to flares in emission (e.g., \citealt{ulrich_1997_VariabilityActiveGalactic}).

As TDEs and AGN variability are both produced through accretion of material onto a SMBH it can be difficult to distinguish between the two. There are, however, differences in properties that can be used to separate the two. TDEs are distinct events rather than ongoing processes and as such have defined beginnings, peaks, and ends, compared to AGN, which will show repeated increases and decreases in luminosity over time. TDEs also tend to produce brighter peaks in luminosity (with outbursts several magnitudes brighter than the quiescent host galaxy flux), whilst the variability in AGN tends to be on the level of 0.1~mag (though some AGN have been seen to also produce bright flares e.g., \citealt{neustadt_2023_Multipleflareschanginglook}).The types and evolution of spectral features exhibited by TDEs and AGNs can also be used to distinguish between the two. Despite these differences, some observed transients remain difficult to definitively classify; one such example, \dyk, is the subject of this work.

\dyk\ was first detected on 2018 May 31 (MJD~58269.31) by the Zwicky Transient Facility \citep[ZTF;][]{bellm_2019_ZwickyTransientFacility} with a \textit{g}-band magnitude of 19.41 \citep{fremling_2023_ZTFTransientDiscovery}. The host galaxy was determined to have a redshift of 0.0367 (Table~\ref{tab:basic_info}). It was classified as a TDE due to the presence of broad Balmer and \Heii\ emission lines \citep{arcavi_2018_FLOYDSClassification2018dyk}, though was noted to have narrow emission lines that had not been previously observed in TDEs and was fainter than most such transients.

\citet{frederick_2019_NewClassChanginglook} found that the host galaxy of \dyk\ transitioned from a low-ionization nuclear emission-line region (LINER) galaxy to a narrow-line Seyfert 1 (NLSy1) galaxy and so classified this event as the turn-on phase of this transition rather than being the result of a TDE, with \dyk\ being included in their sample of `changing look' LINERs (CL-LINERs). LINERs present an additional difficulty for distinguishing between true TDEs and AGN activity, as the source of the emission lines is also ambiguous, with a weak underlying AGN or evolved stellar populations in the nucleus of a galaxy containing a quiescent SMBH both being possible sources. This is in comparison to Seyfert galaxies, whose spectral properties are conclusively the result of an AGN. 

\dyk\ was also included in the sample of Ambiguous Nuclear Transients (ANTs) investigated by \citet{hinkle_2024_MidInfraredEchoesAmbiguous}. ANTs are objects displaying narrow UV/optical emission features and smooth photometric evolution but unclear, likely multiple possible origins \citep[e.g.,][]{wiseman_2023_Multiwavelengthobservationsextraordinary,wiseman_2025_systematicallyselectedsampleluminous}. In that work, \dyk\ was determined to have a high dust covering fraction (0.42~$\pm$~0.15 : a value much larger than typical TDEs) and mid-infrared (MIR) behaviour consistent with a dust reprocessing echo. ANTs were generally found to occur in hosts currently displaying active galactic nucleus (AGN) activity; that recently hosted an AGN; or whose nuclei had a high dust fraction due to other processes, such as significant star formation. \citet{hinkle_2024_MidInfraredEchoesAmbiguous} also noted that whilst the dust covering factors of the ANT sample are similar to those of AGNs, the evolution of the events are unlike AGN outbursts.

\dyk\ was revisited by \citet{huang_2023_AT2018dykRevisitedTidal}, who used further multi-wavelength observations to reassess the changing-look LINER classification. The variability of \dyk\ was found to be shorter lived than typical variability in AGNs, and it showed none of the characteristic trends between colour and brightness that have previously been seen in such objects. The optical light curve was found to follow the characteristic $t^{-\frac{5}{3}}$ power-law expected of TDEs, and a TDE scenario was also able to explain the $\sim140$ d lag between the optical and X-ray light curve peaks. They concluded that \dyk\ was a TDE that had occurred in a LINER. However, they also noted that the mass of the black hole in the host galaxy of \dyk\ was high for a TDE host, with mass estimates based on luminosity, bulge mass, and velocity dispersion measured by \citet{frederick_2019_NewClassChanginglook} ranging from 7.6--8.0~log$_{10}$(\msol), though we note here that the mass estimate provided by the virial method is much lower at 5.5~log$_{10}$(\msol) as derived from the low measured velocity of the H~$\beta$ line. The highest mass estimates are close to the Hills mass \citep{hills_1975_PossiblePowerSource} for Solar type stars - the upper mass limit for SMBHs for which the tidal radius is outside the event horizon and thus is capable of producing a visible TDE. \citet{huang_2023_AT2018dykRevisitedTidal} also noted the presence of a dusty torus around the nucleus of the host galaxy.

Designated WTP~18aamced, \dyk\ was also included in the silver sample of MIR TDEs selected from Near-Earth Object Wide-field Infrared Survey Explorer (NEOWISE) flares by \citet{masterson_2024_NewPopulationMidinfraredselected}. The authors excluded it from their gold sample due to its pre-flare LINER emission features, which may have marked it as an AGN contaminant.

The optical spectra following the discovery of \dyk\ exhibited the high-ionization (coronal) iron emission lines \fspectralline{Fe}{vii}{5722}, \fspectralline{Fe}{vii}{6088}, \fspectralline{Fe}{x}{6376}, and \fspectralline{Fe}{xiv}{5314}. The presence of these lines requires the presence of a extreme UV / soft X-ray continuum, with required photon energies exceeding $\sim$~100~eV, which identifies \dyk\ as an extreme coronal line emitter (ECLE). In objects where these emission lines are transient in nature, they are thought to be produced by TDEs, where X-rays generated by the accretion of the tidally-disrupted star's matter are absorbed and reprocessed by the surrounding interstellar medium (ISM; \citealt{komossa_2008_DiscoverySuperstrongFading,wang_2011_TransientSuperstrongCoronal,wang_2012_EXTREMECORONALLINE, yang_2013_LONGTERMSPECTRALEVOLUTION, hinkle_2024_CoronalLineEmitters,clark_2024_Longtermfollowupobservations, callow_2024_RateExtremeCoronal, callow_2025_rateextremecoronal}). Such coronal lines have also been observed in some Type IIn supernovae (SNe), but the lines were much weaker than those seen in either type of ECLE \citep{smith_2009_CoronalLinesDust, izotov_2009_TYPEIInSUPERNOVA}.

Where these ECLE line signatures do not vary on a timescale of years to decades, such objects are thought to be exotic AGNs in gas-rich environments that produce unusually strong coronal line signatures. High-ionization Fe coronal lines are present in some Seyfert galaxies but only at the level of a few percent the line strength of \fspectralline{O}{iii}{5007} \citep{nagao_2000_HighIonizationNuclearEmissionLine}. As discussed by \citet{frederick_2019_NewClassChanginglook}, the spectra of \dyk\ obtained around peak brightness match the criteria set out by \citet{wang_2012_EXTREMECORONALLINE} to classify it as an ECLE (the [\ion{Fe}{x}] line was stronger than \fspectralline{O}{iii}{5007} where \citealt{wang_2012_EXTREMECORONALLINE} require coronal line strength greater than 20\% that of \fspectralline{O}{iii}{5007}). 

We utilise a new spectrum of \dyk, obtained several years post outburst, obtained by the Dark Energy Spectroscopic Instrument \citep[DESI;][]{levi_2013_DESIExperimentwhitepaper, desicollaboration_2016_DESIExperimentPart, desicollaboration_2016_DESIExperimentParta, collaboration_2022_OverviewInstrumentationDark, collaboration_2024_EarlyDataRelease}, to explore the long-term evolution of \dyk\ and conclusively identify it as a transient generated by a TDE rather than AGN activity.

In this paper, we perform a detailed analysis of \dyk\ and seek to settle the discussion on its origin. In Section~\ref{Sec:ObservationsAndData}, we discuss the spectroscopic, photometric, and comparative datasets constructed for the analysis. In Section~\ref{Sec:Analysis}, we first analyze the optical spectroscopic evolution of \dyk\ and its host galaxy over a timespan of more than 20 yr, placing this evolution in context with other transient populations through comparisons with similar events. We then move on to an analysis of its MIR evolution and include a detailed study of its host galaxy's properties, with a focus on determining the source of its LINER emission features. In Section~\ref{Sec:Discussion}, we discuss the implications of the evolution of \dyk's oxygen emission behaviour, as well as the MIR outburst properties of the transient in the context of ECLEs, TDEs and CL-LINERs. Finally, in Section~\ref{Sec:Conclusions} we conclude that the evolution of \dyk\ is well matched by variable or ECLEs assumed to be caused by TDEs (TDE-ECLEs) though on a faster evolutionary timescale \citep{clark_2024_Longtermfollowupobservations, callow_2024_RateExtremeCoronal}. To minimise potential confusion between `coronal line' and `changing look' objects, we abbreviate `coronal line' to `CrL' and `changing look' to `CL'.

Throughout, we assume a Hubble-Lema\^ itre constant H$_0 = 73$ \kms\ Mpc$^{-1}$ and adopt a standard cosmological model with $\Omega_M=0.27$ and $\Omega_{\Lambda}=0.73$.

\begin{table}
\caption{Properties of AT~2018dyk.}
\label{tab:basic_info}
\begin{adjustbox}{width=1\columnwidth}
\begin{tabular}{lll}
\hline
\textbf{Parameter} & \textbf{Value} & \textbf{Source} \\ \hline
Host Galaxy & SDSS J153308.01+443208.4 &\\
Right Ascension & 233.2833955 & $1$\\
Declination & +44.5356122 & $1$\\
Redshift & 0.0367\textsuperscript{\textdagger} & $1$\\ 
E(B-V) - Milky Way & 0.0164 mag & $2$\\ 
Alternative identifiers & ZTF18aajupnt, WTP~18aamced & \\
\hline
\end{tabular}
\end{adjustbox}
\begin{flushleft}
\textbf{Sources:}\\
$1$: Transient Name Server (TNS): \url{https://www.wis-tns.org/object/2018dyk}\\
$2$: IRSA Galactic Dust Reddening and Extinction: \url{https://irsa.ipac.caltech.edu/applications/DUST/}\\
\smallskip
\textsuperscript{\textdagger}Retrieved from the TNS and consistent with the pre-transient SDSS spectrum. The redshift value returned by the DESI redshift measuring pipeline \textsc{Redrock} (\citealt{guy_2023_SpectroscopicDataProcessing}, S. J. Bailey et al. 2025, in prep.) returns a slightly different redshift value of 0.0368. We believe this offset is due to the increasing strength of the anomalously redshifted \Oiii\ lines (see Section~\ref{subsec:emission_line_behaviour}) at the time of the DESI spectrum and as such we adopt the value determined before these lines strengthened post outburst.
\end{flushleft}
\end{table}
\raggedbottom

\section{Observations and Data Reduction}
\label{Sec:ObservationsAndData}

\begin{figure*}
    \centering
    \includegraphics[width=\textwidth]{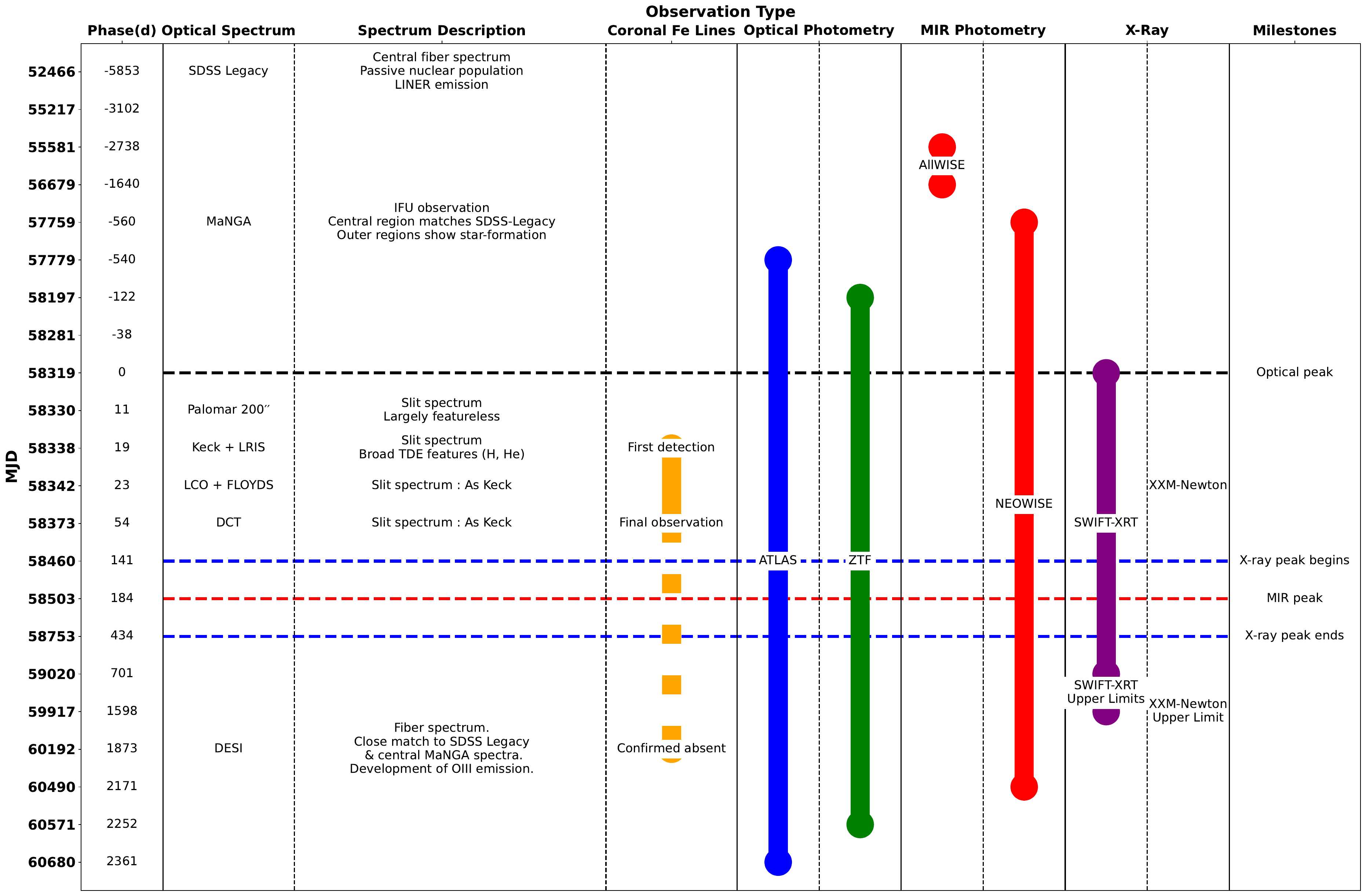}
    \caption{A visual summary of the available observations of \dyk. Additional optical spectra displaying typical TDE-like evolution and broad features exist between the LCOGT+FLOYDS spectrum and the DCT spectrum. All show similar coronal and other emission features. As these have not been made available publicly, we do not utilise them further in this work. Two observations of HST UV spectra were also obtained at phases of +182 and +226~d. We do not make use of these spectra within this work as there are no comparable observations for the other objects in the sample. The additional optical and UV spectra can be viewed in \citet{frederick_2019_NewClassChanginglook}.}
    \label{fig:Observations_Summary}
\end{figure*}

Here we describe the collection and reduction of the spectroscopic and photometric datasets used in this work. Additionally, a summary of all the various observations of \dyk\ across all wavelength regimes is presented visually in Fig~\ref{fig:Observations_Summary}.

\subsection{Optical spectroscopy}
\label{subsec:Optical_Spectroscopy}

The DESI spectrum of \dyk\ presented in this work (Fig.~\ref{fig:18dyk_spec_evolution}) was obtained on 2023 May 05 (MJD~60192) as part of the Bright Galaxy survey \citep[BGS;][]{hahn_2023_DESIBrightGalaxy} during main survey operations \citep{schlafly_2023_SurveyOperationsDark}. The spectrum was processed by the custom DESI spectroscopic pipeline, which includes a full suite of processing and correction steps to provide fully flux- and wavelength-calibrated spectra \citep{guy_2023_SpectroscopicDataProcessing}. DESI itself is designed primarily as a cosmological experiment, and whilst not the focus of this work, Data Release 1 provides a range of state-of-the-art cosmological analyses, including two-point clustering measurements and validation \citep{desicollaboration_2024_DESI2024II}, baryon-acoustic oscillation (BAO) measurements from galaxies and quasars \citep{desicollaboration_2024_DESI2024III}, and from the Lya forest\citep{desicollaboration_2024_DESI2024IV}, as well as a full-shape study of galaxies and quasars \citep{desicollaboration_2024_DESI2024FullShape}. There are Cosmological results from the BAO measurements \citep{desicollaboration_2024_DESI2024VI} and the full-shape analysis \citep{desicollaboration_2024_DESI2024VII}, as well as constrains on primordial non-gaussianities \citep{desicollaboration_2024_DESI2024VIIa}.

We compare our DESI spectrum to the archival Sloan Digital Sky Survey \citep[SDSS;][]{york_2000_SloanDigitalSky} Legacy spectrum of the host of AT~2018dyk, \dykhost, retrieved from the 17th data release \citep[DR17;][]{abdurrouf_2022_SeventeenthDataRelease}. This spectrum was obtained on 2002 July 11 (MJD 52466), providing a separation of 7726~d (21.15~yr) between the earliest archival spectrum of this object and the most recent DESI spectroscopic observation. This timescale provides a long baseline to compare the prior behaviour of the transient's host galaxy to the outburst itself, which returned to photometric quiescence over $\sim$~2~yr. This is comparable to the wider population of TDEs, which primarily evolve on similar timescales \citep{charalampopoulos_2022_detailedspectroscopicstudya}, with the longer term presence of UV-optical plateaus \citep{vanvelzen_2019_LatetimeUVObservations, mummery_2024_Fundamentalscalingrelationships}.

The host galaxy of \dyk\ was observed as part of the SDSS-IV Mapping Nearby Galaxies at Apache Point Observatory survey \citep[MaNGA;][]{smee_2013_MultiobjectFiberfedSpectrographs, bundy_2015_OverviewSDSSIVMaNGA, yan_2016_SDSSIVMaNGAIFS}. This observation was obtained on 2017 January 6 (MJD~57759), 5293~d following the original spectrum and 560~d prior to the optical peak of \dyk. The MaNGA spectrum provides spatially resolved IFU spectroscopy of \dykhost. In this work, we use the reduced data processed by the MaNGA data analysis pipeline \citep{belfiore_2019_DataAnalysisPipeline, westfall_2019_DataAnalysisPipeline}, accessed via the \marvin\ toolkit \citep{cherinka_2019_marvintoolkit}.

Finally, we retrieve two publicly-available spectra from the Weizmann Interactive Supernova Data Repository (WISeREP) online archive \citep{yaron_2012_WISeREPInteractiveSupernova}\footnote{\url{https://wiserep.weizmann.ac.il/}} obtained close to the optical peak of the transient. The first spectrum was obtained with the low-resolution imaging spectrometer \citep[LRIS;][]{oke_1995_KeckLowResolutionImaging} on the Keck telescope by the ZTF team on 2018 August 8 (MJD~58338), 19~d following the observed peak. The second spectrum was obtained with the FLOYDS spectrograph\footnote{\url{https://lco.global/observatory/instruments/floyds}} on the 2-m Las Cumbres Observatory telescope on Haleakala, Hawai'i, as part of the `Transients in Galaxy Centers' observational program (PI: I. Arcavi) on 2018 August 12 (MJD~58342), 23~d following maximum light. A summary of the spectroscopic data used in this work is given in Table~\ref{tab:spec_data}.

An additional set of optical spectra was obtained in the phase range +11--54~d and described by \citet{frederick_2019_NewClassChanginglook}. Spectra observed prior to the +19~d Keck+LRIS spectrum are similar to the the archival SDSS Legacy and MaNGA spectra with the exception of increased Balmer emission. Spectra following the +19~d Keck+LRIS spectrum in this set displayed similar emission features (including the coronal lines) and minimal evolution. Two \textit{Hubble Space Telescope} (HST) UV spectra were also obtained as part of the follow-up program conducted by \citet{frederick_2019_NewClassChanginglook} at phases of +182 and +226~d. As we have no additional observations covering this wavelength regime, or additional comparison spectra from other similar objects, we do not explore this wavelength regime further in this work.

\begin{table*}
\centering
\caption{Spectroscopic observations of AT~2018dyk and its host galaxy (\dykhost) used in this work.}
\label{tab:spec_data}
\begin{tabular}{lccccc}
\hline
Source & Type & MJD & Phase (d) $^1$ & R $^2$ & Source \\ \hline
SDSS Legacy & Fibre & 52466 & -5853 &1500~--~2500 & SDSS DR17 \\
MaNGA & IFU & 57759 & -560 &2000 & SDSS DR17 \\
Keck+LRIS & Slit & 58338 & +19 & $\sim$ 4760 & WISeREP \\
LCOGT+FLOYDS & Slit & 58342 & +23 & 400~--~700 & WISeREP \\
DESI & Fibre & 60192 & +1873 & 1500~--~4000 & This work \\ \hline
\end{tabular}
\begin{flushleft}
$^1$ Here and throughout the paper, phase is quoted relative to the optical peak (MJD~58319) as measured by \citet{huang_2023_AT2018dykRevisitedTidal} in ZTF observations. \\
$^2$ Spectral resolving power. Quoted based on instrumental specifications. \\
\end{flushleft}
\end{table*}
\raggedbottom

\subsection{Photometry}
\label{subsec:Photometry}

In addition to spectroscopy, we make use of photometric observations from both optical and infrared surveys. These are summarised in Table~\ref{tab:Phot_summary}. In the optical regime we use observations taken by the
Asteroid Terrestrial-impact Last Alert System \citep[ATLAS;][]{tonry_2018_ATLASHighcadenceAllsky, smith_2020_DesignOperationATLAS} and Zwicky Transient Facility \citep[ZTF;][]{bellm_2014_ZwickyTransientFacility, bellm_2019_ZwickyTransientFacility}.

ATLAS data were retrieved using the ATLAS forced-photometry server \citep{shingles_2021_ReleaseATLASForced}.\footnote{\url{https://fallingstar-data.com/forcedphot/}} ATLAS uses two broad-band filters: `cyan' (\textit{c}; approximately equivalent to \textit{g} + \textit{r}) and `orange' (\textit{o}; approximately equivalent to \textit{r} + \textit{i}). ATLAS observations are available over the MJD range (following the removal of some early unreliable observations) 57779~--~60680 and thus cover a time range prior to, during, and post outburst and were processed using a modified version of \textsc{plot\_atlas\_fp.py} \citep{young_2024_plot_atlas_fppy}.

ZTF observations were made using the \textit{gri} filters, and retrieved using the ZTF Forced Photometry Service\citep[ZFPS;][]{ masci_2023_NewForcedPhotometry}. These observations cover an MJD range of 58197--60571. 

In Section~\ref{subsec:Transient_Search}, we use both ATLAS and ZTF photometry to look for post-transient variability in the host galaxy, such as repeating outburst behaviour. 

In an identical manner to \citet{clark_2024_Longtermfollowupobservations}, we retrieve and process the available MIR photometry of \dyk\ and its host from both the AllWISE Data Release and the final NEOWISE Reactivation Release (NEOWISE-R) from the through the NASA/IPAC infrared science archive (IRSA).\footnote{\url{https://irsa.ipac.caltech.edu/}}. This processing includes the removal of any individual observations suffering from potential problems (such as contamination by excess moonlight, or being obtained when the spacecraft was close to the south Atlantic anomaly) and combining the individual observations from each visit to provide a single weighted average. The MIR behaviour of AT~2018dyk was previously investigated by \citet{huang_2023_AT2018dykRevisitedTidal} and \citet{masterson_2024_NewPopulationMidinfraredselected}. In this work, we compare the MIR behaviour of \dyk\ to that of TDEs and ECLEs (Section~\ref{subsec:MIR_Evolution}) and include more recent observations available in the final NEOWISE data release.

Throughout this work, apparent magnitudes are given as observed. For absolute magnitudes, a correction for Milky Way extinction was applied using the appropriate photometric extinction coefficients which, unless specified otherwise, were retrieved from \citet{schlafly_2011_MeasuringReddeningSloan}. To match the preferred extinction parameters of \citet{schlafly_2011_MeasuringReddeningSloan}, we use the extinction law of \citet{fitzpatrick_1999_CorrectingEffectsInterstellar} and assume $R_{V} = 3.1$. 

\begin{table*}
\caption{Photometric observations used in this work}
\label{tab:Phot_summary}
\begin{tabular}{llllll}
\hline
Survey & MJD & Phase (d) $^1$ & Filters & Reference\\ \hline
\textbf{Optical}&&&\\
ATLAS & 57779~--~60680 & -540~--~2361 & \textit{c}, \textit{o} $^2$
&\citet{tonry_2018_ATLASHighcadenceAllsky, shingles_2021_ReleaseATLASForced}\\
ZTF & 58197~--~60571 & -122~--~2252 & \textit{g}, \textit{r}, \textit{i}
&\citet{bellm_2019_ZwickyTransientFacility, masci_2023_NewForcedPhotometry}\\
\\
\textbf{MIR}&&\\
AllWISE & 55217~--~55581 & -3102~--~-2738 & \textit{W}1, \textit{W}2, \textit{W}3&\citet{wright_2010_WIDEFIELDINFRAREDSURVEY}\\
NEOWISE & 56679~--~60490 & -1640~--~2171& \textit{W}1, \textit{W}2&\citet{mainzer_2014_INITIALPERFORMANCENEOWISE}\\
\hline
\end{tabular}
\begin{flushleft}
$^1$ Phase quoted relative to ZTF optical peak as measured by \citet{huang_2023_AT2018dykRevisitedTidal} : MJD~58319. \\
$^2$ ATLAS observations were made using two broad-band filters; \textit{c} (cyan) is approximately equivalent to \textit{g} + \textit{r} and \textit{o} (orange) is roughly \textit{r} + \textit{i}. \\
\end{flushleft}
\end{table*}
\raggedbottom

\subsection{Comparison Objects}
\label{subsec:Comparison_Objects}

Whilst this work focuses extensively on the behaviour and properties of \dyk\ itself, it also makes use of comparisons to other ECLEs and traditionally identified TDEs that are known to have displayed Fe coronal emission. 

In comparisons between AT~2018dyk and other objects, this paper makes a distinction between coronal-line TDEs (CrL-TDEs), which were spectroscopically confirmed as TDEs prior to the development of Fe coronal emission features, and TDE-linked ECLEs (TDE-ECLEs), which were identified initially through their coronal emission lines after the traditional UV/optical signatures had faded. Physically, however, these objects are fundamentally identical, with TDE-ECLEs representing a sub-population with long durations that happened to be observed at a late stage in their evolution.

Additionally, several objects show evidence for multiple epochs of TDE associated MIR emission, either through distinct outburst episodes or through extended and multi-peaked MIR rises. We refer to these as `multi-epoch' events and include them as comparison objects only if the MIR light curve suggests one extended period of activity. All of the selected comparison objects are summarised in Table~\ref{tab:cl_transient_comparison_object_summary_table}. 

To better place these objects in context with the wider transient population, we also make use of the WISE MIR photometry for the optically selected TDE TDE~2019azh, which displayed a weak MIR outburst but no Fe coronal line emission \citep{faris_2024_LightcurveStructureHa} along with the two known AGN-ECLEs to compare and contrast their MIR behaviour. An optical spectral comparison between \dyk\ and the AGN-ECLEs has also been conducted.

Furthermore, we also compare the MIR behaviour of \dyk\ to the other members of the CL-LINER classification of \citet{frederick_2019_NewClassChanginglook}.

\begin{table*}
\caption{Summary of TDE-ECLEs, CrL-TDEs and other objects used as comparison objects to \dyk.}
\label{tab:cl_transient_comparison_object_summary_table}
\begin{tabular}{lllp{5cm}}
\hline
\textbf{Object} & \textbf{Classification} & \textbf{Data Used} & \textbf{References for coronal line classification} \\ \hline
SDSS J0748+4712 & TDE-ECLE & MIR Photometry, SDSS Spectrum & \citet{wang_2011_TransientSuperstrongCoronal, wang_2012_EXTREMECORONALLINE, yang_2013_LONGTERMSPECTRALEVOLUTION, dou_2016_LONGFADINGMIDINFRARED, clark_2024_Longtermfollowupobservations} \\ 
SDSS J0952+2143 & TDE-ECLE & MIR Photometry, SDSS Spectrum & \citet{komossa_2008_DiscoverySuperstrongFading, komossa_2009_NTTSpitzerChandra, wang_2012_EXTREMECORONALLINE, yang_2013_LONGTERMSPECTRALEVOLUTION, dou_2016_LONGFADINGMIDINFRARED, clark_2024_Longtermfollowupobservations} \\ 
SDSS J1241+4426 & TDE-ECLE & MIR Photometry, SDSS Spectrum & \citet{wang_2012_EXTREMECORONALLINE, yang_2013_LONGTERMSPECTRALEVOLUTION, dou_2016_LONGFADINGMIDINFRARED, clark_2024_Longtermfollowupobservations} \\ 
SDSS J1342+0530& TDE-ECLE & MIR Photometry, SDSS Spectrum & \citet{wang_2012_EXTREMECORONALLINE, yang_2013_LONGTERMSPECTRALEVOLUTION, dou_2016_LONGFADINGMIDINFRARED, clark_2024_Longtermfollowupobservations} \\ 
SDSS J1350+2916 & TDE-ECLE & MIR Photometry, SDSS Spectrum & \citet{wang_2012_EXTREMECORONALLINE, yang_2013_LONGTERMSPECTRALEVOLUTION, dou_2016_LONGFADINGMIDINFRARED, clark_2024_Longtermfollowupobservations} \\ 
\hline
AT~2017gge & CrL-TDE & MIR Photometry, Optical Spectra & \citet{onori_2022_NuclearTransient2017gge} \\
AT~2018gn $^1$ & CrL-TDE & MIR Photometry & \citet{wang_2024_ASASSN18apDustyTidal}\\
AT~2018bcb & CrL-TDE & MIR Photometry & \citet{neustadt_2020_TDENotTDE}\\
AT~2021dms& CrL-TDE $^2$ & MIR Photometry & \citet{hinkle_2024_CoronalLineEmitters}\\
TDE~2021qth& CrL-TDE & - $^3$ & \citet{yao_2023_TidalDisruptionEvent}\\
AT~2021acak& CrL-TDE $^4$ & MIR Photometry & \citet{li_2023_AT2021acakCandidateTidal}\\
TDE~2022fpx & CrL-TDE & MIR Photometry & \citet{koljonen_2024_ExtremeCoronalLine}\\
TDE~2022upj $^5$ & CrL-TDE & MIR Photometry & \citet{newsome_2022_2022upjZTF22abegjtxDiscovery, newsome_2024_MappingInner01} \\
TDE~2024mvz & CrL-TDE & - $^6$ & \citet{shitrit_2024_ePESSTOTransientClassification}\\
SDSS J0113+0937 & CrL-TDE & - $^7$ & \citet{callow_2025_rateextremecoronal}\\ 
\hline
AT~2019avd & Multi-epoch CrL-TDE / CrL-AGN $^7$ & MIR Photometry & \citet{malyali_2021_2019avdNovelAdditiona} \\ 
TDE~2019qiz & Multi-epoch CrL-TDE $^8$ & MIR Photometry & \citet{short_2023_DelayedAppearanceEvolution} \\ 
AT~2019aalc & Multi-epoch CrL-TDE $^9$ & - & \citet{veres_2024_BackdeadAT2019aalc}\\
TDE~2020vdq& Multi-epoch CrL-TDE? $^{10}$ & - & \citet{somalwar_2023_VLASSTidalDisruption} \\
\hline
VT~J154843.06+220812.6 $^{11}$ & CrL-TDE / CrL-AGN & MIR Photometry & \citet{somalwar_2022_NascentMilliquasarVT} \\
\hline
SDSS~J0938+1353 & AGN-ECLE & SDSS Spectrum, MIR Photometry & \citet{wang_2012_EXTREMECORONALLINE, yang_2013_LONGTERMSPECTRALEVOLUTION, clark_2024_Longtermfollowupobservations} \\ 
SDSS~J1055+5637& AGN-ECLE & SDSS Spectrum, MIR Photometry & \citet{wang_2012_EXTREMECORONALLINE, yang_2013_LONGTERMSPECTRALEVOLUTION, clark_2024_Longtermfollowupobservations} \\ 
\hline
TDE~2019azh & NonCrL-TDE & MIR Photometry & \citet{faris_2024_LightcurveStructureHa} \\ \hline 
iPTF16bco & CL-LINER & MIR Photometry & \citet{frederick_2019_NewClassChanginglook} \\ 
AT~2018aij (ZTF18aahiqfi) $^{12}$ & CL-LINER & MIR Photometry & \citet{frederick_2019_NewClassChanginglook} \\
AT~2018gkr (ZTF18aaabltn) & CL-LINER & MIR Photometry & \citet{frederick_2019_NewClassChanginglook} \\
AT~2018ivp (ZTF18aaidlyq) & CL-LINER & MIR Photometry & \citet{frederick_2019_NewClassChanginglook} \\
AT~2018lnh (ZTF18aasszwr) & CL-LINER & MIR Photometry & \citet{frederick_2019_NewClassChanginglook} \\
ZTF18aasuray & CL-LINER & MIR Photometry & \citet{frederick_2019_NewClassChanginglook} \\
\hline
\end{tabular}
\begin{flushleft}

$^1$ Has a `SN' rather than an `AT' or `TDE' designation on the TNS due to an initial Type II SN classification by \citet{falco_2018_ASASSNTransientClassification}. We refer to it by an `AT' designation here to avoid confusion.\\
$^2$ Final epoch of MIR photometry shows re-brightening in both bands. MIR colour evolution at this epoch remains consistent with previous observations. As such, we treat this object as a single epoch event for the purposes of this analysis. Additional observations are required to determine if this re-brightening is a single epoch outlier or a longer term trend in evolution.\\
$^3$ The host galaxy of TDE~2021qth is in very close proximity to another galaxy, making reliable photometric separation at the resolution of WISE very difficult. As such, we exclude it from the photometric comparison given the significant contamination that would result.\\
$^4$ Host galaxy also likely hosts an AGN.\\
$^5$ Now known to display quasi-periodic x-ray eruptions (QPEs) as described by \citet{chakraborty_2025_DiscoveryQuasiperiodicEruptions}.\\
$^6$ Transient occurred too close to the conclusion of the NEOWISE-R mission for useful inclusion in the MIR analysis.\\
$^7$ Excluded from the MIR comparisons due a poorly constrained time of MIR peak.\\
$^8$ Displays an extended, multi-peaked MIR rise. TDE~2019qiz also displays QPEs \citep{nicholl_2024_QuasiperiodicXrayEruptions}. \\
$^9$ Two overlapping MIR outbursts. As the transient did not return to MIR quiescence between the outbursts, this object is excluded from the MIR comparisons. This transient occurs within a galaxy hosting an AGN. \\
$^{10}$ Two distinct flaring epochs, with only the first displaying coronal emission. As this is the only such object to show multiple possible TDE-linked outbursts with differing behaviour, we do not directly compare TDE~2020vdq to \dyk.\\
$^{11}$ Referred to as VT~J1548 in the remainder of this work. Based on a lack of pre-outburst AGN activity and similarity in MIR evolution to other CrL-TDEs, we treat VT~J1548 as a CrL-TDE in our analysis but note its classification uncertainty in upcoming plots.\\
$^{12}$ Recent observations show a potential second MIR outburst.\\
Further information on each object for which data has been used in this work, including coordinates and redshifts, are given in Table~\ref{Tab:Object_Summary_Info}.
\end{flushleft}
\end{table*}
\raggedbottom

\section{Analysis}
\label{Sec:Analysis}

First, we summarise the optical spectroscopic evolution of \dyk\ with a focus on the evolution displayed in our new DESI spectrum (Section~\ref{subsec:Spectroscopic_Evolution}). Next, we compare the optical spectra of \dyk\ at various phases to a range of other astrophysical objects to provide context on the behaviour of \dyk\ before, during, and post outburst (Section~\ref{subsec:Spectral_Comparisons}). We then conduct a search for signs of more recent transient activity in optical photometric observations (Section~\ref{subsec:Transient_Search}). Finally, we analyse the MIR behaviour of \dyk\ including a comparative analysis to other transients) before completing the analysis with an in-depth study of the properties of its host galaxy (Section~\ref{subsec:MIR_Evolution}).

\subsection{Spectroscopic evolution}
\label{subsec:Spectroscopic_Evolution}

\begin{figure*}
    \centering
    \includegraphics[width=\textwidth]{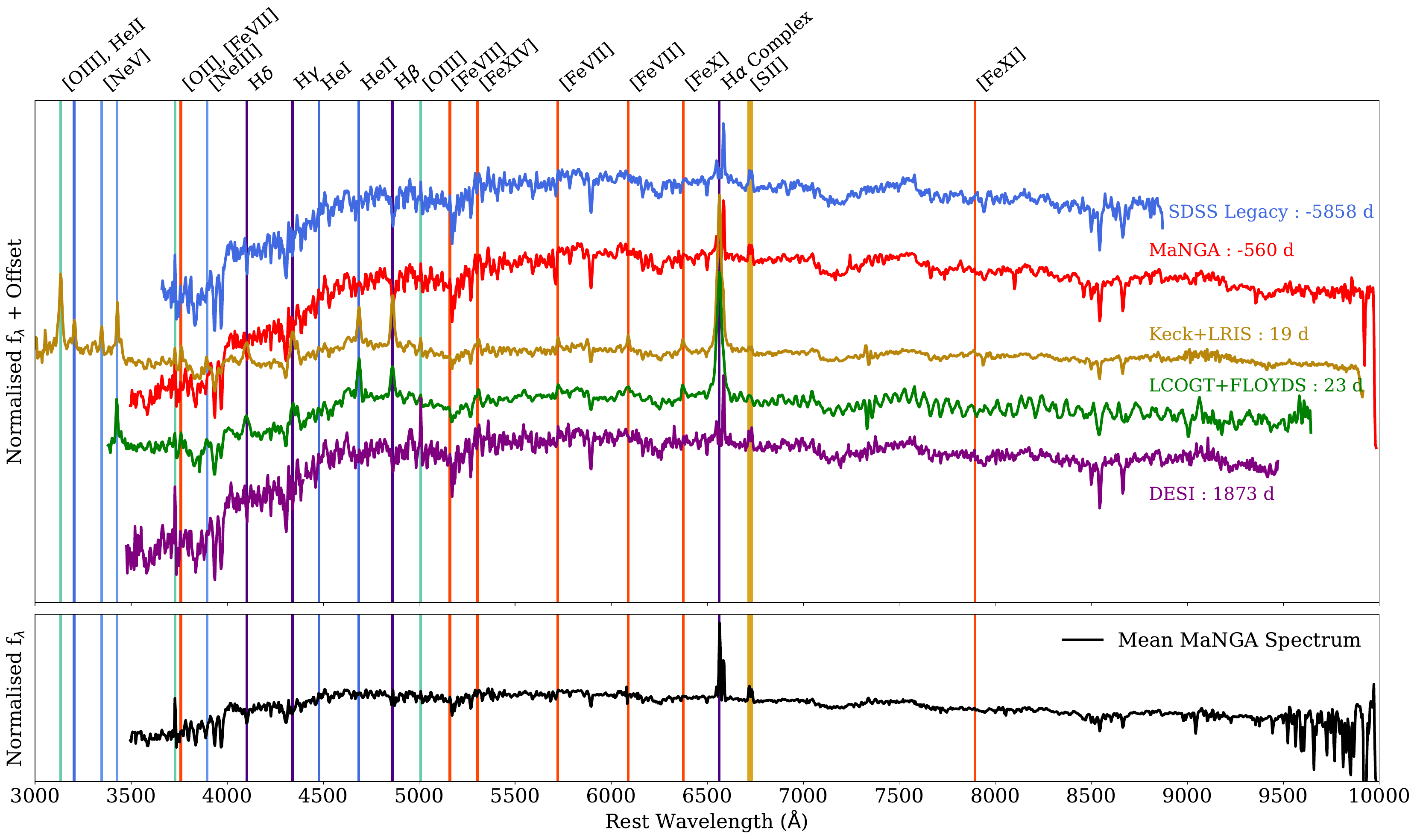}
    \caption{Spectroscopic evolution of \dyk\ with observations from 16~yr before and more than 5~yr following optical peak. \textit{Top Panel:} Comparison between the normalised fibre and long slit spectra of \dyk\ showing the emergence and subsequent fading of broad H and He features along with Fe coronal lines. The MaNGA spectrum shown here is from the local region of \dyk\ and consists of one spaxel (31,31). The phase of each spectrum relative to the peak of the optical outburst is indicated. \textit{Bottom Panel:} Normalised mean MaNGA spectrum covering all spaxels, showing the starforming nature of \dykhost\ as a whole. In all cases, spectra have been Gaussian smoothed ($\sigma$ = 1) and rebinned to a resolution of 4~\AA. For all plots in this work, solid vertical lines are used to indicate the location of important spectral features. Additionally, these have a shared colour scheme where lines from the same element are displayed using the same colour, e.g., orange for all Fe emission lines.}
    \label{fig:18dyk_spec_evolution}
\end{figure*}

The pre-outburst SDSS Legacy optical spectrum was obtained just over 16~yr prior to the optical peak of \dyk\ and displays the typical spectral features of a largely passive stellar population with LINER emission line diagnostics. The SDSS Legacy spectrum was obtained through a nuclear targeted fibre 3~arcsec in diameter. It covers the central and bar regions of \dykhost\ but does not capture the outer starforming regions, which when explored using the MaNGA data show much bluer spectra with the prominent H$\alpha$ emission expected from starforming regions. Indeed, the mean MaNGA spectrum produced from a combination of all observed spaxels (covering a large fraction of \dykhost, with each having a width of 0.5~arsec) has the expected strong H$\alpha$ emission and non-LINER emission line diagnostics of a normal starforming galaxy (see bottom panel of Fig.~\ref{fig:18dyk_spec_evolution} and Section~\ref{subsec:Host_Analysis}). The local environment of \dyk, as revealed by the central MaNGA spaxel coincident with its location (MJD~57759), consists of an older - and redder - stellar population with little active star-formation but prominent $\Nii$ emission lines observed in LINERs.

Following the outburst, broad H and He typical of optically selected TDEs developed, as seen in the Keck and LCOGT spectra included in Fig.~\ref{fig:18dyk_spec_evolution}. As noted by \citet{frederick_2019_NewClassChanginglook}, in addition to these typical features, Fe coronal lines had emerged by the time of the Keck spectrum (MJD~58338, 19~d post optical peak) and persisted until at least 2018 September 12 (MJD~58373, 54~d post peak for a minimum duration of 35~d) as observed in the Discovery Channel Telescope (DCT) spectrum presented by \citet{frederick_2019_NewClassChanginglook}. However, given the slow evolution of the coronal lines observed in other TDEs, it is reasonable to assume their true duration is longer than this lower limit; see, for example, AT~2017gge \citep{onori_2022_NuclearTransient2017gge} and the original TDE-ECLE sample \citep{wang_2012_EXTREMECORONALLINE, yang_2013_LONGTERMSPECTRALEVOLUTION, clark_2024_Longtermfollowupobservations}. Additionally, since all Fe coronal lines had faded prior to the DESI observation, we can derive an (albeit loose) upper limit on the duration of the coronal line emission of <1819 d (time between the DCT and DESI spectra). In summary, Fe coronal line emission in \dyk\ commenced 19~d following the peak of optical emission and persisted for between 35 and 1819~d. In addition to the Fe coronal lines, coronal emission lines from \Nev\ are present in the Keck, LCOGT, and DCT spectra. These lines have very similar energy requirements to \Fevii\ and independently confirm the existence of a high energy continuum. Unfortunately neither the pre-outburst nor the DESI spectra extend blueward enough for a comparative analysis, though a similar evolution to \Fevii\ is expected.

As seen in other CrL-TDEs, the start of an X-ray flaring event was also observed prior to the emergence of the coronal lines. This is consistent with the coronal lines resulting from X-ray reprocessing by material close to the SMBH Despite the small number of CrL-TDEs currently known (15 at time of writing based on public reports of coronal line emission in classified TDEs), there is already a significant range in observed delays between the optical peak and the ensuing X-ray emission. In the case of \dyk, this X-ray emission peaked $\sim$~140~d following the optical peak as reported by \citet{huang_2023_AT2018dykRevisitedTidal}. Whilst the X-ray peak is not covered by the available optical spectra, the Keck and LCOGT spectra were taken whilst the X-ray emission was on the rise and the high ionisation potential of the Fe coronal lines requires the presence of a strong X-ray continuum which, given the previous X-ray non-detections, was not previously present in this object.

\begin{figure*}
    \centering
    \includegraphics[width=0.95\textwidth]{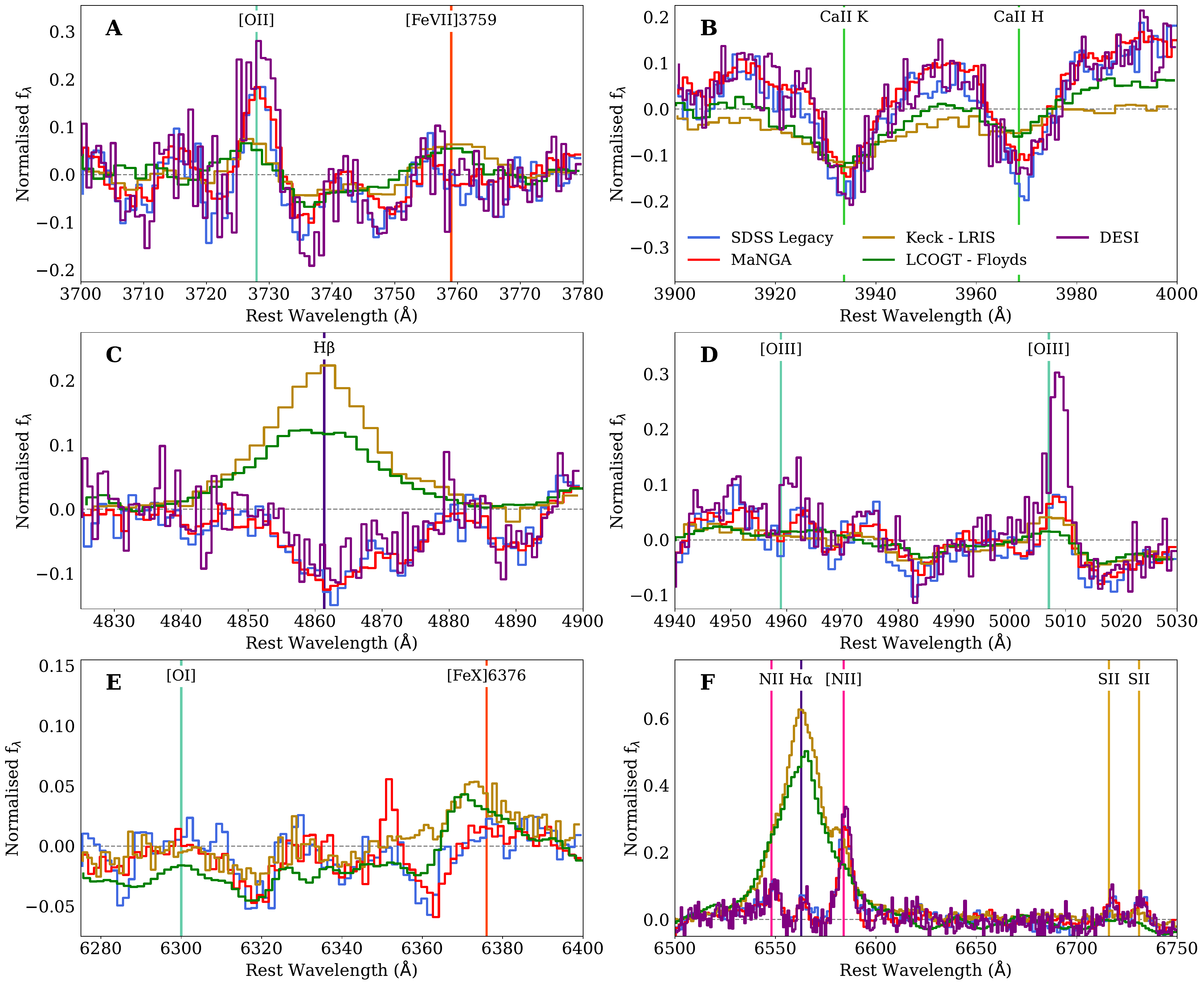}
    \caption{Emission line evolution of \dyk\ across the full range of spectral observations. Spectral normalisation and scaling is shared along rows with all spectra first normalised and then scaled to the mean of a local region clear of spectral features: \textit{A\&B} scaled to the 3650~--~3700~\AA\ region, \textit{C\&D} scaled to the 4900~--~4950~\AA\ region, and \textit{E\&F} scaled to the 6400~--~6500~\AA\ region. All selected scaling regions are free from spectral features. Additionally, no rebinning or smoothing has been applied. A dashed horizontal line marking the scaled continuum level is included for reference.
    \textit{A:} Spectral evolution of the \Oii~$\lambda$~3728~\AA\ line region including the weak coronal line \Fevii~$\lambda$~3759~\AA. The increase in the \Oii~$\lambda$~3728~\AA\ line strength post outburst relative to spectra obtained both pre and during the outburst is clear. \textit{B:} Spectral evolution of the \CaII\ H \& K absorption features. 
    \textit{C:} Spectral evolution of the H$\beta$ line region. Transient and broad H$\beta$ emission generated by the TDE is clearly visible. This feature has completely faded away, with the most recent DESI spectrum now matching the pre-outburst SDSS Legacy and local MaNGA spectra.
    \textit{D:} Spectral evolution of the \Oiii~$\lambda$~4959~\AA\ and \Oiii~$\lambda$~5007~\AA\ lines. Emergence of the \Oiii~$\lambda$~4959~\AA\ and the strengthening of the narrow\Oiii~$\lambda$~5007~\AA\ emission line in the latest DESI spectrum is observed. This emission, whilst slightly redshifted from the zero velocity position, has not displayed velocity evolution across the available observations (see Fig.~\ref{fig:18dyk_oxygen_velocity_comparison} and Section~\ref{subsec:emission_line_behaviour}).
    \textit{E:} Spectral evolution of the \Oiii~$\lambda$~6300~\AA\ emission feature (which remains undetected at all phases) and the coronal \Fex~$\lambda$~6376~\AA\ emission line which is only present in the Keck+LRIS and LCOGT+FLOYDS spectra. The DESI spectrum is excluded from this subplot as its lower SNR would otherwise mask the \Fex~$\lambda$~6376~\AA\ feature. As with other regions, the DESI spectrum closely matches the SDSS-Legacy and MaNGA spectra.
    \textit{F:} Spectral evolution of the H$\alpha$ complex region. As with H$\beta$, pronounced evolution is seen in the development and subsequent fading of the broad H feature.
    \label{fig:18dyk_zoom_spec_evolution}
    }
\end{figure*}

The most recent spectrum of \dyk\ is the DESI spectrum obtained more than 5~yr post optical peak (MJD~60192). This spectrum now closely resembles the archival pre-outburst SDSS spectrum (Fig.~\ref{fig:18dyk_spec_evolution}). The broad H and He features and coronal lines are no longer present, and the original continuum shape is now restored. In Fig.~\ref{fig:18dyk_zoom_spec_evolution}, we show the evolution of the H$\alpha$ complex and \Oiii~$\lambda$~5007~\AA\ line region (with local normalization and no smoothing or rebinning). Whilst H$\alpha$, H$\beta$, \Nii, and \Sii\ emission lines have now returned to their quiescent states, the same cannot be said for \Oiii~$\lambda$~5007~\AA, which has significantly strengthened compared to the pre-outburst SDSS spectrum. This evolution is also seen in the behaviour of the \Oii~$\lambda$~3727~\AA\ emission line. We discuss this evolution in Section~\ref{subsec:emission_line_behaviour}.

\begin{figure*}
    \centering
    \includegraphics[width=\textwidth]{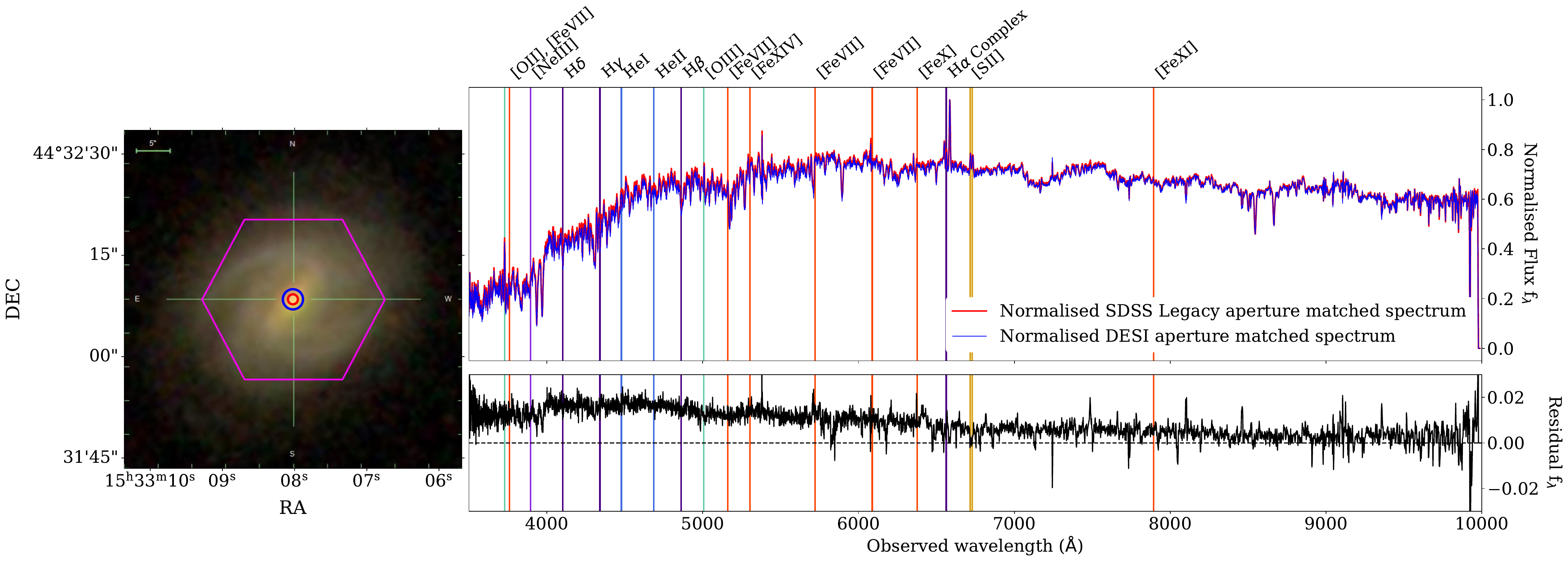}
    \caption{Comparison between the synthetic aperture spectra generated from the MaNGA observation and configured to match the fibre sizes of the SDSS Legacy survey and DESI observation. The overall extent of the MaNGA data is shown by the outer purple hexagon, with the regions used to construct the synthetic apertures shown by circles with colours that match those used for the spectra. The inner DESI matched aperture has a diameter of 1.5~arcsec, with the outer SDSS Legacy matched aperture having a diameter of 3~arcsec. The choice to use synthetic aperture spectra here rather than a direct comparison is made to isolate any temporal variation from the comparison and highlight any potential aperture effect. No significant differences are observed, with a mean absolute difference of one per cent, confirming that the changes observed between the SDSS and DESI spectra are not the result of aperture size and reflect the residual physical effects of \dyk. The aperture spectra presented here have been de-redshifted and corrected for Milky-Way extinction, but have not undergone any smoothing for this comparison.}
    \label{fig:manga_ap_spec_comp}
\end{figure*}

We note here that the SDSS Legacy and DESI spectra were obtained with fibres of differing diameter (3 and 1.5 arcsec respectively). This results in the respective spectra sampling different regions of \dykhost\, which could influence the obtained spectra. We investigate the potential effect of this difference in fibre size using the MaNGA IFU observation which covers a much larger fraction of \dykhost. We obtain synthetic aperture spectra by applying circular apertures of radii matched to both the SDSS Legacy and DESI spectra, centred on the nucleus of \dykhost\ (spaxel 31,31) and obtaining a mean of all spaxels that are at least 80 per cent within the aperture region. Following this we normalise each resultant spectrum and generate a residual by subtracting the smaller aperture (DESI-like) spectrum from the larger (SDSS-like). Using synthetic aperture spectra constructed from the MaNGA IFU observation ensures that no temporal variation is present within the comparison, allowing any aperture size related effects to be fully isolated. This comparison (Fig.~\ref{fig:manga_ap_spec_comp}) reveals that the synthetic aperture spectra are almost identical, with the normalised residual spectrum having a mean absolute difference of one per cent. Additionally, neither the \Oii~$\lambda$~3728~\AA\ nor \Oiii~$\lambda$~5007~\AA\ emission lines are outliers in the residual spectrum, confirming that their increase in strength between the SDSS Legacy and DESI spectra is the result of true evolution rather than the differing aperture sizes between the observations.

\subsection{Spectral comparisons}
\label{subsec:Spectral_Comparisons}

\begin{figure*}
    \centering
    \includegraphics[width=\textwidth]{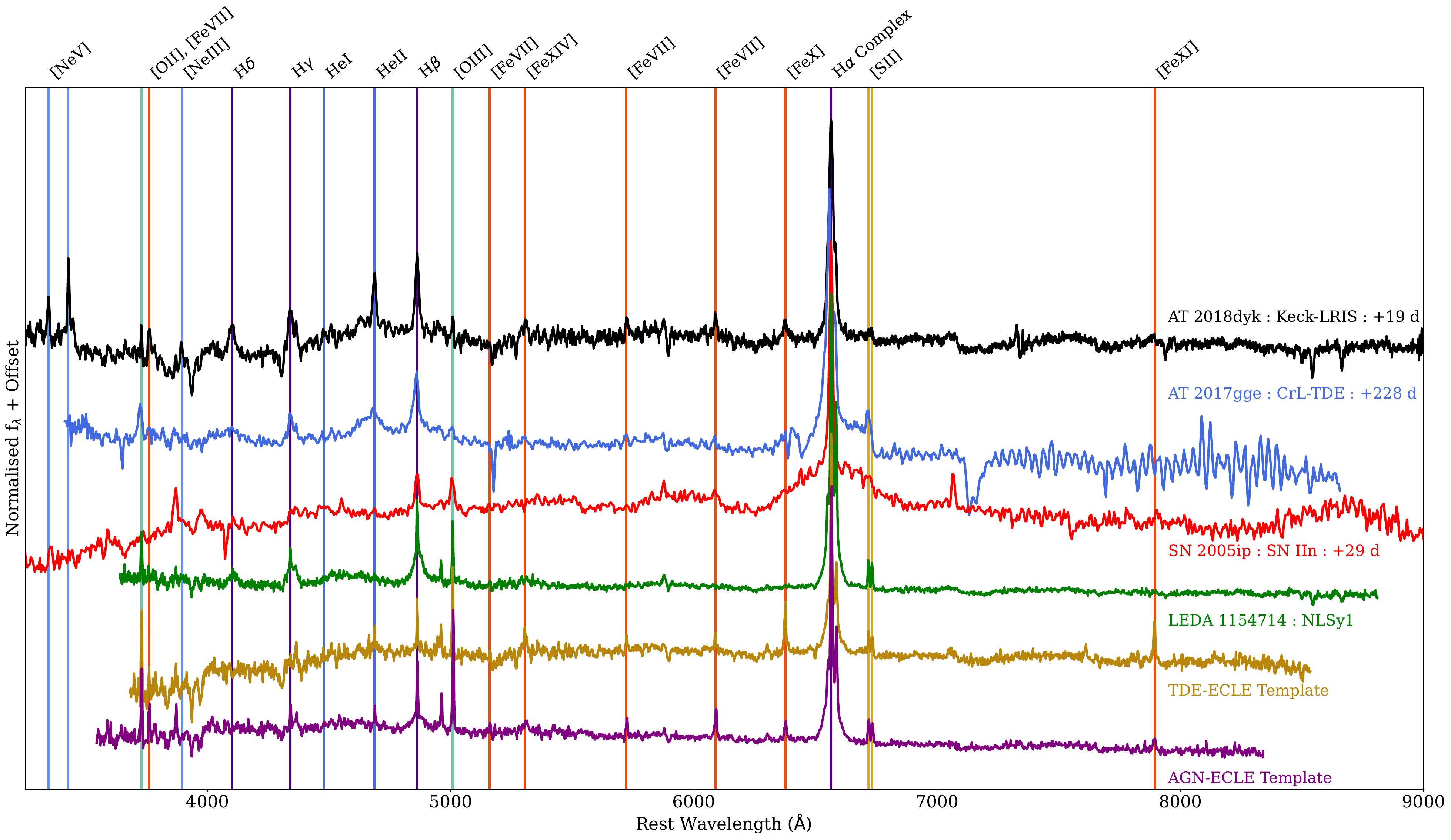}
    \caption{Comparison between the Keck+LRIS spectrum (obtained 19~d following optical peak) to a range of comparison objects. The \dyk\ spectrum shows similar coronal line features as those seen in the CrL-TDE AT~2017gge and in the ECLE templates (composite spectra obtained by combining the SDSS Legacy spectral observations of the sample) from \protect\citet{clark_2024_Longtermfollowupobservations}, though at lower relative strength than the ECLE templates. The broad TDE features of H, \Hei, and \Heii\ are also shown by both \dyk\ and AT~2017gge. The spectrum of AT~2017gge is from a later phase relative to maximum light compared to \dyk\ as the coronal line features of AT~2017gge developed $\sim$~200~d post maximum light (as opposed to the near maximum light line emergence in \dyk). Such features are absent from the TDE-ECLE template as this is composed of spectra at much later relative phases (several years). The Type IIN SN~2005ip displays coronal emission at a much lower intensity than \dyk\ and the rest of the comparison sample. It also has a much broader H~$\alpha$ emission component but lacks the \Hei, \Heii, H$\gamma$, and H$\delta$ emission lines.
    }
    \label{fig:18dyk_spectrum_comparison}
\end{figure*}

Here, we compare \dyk\ with other types of transients that exhibit coronal iron lines in their spectra. In these comparisons, the spectra are first rebinned to 2~\AA\ or the dispersion resolution of the comparison spectrum (whichever is larger) to improve the signal-to-noise ratio (SNR). Given the close separation in phase between the two spectra of \dyk\ that display coronal lines, we only use the higher-quality Keck+LRIS spectrum for this analysis. The spectra used in these comparisons are shown in Fig.~\ref{fig:18dyk_spectrum_comparison}.

We first compare the Keck spectrum of \dyk\ spectrum to the spectral sequence of the CrL-TDE AT~2017gge \citep{onori_2022_NuclearTransient2017gge}. We find good matches -- as determined through the use of the Akaike information criterion \citep[AIC;][]{akaike_1974_NewLookStatistical} -- to the spectra of AT~2017gge from 2018 April 8~--~2018 June 29. The overall spectral shape and coronal line features are very similar, though the Balmer emission features of AT~2017gge are significantly broader. These spectra are from a much later phase in the evolution of AT~2017gge: 218~--~321~d post optical peak compared to 19~d post optical peak for the \dyk\ Keck spectrum, highlighting the significant diversity in evolution timescales of these objects.

As previously discussed, Type IIn SNe have also been observed to display weak Fe coronal line features. For comparison, we examine the spectral sequence of the Type IIn SN~2005ip \citep{stritzinger_2012_MultiwavelengthObservationsEnduring}.

In Fig.~\ref{fig:18dyk_spectrum_comparison}, we show the spectrum of SN~2005ip at a phase of +29~d, due to the similar phase and wavelength coverage when compared to the \dyk\ Keck spectrum. As expected, the coronal line features of \dyk\ are significantly stronger than those of SN~2005ip, with SN~2005ip also lacking the strong \Hei, \Heii, and higher order Balmer emission lines present in \dyk. Additionally, the \Neiii\ and \Nev\ emission lines at wavelengths < 4000~\AA\ displayed by \dyk\ are not observed in SN~2005ip. Finally, whilst \dyk\ does have a broad H$\alpha$ emission component at this phase of evolution, the equivalent emission in SN~2005ip is much stronger and broader overall.

Following these single object comparisons, we also compare \dyk\ to the TDE-ECLE and AGN-ECLE templates constructed by \citet{clark_2024_Longtermfollowupobservations}, though due to observational limitations, these represent a much later phase in the evolution of these objects. These template spectra were constructed by averaging the SDSS Legacy spectra of the TDE-ECLE and AGN-ECLEs originally identified in the search of SDSS DR7 by \citet{wang_2012_EXTREMECORONALLINE}.

Whilst similar in overall spectral shape, the Fe coronal emission lines exhibited by \dyk\ are weaker than those displayed in either ECLE template. However, both \dyk\ and the TDE-ECLE template lack the strong \Oiii\ emission typical of AGN activity. The comparative weakness of, the coronal emission features of \dyk\ when compared to the existing TDE-ELCE sample is explored in the context of its MIR behaviour in Section~\ref{subsec:MIR_Evolution}.

\subsection{Search for additional transient outburst activity}
\label{subsec:Transient_Search}
\begin{figure*}
    \centering
    \includegraphics[width=0.9\textwidth]{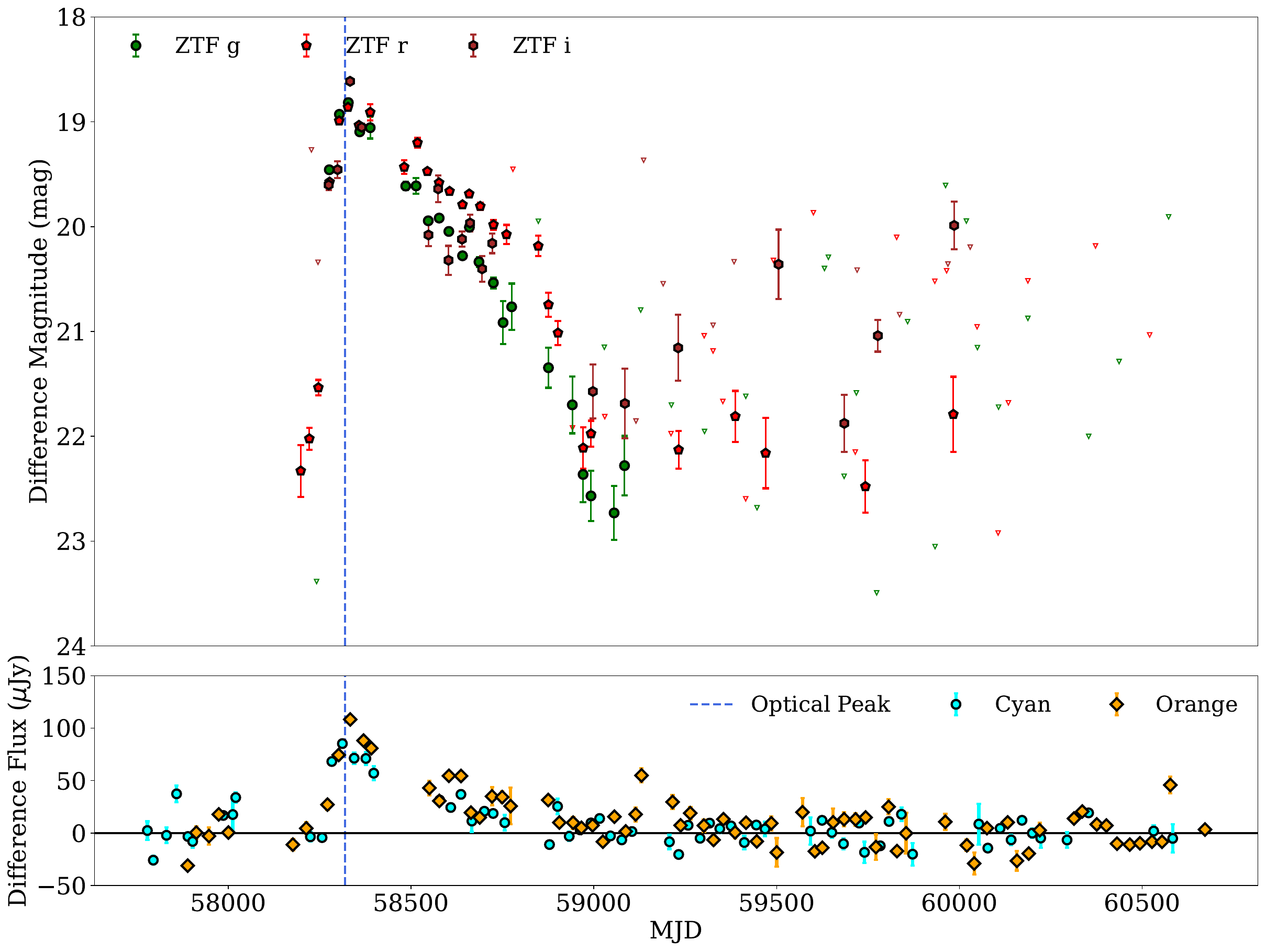}
    \caption{\textit{Top:} ZTF forced photometry difference light curves for \dyk. Upper limits for non-detections are shown by empty triangles. Late time detections are likely spurious as evidenced by their erratic nature and deeper proximal upper detection limits. \textit{Bottom:} ATLAS difference light curves for \dyk. For both data sources, observations in each band have been stacked to obtain a 30~d cadence and are displayed following the trimming of unreliable observations. The time of optical peak observed in ZTF photometry (as detailed by \protect\citealt{huang_2023_AT2018dykRevisitedTidal}) is shown for reference by the dashed blue line. As some flux from \dyk\ is present in the reference ranges used by the forced photometry systems of both telescopes, all observations have been rescaled based on the per-band mean difference flux of observations made at MJDs~$>$~59500. ATLAS observations are shown in flux rather than magnitude space due to the lower overall signal-to-noise ratio of the observations. In both datasets, \dyk\ is clearly a single-epoch, non repeating event.}
    \label{fig:ZTF_and_ATLAS_LC}
\end{figure*}

As stated in previous works, \dyk\ was visible in optical photometric observations with a rapid rise and decline consistent with a power-law with a best fitting index of -1.58 \citep{huang_2023_AT2018dykRevisitedTidal}. We use recent ZTF and ATLAS observations to investigate whether any additional flares occurred since 2018, which could point to \dyk\ resulting from recurrent AGN flaring. The resulting multi-filter light curves generated from forced photometry are shown in Fig.~\ref{fig:ZTF_and_ATLAS_LC}. No flaring activity has been observed in the $\sim$ 2000~d since the transient returned to quiescent optical flux, with a stable flux level observed following the singular outburst and decline. There have also been no new reports of transient activity at the location of \dyk\ or elsewhere within its host galaxy.

\subsection{MIR photometric evolution}
\label{subsec:MIR_Evolution}

\subsubsection{Light curve analysis}
\label{subsubsec: MIR_LightCurves}

The MIR evolution of TDEs (and AGNs) reveals the presence of circumnuclear material in the region around the SMBH. Higher energy photons are absorbed by this material, reprocessed, and re-radiated in the MIR with the luminosity of this MIR, emission directly linked to the incoming flux. As such, where circumnuclear material is present in sufficient quantities and in a suitable physical configuration, a TDE is expected to generate a MIR outburst (delayed from the peak of the direct emission from the disruption itself due to distance between the SMBH and surrounding material) which will then fade as the higher energy emission from the TDE also fades and thus provides less incident radiation for reprocessing. In contrast, as AGN are not single epoch events, but rather variable objects, their MIR emission is expected to show smaller amplitude but repeated variation as their overall energy output varies. Colour evolution is produced by varied strengthening of specific spectral emission features, with AGN seen to be redder than quiescent galaxies in \textit{W1}-\textit{W2} colour space.

As described by \citet{huang_2023_AT2018dykRevisitedTidal} and \citet{masterson_2024_NewPopulationMidinfraredselected}, prior to the observed outburst the MIR luminosity of the host galaxy of \dyk\ was largely constant in both \textit{W1} and \textit{W2}, with the exception of low level stochastic variability. Specifically, the standard deviation of the NEOWISE observations prior to outburst was 0.01~mag in both \textit{W1} and \textit{W2}, while standard deviations of 0.01 mag in \textit{W1} and 0.03 mag in \textit{W2} were measured after the end of the outburst. This is in good agreement with the pre-outburst level of variability observed in the CrL-TDE sample, which averaged 0.03~mag in \textit{W1} and 0.05~mag in \textit{W2}. It is also lower than the variability displayed by either the AGN-ECLEs or CL-LINERs, with mean standard deviations of 0.08~mag and 0.15~mag in \textit{W1} and 0.09~mag and 0.20~mag in \textit{W2}, respectively. Additionally, the overall maximum changes in pre-outburst magnitude for \dyk\ and the other Crl-TDEs are also smaller (0.04~mag and 0.11~mag in \textit{W1} and 0.03~mag and 0.17~mag in \textit{W2}, respectively) compared to the variability displayed by AGN-ECLEs or CL-LINERs (0.33~mag and 0.49~mag in \textit{W1} and 0.36~mag and 0.62~mag in \textit{W2}, respectively). This serves to highlight that \dyk\ and the other CrL-TDEs display pre-outburst MIR variability at a lower level than observed in otherwise potentially contaminating AGN. A full detailing of this variability analysis can be found in Appendix Tables~\ref{Tab:Varibilty_Sigmas} and \ref{Tab:Varibilty_Deltas}. 

We extend previous light curve analyses to include the most recent NEOWISE data release, which shows that \dyk\ has now returned to its pre-outburst quiescent brightness in both \textit{W1} and \textit{W2} (Fig.~\ref{fig:18dyk_mir_evolution}). The observed emission peaked on 2019 January 20 (MJD~58503) with an apparent delay of $\sim$~180~d with respect to the optical peak (approximately coincident with the observed peak of X-ray emission). However, given the lower cadence of WISE observations ($\sim$~6~months) relative to the optical observations, the true MIR peak was likely brighter than what is captured by the available observations.

Prior to outburst, the \textit{W1} - \textit{W2} colour of \dyk\ was close to 0~mag, well below the \citet{stern_2012_MIDINFRAREDSELECTIONACTIVE} AGN colour cut of \textit{W1}~-~\textit{W2}~>~0.8 mag. The \textit{W1}~-~\textit{W2} colour of \dyk\ changes significantly with the onset of the MIR flare and reaches a peak value of 0.4~mag, coincident with peak MIR luminosity, before returning to its original colour over the next $\sim$~1000~d. Additionally, at no point during its evolution does \dyk\ meet the \citep{assef_2018_WISEAGNCatalog} \textit{W1}~-~\textit{W2} vs \textit{W2} classification for AGN activity, at either the 90 per cent or 50 per cent confidence levels. Further observations will be required to confirm that \dyk\ has returned to a long-term stable post-outburst flux (i.e., that no additional outbursts occur). However, both the lack of pre-outburst AGN activity and the return to quiescent behaviour following the end of the outburst are consistent with the absence of any significant AGN activity prior to, or following the end of the outburst.

When compared to the TDE-ECLEs and CrL-TDEs, the MIR evolution of \dyk\ shows the same overall behaviour, though has some notable distinctions. The variability displayed by \dyk\ has a smaller amplitude (both in the individual MIR bands and in the \textit{W1} - \textit{W2} colour) and a shorter duration. For example, whilst \dyk\ returned to MIR quiescence $\sim$~3.5~yr following outburst, several TDE-ECLEs are still in their decline phase more than 20~yr following their initial outburst. Additionally, whilst the beginning of the ECLE MIR outbursts were not observed (as the outbursts pre-date the start of WISE observations), they all initially had \textit{W1} - \textit{W2} colours at or above the \citet{stern_2012_MIDINFRAREDSELECTIONACTIVE} 0.8~mag AGN colour cut; \dyk, however, remains much bluer than this cut at all stages of its evolution, though it does trend towards it during outburst.

We investigate whether the MIR post-outburst behaviour of \dyk\ can be modelled using a power-law decline in a similar manner as TDE-ECLEs. Following \citet{dou_2016_LONGFADINGMIDINFRARED} and \citet{clark_2024_Longtermfollowupobservations}, we fit both the \textit{W1} and \textit{W2} bands separately in flux space and compare to the other objects in the comparison sample (lower right of Fig.~\ref{fig:18dyk_mir_evolution}). Given the low cadence of the WISE observations, for the purposes of the power-law fitting we assume that true peak MIR luminosity occurred 100~d prior to the observed peak for \dyk\ and the other CrL-TDEs (i.e., at approximately half the time between the observed peak and the prior observation), whilst adopting the peak times of the TDE-ECLEs used by \citet{dou_2016_LONGFADINGMIDINFRARED} and \citet{clark_2024_Longtermfollowupobservations}. We note here that the time of peak luminosity and the power-law index are highly degenerate, with tests showing that changing the peak date by 50~d produces a change of $\sim$~23 per cent in the determined values of the power-law indices. Due to the cadence of WISE observations, the true time of peak cannot be constrained further, and we include a 25 per cent statistical uncertainty for the power law index values plotted in Fig.~\ref{fig:18dyk_mir_evolution}. This statistical uncertainty is not included in the reporting of the direct fitting results within the following text or in Appendix~\ref{Appendix:MIR_Power_Law_Fits}, which detail the values obtained directly from the fitting.

Additionally, unlike in previous analyses \citep{dou_2016_LONGFADINGMIDINFRARED, clark_2024_Longtermfollowupobservations}, we are able to constrain the expected quiescent flux of the underlying galaxies of \dyk\ and the comparison CrL-TDEs using the available pre-outburst photometry by setting the floor of the power law fit to reflect this baseline flux. This was not possible for studies into the TDE-ECLEs as pre-outburst MIR observations for these objects do not exist.

For \dyk\ there are notable early flux excesses immediately following peak in both MIR bands (also observable in the light curves as noticeable shoulders). When these excesses are removed the overall fits to the remaining photometry (as determined by an AIC comparison) are significantly improved. This is the first time such features have been identified in the light curves of CrL-TDEs, with the presence of such excesses are indications of multiple emission components likely produced by complex dust configurations. Whilst deconvolving these components and modelling of their corresponding physical configurations is beyond the scope of this work, we report the results of the power-law fits with and without excluding these early data-points. The full results of these fits are given in Table~\ref{Appendix:MIR_Power_Law_Fits} and the lower right panel of Fig.~\ref{fig:18dyk_mir_evolution}.

When including all post-peak data, the measured power-law indices for \dyk\ are -1.25~$\pm$~0.04 in \textit{W1} and -1.02~$\pm$~0.03 in \textit{W2}. When the two observations comprising the excess are excluded, the overall fit to both the early peak and late-time decline tails are improved, particularly for the \textit{W2} band with decline indices of -1.92~$\pm$~0.19 and -1.25~$\pm$~0.05 for \textit{W1} and \textit{W2}, respectively.

For comparison, we perform the same fitting procedure on the existing sample of TDE-ECLEs, updating the work of \citet{clark_2024_Longtermfollowupobservations} to include the data from the latest NEOWISE-R release, whilst also performing new fits on the CrL-TDE sample. 

As only a fraction of the CrL-TDEs have passed their MIR peaks and are now declining, we only fit AT~2017gge, AT~2018gn, AT~2018bcb, and VT~J1548, which all have at least five observations following their MIR peaks. Additionally, we do not include AT~2019avd in this comparison due to its differing overall multi-epoch MIR behaviour, despite now being in an established decline phase. The resultant power-law indices are compared to a range of models of different types of SMBH accretion: standard fallback \citep[e.g.,][]{evans_1989_TidalDisruptionStar,phinney_1989_CosmicMergerMania}, viscous disk accretion \citep{cannizzo_1990_DiskAccretionTidally}, disk emission \citep{lodato_2011_MultibandLightCurves}, and advective super-Eddington thin-disk accretion \citep{cannizzo_2009_NewParadigmGammaray, cannizzo_2011_GRB110328ASwift}. As with \dyk, AT~2017gge, AT~2018gn and AT~2018bcb show evidence of complex circumnuclear dust configurations with post-peak excesses / light-curve shoulders in both WISE bands, as such we report the fitting results including and excluding these excesses (Table~\ref{Appendix:MIR_Power_Law_Fits}). For these objects a visual inspection of the light-curves indicates that fits to both bands are improved when the excesses are removed, with an AIC comparison confirming this for AT~2018bcb. However, retaining the excesses for AT~2017gge and AT~2018gn is statistically preferred. The resulting power-law indices for all objects are found within the parameter space consistent with other previous works, with the range of index values spanning -2.51 to -0.42 in \textit{W1} and -1.97 to -0.32 in \textit{W2}. 

The results of the fitting excluding these excesses are shown in full in Figure~\ref{fig:CLTDES_MIR_Power_Law_Fit_All_Points}, with the corresponding fits removing the early excesses shown in Figure~\ref{fig:CLTDES_MIR_Power_Law_Fit_Early_Excess_Excluded}.

\afterpage{
\begin{landscape}
\centering

\begin{figure}
    \includegraphics[width=0.9\columnwidth]{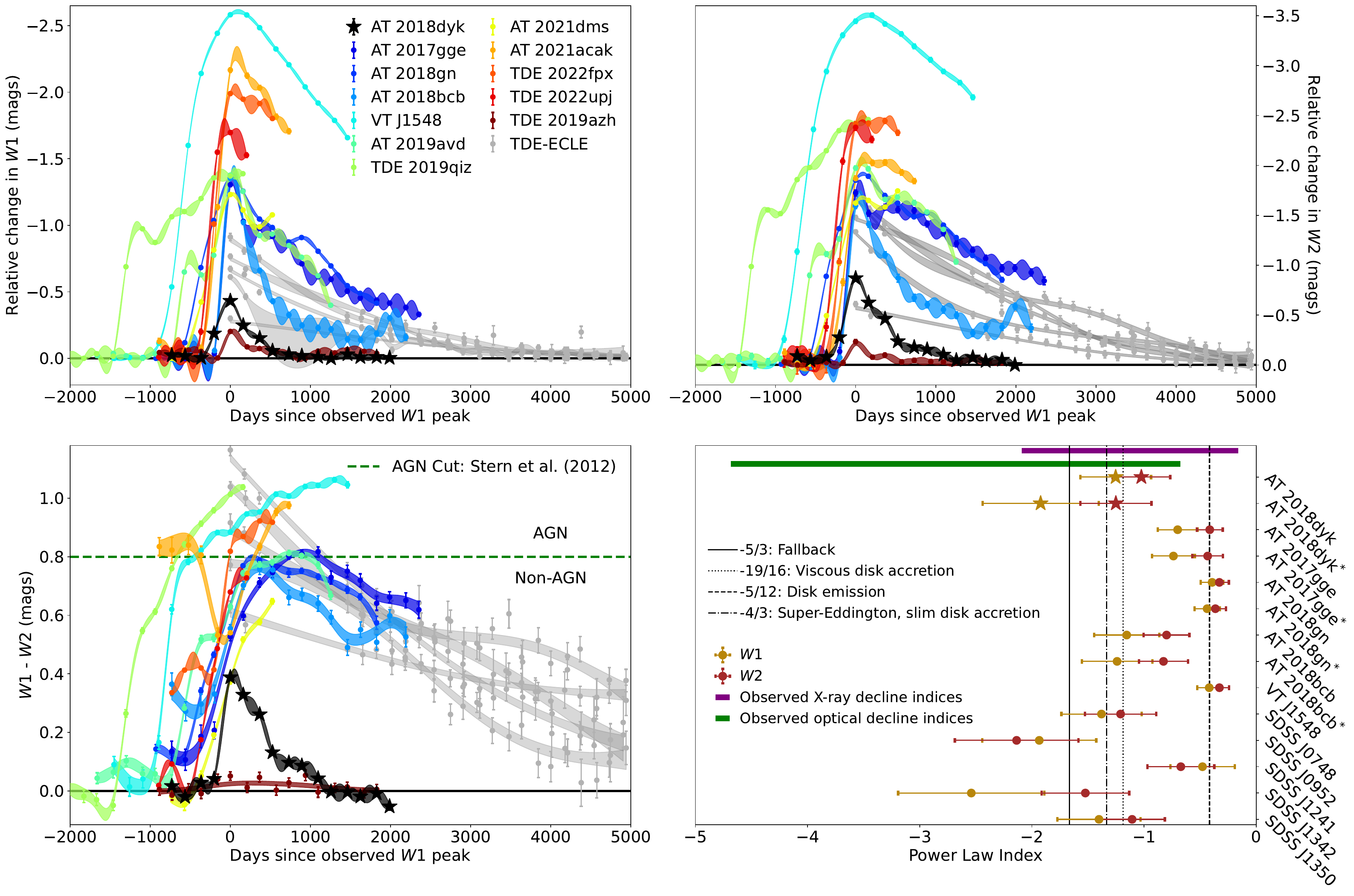}
    \caption{{\it Top left:} Relative change in \textit{W1} compared to observed \textit{W1} peak. 
    {\it Top Right:} Relative change in \textit{W2} compared to observed \textit{W2} peak. 
    {\it Bottom left:} \textit{W1}--\textit{W2} colour evolution. The dashed horizontal line shown is the AGN/non-AGN dividing line from \protect\citet{stern_2012_MIDINFRAREDSELECTIONACTIVE}. Objects with a \textit{W1}--\textit{W2} colour greater than this value display AGN-like behaviour. In all panels, fits displayed are obtained through Gaussian processes, with the shaded regions indicating the 1$\sigma$ fitting uncertainties. {\it Bottom right:} Comparison between the MIR power-law decline indices for AT~2018dyk, the TDE-linked ECLEs \protect\citep{clark_2024_Longtermfollowupobservations}, and the coronal line TDEs AT~2017gge and AT~2018bcb, along with the range of X-ray power-law decline indices determined from the \protect\citet{auchettl_2017_NewPhysicalInsights} TDE sample (purple shading) and the optical decline indices of the \protect\citet{hammerstein_2023_FinalSeasonReimagined} TDE sample (green shading). Vertical lines indicate the expected values for a range of accretion models.}
    \label{fig:18dyk_mir_evolution}
\end{figure}

\end{landscape}
    \clearpage
}

\FloatBarrier

Earlier AllWISE photometry includes the longer-wavelength \textit{W3} filter, though only for two epochs, both obtained well before the outbursts of any of the objects explored in this work. This can be used to better differentiate between various galaxy types or AGN activity, obscured or otherwise. We present these data in Fig.~\ref{fig:AllWISE_Fig} for \dyk\ and for selected comparison objects, including the original TDE-ECLE sample and other CrL-TDEs. In this parameter space, \dyk\ sits well within the region occupied by star-forming galaxies and is outside the `Mateos wedge' of AGN-hosting galaxies \citep{mateos_2012_UsingBrightUltrahard} and other regions that would indicate the presence of an obscured AGN. This is similar to other optically selected CrL-TDEs (excluding AT~2021acak, which occupies the edge of several different AGN regions, attributed to its host galaxy also possessing an AGN, as described by \citealt{li_2023_AT2021acakCandidateTidal}), but distinct from the original TDE-ECLE sample, which were observed by the AllWISE survey during their MIR outburst phase and display colours consistent with AGNs (Fig.~\ref{fig:AllWISE_Fig}). As TDE-ECLEs and CrL-TDEs are subpopulations of the same underlying group, whilst, the AllWISE observations are limited in time, they provide a general view of their behaviour i.e., quiescent prior to outburst and AGN-like during outburst. No object within the sample has AllWISE observations post the end of their outbursts, but given the return to quiescence in both \textit{W1} and \textit{W2} bands, it is reasonable to conclude a similar behaviour in the \textit{W3} band, though any potential differences in timescale are of course unobserved.

\afterpage{
\clearpage
\begin{landscape}
\centering

\begin{figure}
    \centering
    \includegraphics[width=0.9\columnwidth]{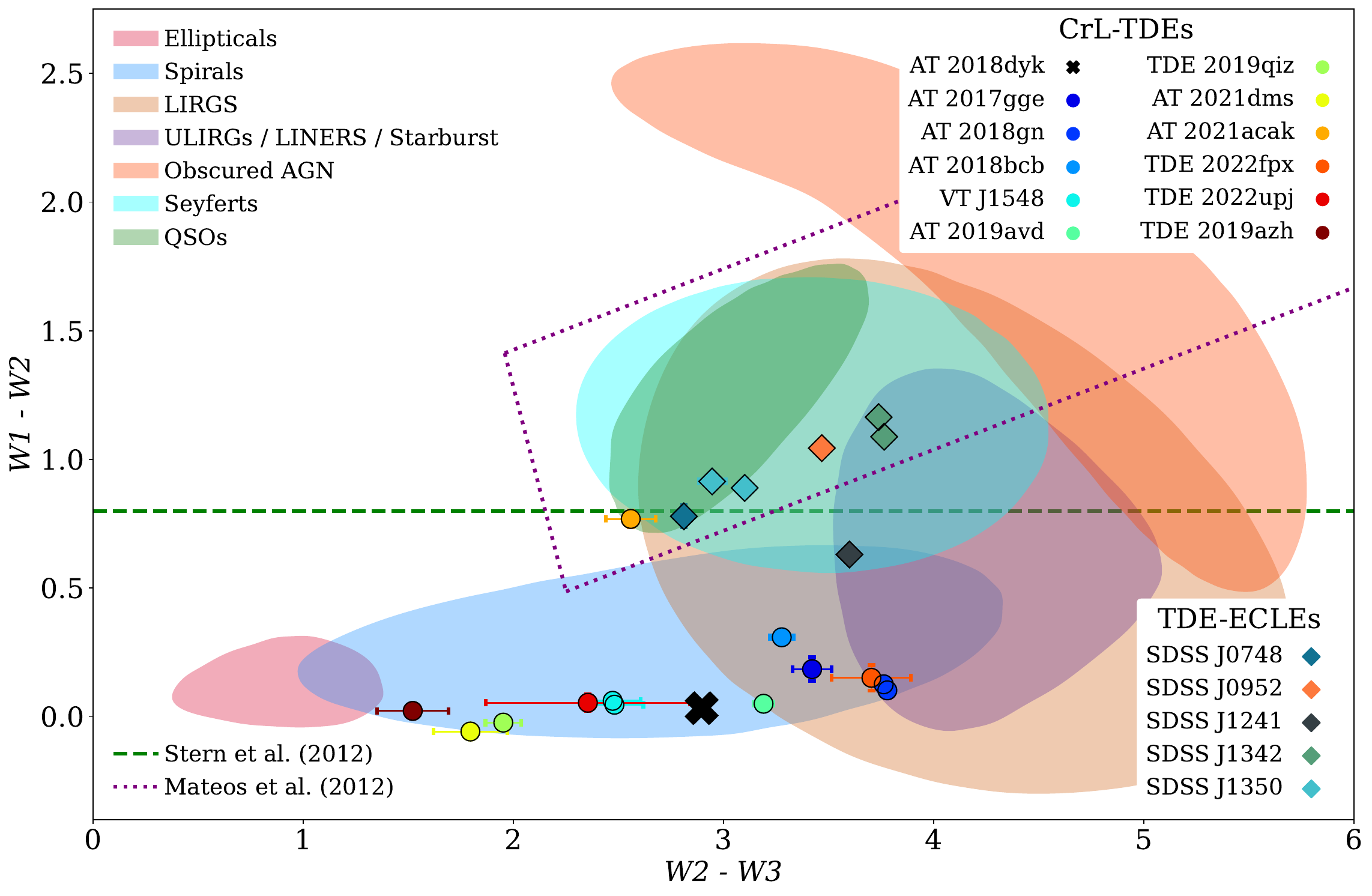}
    \caption{AllWISE colour-colour plot showing \dyk\ (black cross) in comparison to the known sample of TDE-ECLEs (diamonds) and CrL-TDEs (circles). Whilst two epochs of data are included for \dyk, they overlap due to the lack of variability during this time. Regions have been sourced from \protect\citet{wright_2010_WIDEFIELDINFRAREDSURVEY}. The AGN identification cuts from \citet{stern_2012_MIDINFRAREDSELECTIONACTIVE} and \citet{mateos_2012_UsingBrightUltrahard} are included as a green dashed line and purple dotted lines, respectively. Additionally, we note that NLSy1 AGNs occupy the same region of parameter space as more conventional Seyfert galaxies, as shown by \citet{paliya_2024_NarrowlineSeyfert1}. \dyk\ and the other CrL-TDEs that were observed in their pre-TDE quiescent state occupy the parameter space of spiral / starforming galaxies and are outside the regions that would indicate any AGN activity, with the exception of AT~2021acak, which is hosted by a galaxy that also hosts an AGN. In contrast, the SDSS TDE-ECLEs - observed by AllWISE during outburst - are found within or close to the AGN regions. We remind the reader that these classes of objects likely represent the same underlying population observed at different phases in their evolution. Uncertainties in both axes are included but are generally smaller than the points.}
    \label{fig:AllWISE_Fig}
\end{figure}

\end{landscape}
    \clearpage
}

\subsubsection{Outburst properties}
\label{subsubsec:MIR_Outburst_Properties}

We also measure the maximum difference between outburst peak and pre-outburst quiescence ($\Delta$ values) for both MIR bands and colour. For \dyk, these values are $\Delta$\textit{W1}~=~0.40~$\pm$~0.01~mag, $\Delta$\textit{W2}~=~0.79~$\pm$~0.01~mag, and $\Delta$(\textit{W1-W2})~=~0.39~$\pm$~0.01~mag.

\afterpage{
\clearpage
\begin{landscape}
\centering

\begin{figure}
    \centering
    \includegraphics[width=0.9\columnwidth]{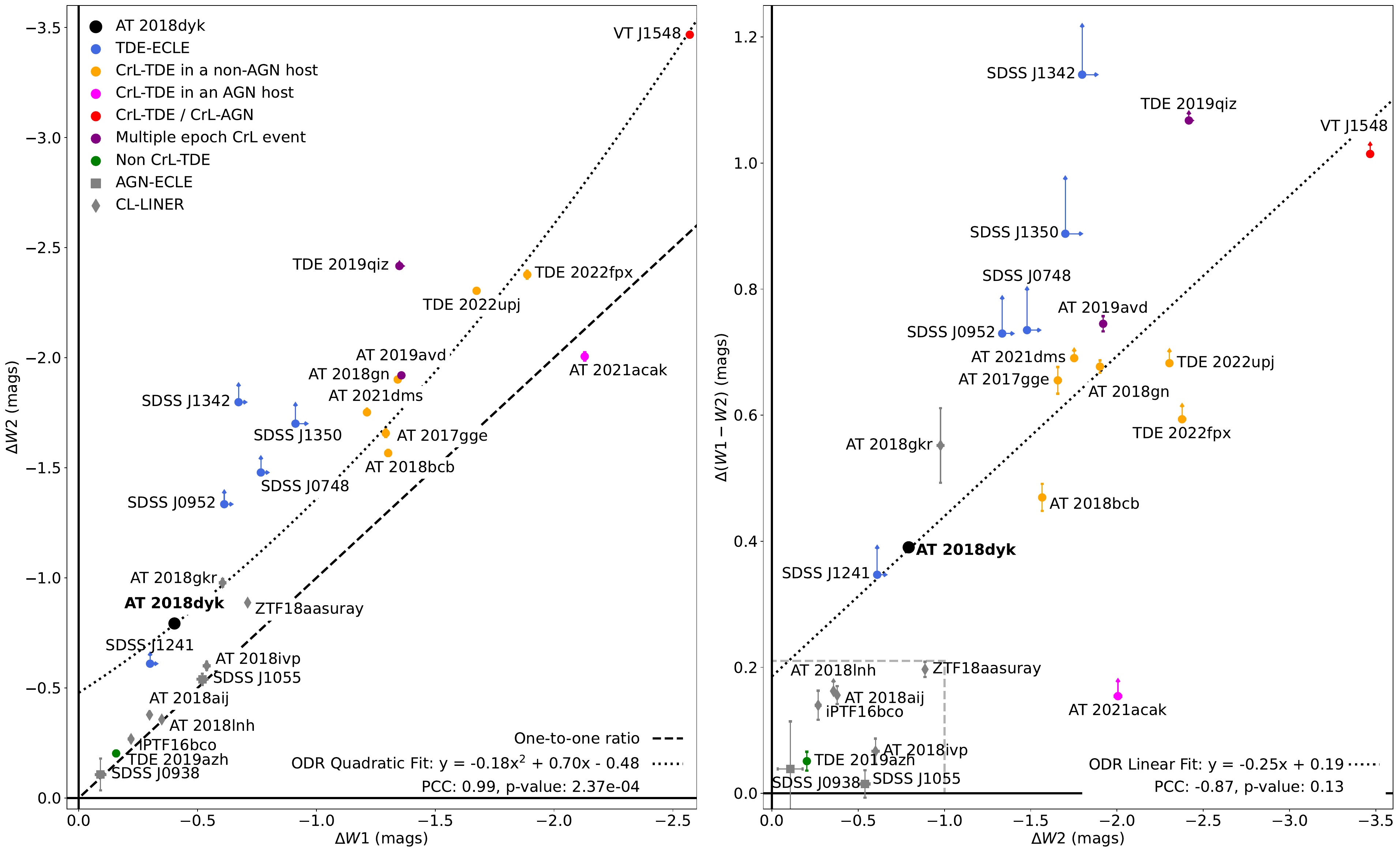}
    \caption{{\it Left:} Comparison between the maximum change in \textit{W1} and \textit{W2} between \dyk\ (black circle) and a range of comparison objects. Dashed black line shows a 1-to-1 relation. Dotted line shows the orthogonal distance regression (ODR) best fitting quadratic for the CL-TDEs excluding objects with only upper limits. A quadratic fit to the data was preferred at a >~5~$\sigma$ level when tested against a constant and a linear fit using a likelihood ratio test, supported through AIC comparisons. A strong positive correlation is also supported by the Pearson's correlation coefficient with a value of 0.99 and corresponding p-value of 2.4~$\times$~10$^{-4}$.
    {\it Right:} Comparison between the maximum change in \textit{W1} - \textit{W2} colour versus maximum change in \textit{W2} brightness. A trend is visible in the behaviour of the CrL-TDEs and TDE-ECLEs with those objects with brighter MIR outbursts (larger $\Delta$\textit{W2} values) also showing larger shifts to MIR redder colours. Dotted line shows the ODR linear fit for this trend after excluding objects that only have upper limits on either variable. The small number of data points available for fitting (4) prevents the statistical confirmation of this trend as seen in the high value of the Pearson's correlation p-value of 0.13. The region of parameter space occupied by the CL-LINERs and AGN-ECLEs (excluding AT~2018gkr) is demarcated by the dashed grey box.}
    \label{fig:MIR_Delta_Fig}
\end{figure}

\end{landscape}
    \clearpage
}

For comparison, we measure the equivalent values for the other objects within the comparison sample, treating these as lower limits for those objects that are still rising. Additionally, as the quiescent states of the TDE-ECLEs were not observed in the MIR, lower limits on the magnitude of the outbursts are obtained using the observed differences in magnitudes between the first and faintest observations. 
We note here that as values are measured independently, they do not necessarily occur at the same MJD, especially when comparing the maximum change in colour, which can lag the peak MIR luminosity by several years (see lower left panel of Fig.~\ref{fig:18dyk_mir_evolution}). We detail the full results of these calculations for all objects in the comparison sample within Table~\ref{tab:MIR_Deltas} and Fig.~\ref{fig:MIR_Delta_Fig}, grouped by object classification. We now highlight the results from this analysis.

All other members of the CL-LINER class described by \citet{frederick_2019_NewClassChanginglook}, with the exception of AT~2018gkr, display $\Delta$\textit{W1} and $\Delta$\textit{W2} at close to a one-to-one ratio, with all having maximum brightening in both bands of less than one magnitude. This behaviour is shared by the AGN-ECLEs and the non-CrL TDE~2019azh. These objects also display small $\Delta$(\textit{W1-W2}) values of less than 0.2~mags. The exception to this general behaviour is AT~2018gkr, which has displayed a long-term rise in luminosity in both WISE bands since the start of observations, though preferentially brightening in \textit{W2}.

In contrast, all observed CrL-TDEs (including \dyk) and TDE-ECLEs display outbursts that are brighter in \textit{W2} than \textit{W1}. Additionally, whilst there is no apparent relation between the overall change in brightness and the colour of the outburst for the CL-LINERs and AGN-ECLEs, a trend is apparent for the CrL-TDEs and TDE-ECLEs with the brighter the outburst, the redder its peak change in colour (see left panel of Fig~\ref{fig:MIR_Delta_Fig}). We further discuss these results and the behaviour of the CrL-TDE sample as a whole in Section~\ref{subsec:MIR_Photometric_Behaviour}.

\begin{table}
\caption{Peak changes in absolute magnitude and colour of the MIR outbursts displayed by \dyk\ and objects of interest from the literature.}
\label{tab:MIR_Deltas}
\begin{adjustbox}{width=1\columnwidth}
\begin{tabular}{lccc}
\hline
Object & $\Delta$\textit{W1} & $\Delta$\textit{W2} & $\Delta$(\textit{W1-W2}) \\ \hline
\textbf{AT~2018dyk}& -0.40 $\pm$ 0.01 & -0.79 $\pm$ 0.01 & 0.39 $\pm$ 0.01\\
\\
\textbf{CrL-TDEs}\\
AT~2017gge& -1.29 $\pm$ 0.01 & -1.66 $\pm$ 0.02 & 0.66 $\pm$ 0.02\\
AT~2018gn & -1.34 $\pm$ 0.01 & -1.90 $\pm$ 0.01 & 0.68 $\pm$ 0.02\\
AT~2018bcb& -1.30 $\pm$ 0.01 & -1.57 $\pm$ 0.01 & 0.47 $\pm$ 0.02\\
AT~2021dms& -1.21 $\pm$ 0.01 & <~-1.75 & >~0.69\\
AT~2021acak $^{1}$& -2.13 $\pm$ 0.01 & -2.01 $\pm$ 0.01 & >~0.15\\
TDE~2022fpx& -1.89 $\pm$ 0.01 & -2.38 $\pm$ 0.02 & 0.57 $\pm$ 0.02\\
TDE~2022upj& -1.67 $\pm$ 0.01 & -2.30 $\pm$ 0.01 & >~0.68\\
\\
\textbf{Multi-epoch CrL-TDEs}\\
AT~2019avd& -1.36 $\pm$ 0.01 & -1.92 $\pm$ 0.01 & 0.74 $\pm$ 0.01\\
TDE~2019qiz& <~-1.35& <~-2.42& >~1.07\\
\\
\textbf{Uncertain CrL-TDEs/AGNs}\\
VT J154843.06+220812.6 & -2.57 $\pm$ 0.01 & -3.47 $\pm$ 0.01 & 1.01 $\pm$ 0.01\\
\\
\textbf{TDE-ECLEs}\\
SDSS~J0748+4712 & <~-0.77& <~-1.48&>~0.73\\
SDSS~J0952+2143 & <~-0.61& <~-1.33&>~0.73\\
SDSS~J1241+4426 & <~-0.30& <~-0.61&>~0.35\\
SDSS~J1342+0530 & <~-0.67& <~-1.80&>~1.14\\
SDSS~J1350+2916 & <~-0.91& <~-1.70&>~0.89\\
\\
\textbf{AGN-ECLEs}\\
SDSS~J0938+1353 $^{2}$ & -0.09 $\pm$ 0.02 & -0.11 $\pm$ 0.07 & 0.03 $\pm$ 0.01 \\
SDSS~J1055+5637 $^{2}$ & -0.52 $\pm$ 0.02 & -0.54 $\pm$ 0.03 & 0.08 $\pm$ 0.02 \\
\\
\textbf{NonCrL-TDEs}\\
TDE~2019azh& -0.16 $\pm$ 0.01 & -0.20 $\pm$ 0.01 & 0.05 $\pm$ 0.02 \\
\\
\textbf{CL-LINERs}\\
iPTF16bco $^{3}$ & -0.22 $\pm$ 0.01 & -0.27 $\pm$ 0.01 & 0.14 $\pm$ 0.02 \\
AT~2018aij $^{4}$ & -0.30 $\pm$ 0.01 & -0.38 $\pm$ 0.01 & 0.16 $\pm$ 0.01\\
AT~2018gkr $^{2}$ & -0.61 $\pm$ 0.01 & -0.98 $\pm$ 0.02 & 0.55 $\pm$ 0.01\\
AT~2018ivp & -0.54 $\pm$ 0.01 & -0.60 $\pm$ 0.02 & 0.08 $\pm$ 0.02\\
AT~2018lnh $^{3}$ & -0.35 $\pm$ 0.01 & -0.36 $\pm$ 0.01 & > 0.16 \\
ZTF18aasuray & -0.71 $\pm$ 0.01 & 0.89 $\pm$ 0.01 & 0.20 $\pm$ 0.01 \\
\hline
\end{tabular}
\end{adjustbox}
\begin{flushleft}
In all cases, values for each band and the overall observed peak colour change are determined independently and do not necessarily occur at the same phase.\\
$^{1}$ Host galaxy also likely hosts an AGN.\\
$^{2}$ These objects have not displayed specific outbursts, with these values describing their level of general variability instead.\\
$^{3}$ MIR outburst occurred significantly after initial behavioural change and classification (i.e., years later).\\
$^{4}$ Recent observations show re-brightening in both \textit{W1} and \textit{W2} bands and a reddening of \textit{W1}-\textit{W2} colour. Values quoted are based on photometry obtained prior to the start of the rebrightening phase.
\end{flushleft}
\end{table}
\raggedbottom

\subsection{Host-galaxy analysis}
\label{subsec:Host_Analysis}

With both the spectroscopic and photometric behaviour of \dyk\ consistent with that of a CrL-TDE, we now explore the properties of its host galaxy (\dykhost) on both global and local scales to determine if it is consistent with the wider TDE host galaxy population. We note than in this section, uncertainties quoted are determined directly from observations and do not include any scatter inherent in scaling relations used.

\subsubsection{Global properties}
\label{subsubsec:Global_Properties}

We retrieve photometrically determined global host galaxy properties for \dykhost\ from the MaNGA Visual Morphology Catalogue (\citealt{vazquez-mata_2022_SDSSIVMaNGA}), the GALEX-SDSS-WISE Legacy Catalog \citep[GSWLC-2;][]{salim_2016_GALEXSDSSWISELegacyCatalog, salim_2018_DustAttenuationCurves}, the MaNGA PyMorph DR17 photometric catalog \citep[MPP-VAC-DR17;][]{fischer_2019_SDSSIVMaNGAPyMorph, dominguezsanchez_2022_SDSSIVDR17Final}, and the NASA Sloan Atlas (NSA). \footnote{In this work we make use of v1\textunderscore0\textunderscore1 of the NSA accessible here: \url{https://www.sdss4.org/dr17/manga/manga-target-selection/nsa/}} The retrieved properties are summarised in Table~\ref{tab:phot_host_info}. Given their specific configuration for galaxies observed by MaNGA and the inclusion of uncertainties, where a parameter has been measured in multiple catalogues we prefer the values included in GSWLC-2 and MPP-VAC-DR17.

To determine if the global galaxy properties of \dykhost\ are consistent with other TDE host galaxies, we explore a range of properties, comparing each in turn to existing samples of optically selected TDE host galaxies from \citet{law-smith_2017_TidalDisruptionEvent}, \citet{graur_2018_DependenceTidalDisruption}, and \citet{hammerstein_2023_IntegralFieldSpectroscopy}. Furthermore, we investigate the local properties of the region in which \dyk\ occurred in more depth in Section~\ref{subsubsec:MaNGA_Analysis}. 

We find that the stellar mass of \dykhost\ is consistent with, though is at the high end of measured TDE host galaxy masses, with a GSWLC-2 mass of 11.02~$\pm$~0.02~log$_{10}$(\msol~yr$^{-1}$). For comparison, \citet{law-smith_2017_TidalDisruptionEvent} found a TDE host galaxy mass range of 9.2~--~10.4~log$_{10}$(\msol), similar to the \citet{graur_2018_DependenceTidalDisruption} range of 8.5~--~11.0~log$_{10}$(\msol) and the \citet{hammerstein_2023_IntegralFieldSpectroscopy} range of 9.49~--~11.23~log$_{10}$(\msol).

We also examined whether the S\'ersic profile of \dykhost\ is consistent with the previously described TDE host samples. Taking a weighted mean of the per-band S\'ersic indices from MPP-VAC-DR17 pure S\'ersic fits, we find a mean index n = 2.20~$\pm$~0.01, consistent within 1.5~$\sigma$ of the median measured by \citet{law-smith_2017_TidalDisruptionEvent} ($4.03^{+0.92}_{-1.55}$) and close to the median of the \citet{hammerstein_2023_IntegralFieldSpectroscopy} sample (1.87).

Next, we explored the stellar surface mass density of \dykhost\ measured using the S\'ersic half-light radius. We find a value from GSWLC-2 of 8.46~$\pm$~0.02 log$_{10}$(\msol~kpc$^{-2}$), which is also consistent with the values measured for the \citet{hammerstein_2023_IntegralFieldSpectroscopy} sample but lower than the average stellar surface mass density of the TDE host galaxies in the \citet{graur_2018_DependenceTidalDisruption} sample. \citet{law-smith_2017_TidalDisruptionEvent} and \citet{hammerstein_2023_IntegralFieldSpectroscopy} find that TDE host galaxies are more centrally concentrated than the general galaxy population, with this attributed to TDEs being more likely to occur in galaxies that have experienced a recent merger event \citep{graur_2018_DependenceTidalDisruption}. Similarly, while the density of the passive TDE host galaxies in the \citet{graur_2018_DependenceTidalDisruption} sample was consistent with the general population of passive galaxies in SDSS, the star-forming host galaxies were significantly denser than the general star-forming galaxy population.

Finally, using the stellar mass and star-formation rate measurements from GSWLC-2, we calculate the specific star-formation rate (sSFR) of \dykhost\ to be -10.97~$\pm$~0.13 log$_{10}$(yr$^{-1}$). This value falls within the `green valley' of galaxies thought to be transitioning between starforming and quenched states. \citet{salim_2014_GreenValleyGalaxies} defines this region as:

\begin{equation}
    -11.8 \geq \mathrm{sSFR(log_{10})} \geq -10.8
\end{equation}

The lower values for stellar mass from the NSA when using either S\'ersic or Petrosian photometry (10.68~$\pm$~0.04 and 10.56~$\pm$~0.04 log$_{10}$(\msol), respectively), correspond to higher overall sSFRs of -10.62~$\pm$~0.13 and -10.51~$\pm$~0.13 log$_{10}$~(yr$^{-1}$). These place \dykhost\ slightly outside the high sSFR limit of the green valley, though both estimates are <3~$\sigma$ of the upper boundary.

All the measured sSFRs are within the low star-formation tail of the observed sample of MaNGA galaxies with spiral morphologies (see, e.g., figure~1 of \citealt{biswas_2024_StructureKinematicsStarforming}). \citet{hammerstein_2023_IntegralFieldSpectroscopy} found that TDE hosts were preferentially hosted by galaxies within or close to the green valley (63 per cent of their TDE host sample were green valley galaxies when classified using rest-frame \textit{u} - \textit{r} colours compared to 13 per cent of their comparison galaxy population). 

As noted by \citet{frederick_2019_NewClassChanginglook}, whilst \dykhost\ was observed by the FIRST VLA survey \citep{becker_1994_VLAsFIRSTSurvey} in the 20~cm radio band, no source was detected to an upper limit of 0.89~mJy beam$^{-1}$, ruling out any radio-loud AGN activity at the time of observation in 1997.

\begin{table*}
\caption{Photometrically determined properties of AT~2018dyk's host galaxy \dykhost. Where parameter values have been obtained from more than one source the preferred value is shown in bold.}
\label{tab:phot_host_info}
\begin{tabular}{llll}
\textbf{Parameter} & \textbf{Unit} & \textbf{Value} & \textbf{Data source} \\ \hline
Morphological classification &  & SBb & 1 \\
Star-formation rate (SFR) & log$_{10}$(\msol yr$^{-1}$) & 0.06 $\pm$ 0.13 & 2 \\
 &  &  &  \\
Stellar mass & log$_{10}$(\msol) &  &  \\
 &  & \textbf{11.02 $\pm$ 0.02} & 2 \\
 &  &  10.68 $\pm$ 0.04$^{*}$ & 3 \\
 &  &  10.56 $\pm$ 0.04$^{*}$ & 4 \\ 
 &  &  &  \\
Specific star-formation rate (sSFR) & log$_{10}$(yr$^{-1}$) &  &  \\
 &  & \textbf{-10.97 $\pm$ 0.13} & Calculated using 2 \\
 &  & -10.62 $\pm$ 0.13 & Calculated using the SFR of 2 and mass from 3 \\
  &  & -10.51 $\pm$ 0.13 & Calculated using the SFR of 2 and mass from 4 \\
 &  &  &  \\
S\'ersic index &  &  &  \\
 &  & 2.30 $\pm$ 0.02 $^{*}$ & 3 \\
 &  & \textbf{2.20 $\pm$ 0.01} & 5 \textsuperscript{\textdagger} \\
Half light radius & arcseconds &  &  \\
 &  & 9.96 $\pm$ 0.06 $^{*}$ & 3 \\
 &  & 8.78 $\pm$ 0.06 $^{*}$ & 4 \textsuperscript{\textsquare}\\
 &  & \textbf{10.20 $\pm$ 0.03} & 5 \textsuperscript{\textdagger} \\
Stellar surface mass density & log$_{10}$(\msol kpc$^{-2}$) &  &  \\
 &  & 8.14 $\pm$ 0.02 & Calculated using 3 \\
  &  & 8.56 $\pm$ 0.02 & Calculated using 4 \\
 &  & \textbf{8.46 $\pm$ 0.02} & Calculated using 5 \\
\hline
\end{tabular}
\begin{flushleft}
\textbf{Sources:}\\
1: MaNGA Visual Morphology Catalogue : \citet{vazquez-mata_2022_SDSSIVMaNGA}\\
2: GALEX-SDSS-WISE LEGACY CATALOG - 2 (GSWLC-2) : \citet{salim_2016_GALEXSDSSWISELegacyCatalog, salim_2018_DustAttenuationCurves}\\
3: S\'ersic based photometry from the NASA Sloan Atlas (NSA) v1\textunderscore0\textunderscore1: \url{https://www.sdss4.org/dr17/manga/manga-target-selection/nsa/}\\
4: Petrosian based photometry the NASA Sloan Atlas (NSA) v1\textunderscore0\textunderscore1: \url{https://www.sdss4.org/dr17/manga/manga-target-selection/nsa/}\\
5: MaNGA PyMorph photometric Value Added Catalogue (MPP-VAC-DR17) : \citet{fischer_2019_SDSSIVMaNGAPyMorph, dominguezsanchez_2022_SDSSIVDR17Final}\\
\smallskip
$^{*}$ No uncertainties for these parameters are quoted within the NSA. As these parameters are required to derive the value of further parameters, we assume a conservative estimate on the error double that of the measurement given in GSWLC-2.\\
\textsuperscript{\textdagger} Value calculated from the weighted average of the \textit{gri} pure S\'ersic profile fits for comparison to \citet{hammerstein_2023_IntegralFieldSpectroscopy}. When using the MPP-VAC-DR17 preferred S\'ersic+Exponential fit, the weighted average of S\'ersic index for the bulge component is 1.36~$\pm$~0.78.\\ 
\textsuperscript{\textsquare} \textit{gri} average.\\
\end{flushleft}
\end{table*}
\raggedbottom

\subsubsection{Spatially resolved spectroscopic host galaxy analysis with MaNGA}
\label{subsubsec:MaNGA_Analysis}

The available pre-outburst MaNGA IFU observation allows for a detailed examination of the spatially resolved properties of \dykhost\ on much smaller physical scales than is possible with global properties. Considering local properties is known to be important for studies of transients as environmental properties can vary widely across a galaxy - a nuance that is lost when considering only the overall averaged properties and can significantly affect the interpretation of transient events (e.g., TDEs: \citealt{nicholl_2019_TidalDisruptionEvent}; Type Ia SNe: \citealt{kelsey_2021_EffectEnvironmentType}; core collapse SNe: \citealt{pessi_2023_CharacterizationASASSNCorecollapse}).

\begin{table}
\caption{Spectroscopically determined properties of AT~2018dyk's host galaxy \dykhost.}
\label{tab:spec_host_info}
\begin{adjustbox}{width=1\columnwidth}
\begin{tabular}{lccc}
\textbf{Parameter} & \textbf{Unit} & \textbf{Value} & \textbf{Data source} \\ \hline
\textbf{Global Properties} &  &  &  \\
Stellar Velocity Dispersion  & \kms & 68.0~$\pm$~1.1 & 1 \\
Central SMBH Mass \textsuperscript{\textdagger} & log$_{10}$(\msol) & 6.86~$\pm$~0.05 & 1 \\
D4000 &  & 1.65~$\pm$~0.01 & 1 \\
 &  &  &  \\
\textbf{Local Properties} &  &  &  \\
Stellar Velocity Dispersion & \kms & 122.5~$\pm$~1.9 & 2 \\
Central SMBH Mass \textsuperscript{\textdagger} & log$_{10}$(\msol) & 7.56~$\pm$~0.04 & 2 \\
D4000 &  & 2.09~$\pm$~0.01 & 2 \\
 &  &  &  \\
\hline
\end{tabular}
\end{adjustbox}
\begin{flushleft}
\textbf{Sources:}\\
1: Measured from a weighted mean of all MaNGA spaxels.\\
2: Measured from the central - (31,31) - spaxel of the MaNGA IFU data cube coincident with the location of \dyk.\\
\smallskip
\textsuperscript{\textdagger} Calculated using the relation from \citet{kormendy_2013_CoevolutionNotSupermassive}. \\
\end{flushleft}
\end{table}

\raggedbottom

\begin{figure*}
    \centering
    \includegraphics[width=0.8\textwidth]{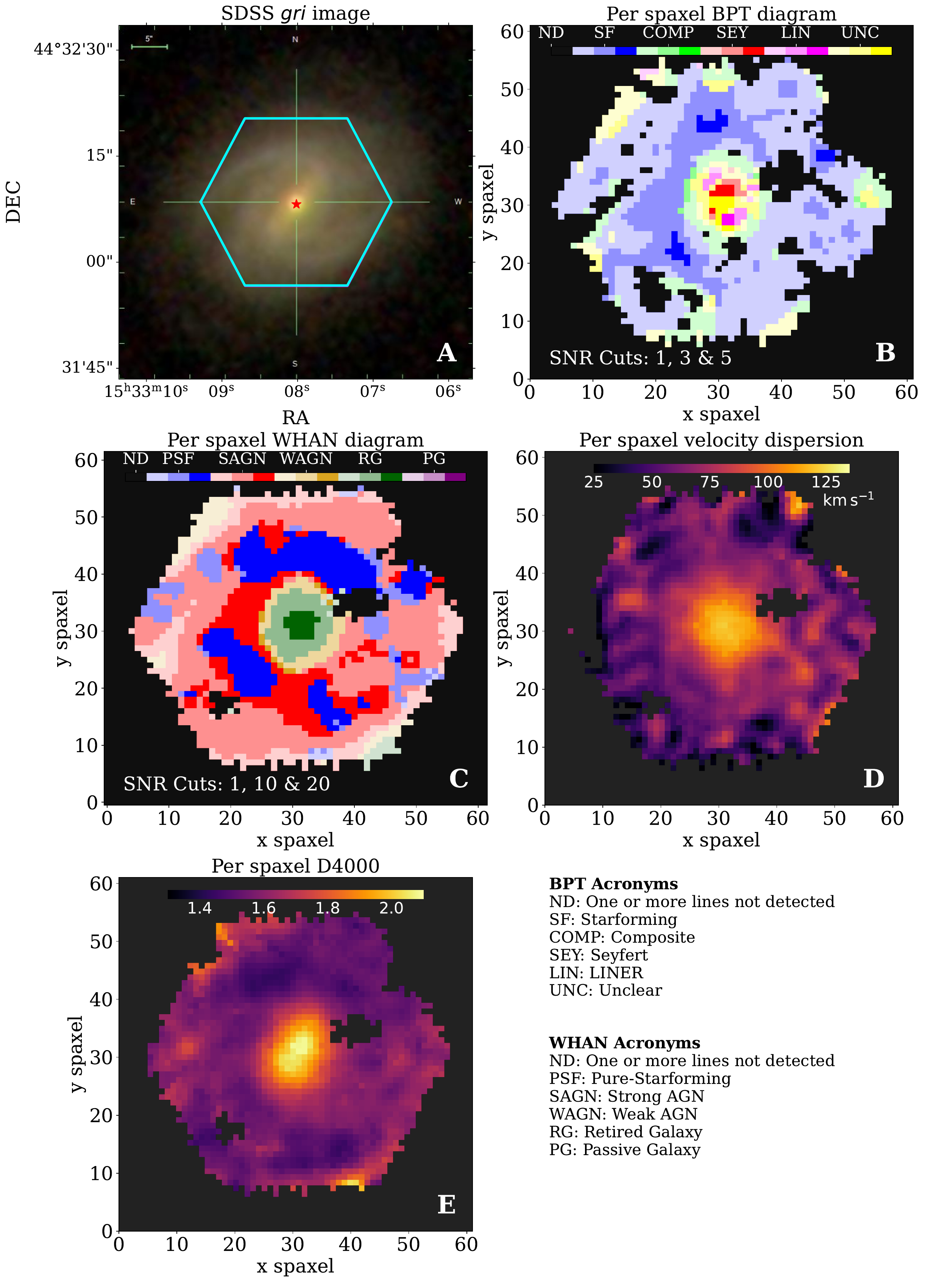}
    \caption{\textit{A}: SDSS \textit{gri} composite image of \dykhost. Hexagonal outline displays the footprint of the MaNGA IFU data. \textit{B}: Per spaxel combined BPT diagram of \dykhost. Colour intensity indicates the SNR of the included emission lines in three bins at SNRs of 1, 3, and 5. Classifications use all four standard BPT diagnostic line ratios. BPT diagnostics cannot provide a clear classification for the emission line properties at the location of \dyk. \textit{C}: Per spaxel WHAN diagram of \dykhost. Colour intensity indicates the SNR of the included emission lines in three bins at SNRs of 1, 10, and 20. The WHAN diagram returns a clear (SNR of all lines >20) `retired galaxy' (RG) classification for the nuclear region of \dykhost, including the location of \dyk, indicating the LINER emission features are produced by an older stellar population rather than AGN activity. \textit{D}: Per spaxel map of the measured stellar velocity dispersion (measured in \kms) of \dykhost. Overall velocity dispersion is within the range, though above the mean, of the velocity dispersions of TDE hosts found by \citet{graur_2018_DependenceTidalDisruption}.
    \textit{E}: Per spaxel map of the measured D4000 spectral indices within \dykhost. The increasing magnitude of this value moving radially towards the centre of the galaxy is consistent with increasing average stellar age. The spiral arms observable in both optical imaging and through BPT diagnostics are visible as the dark coloured regions above and below the galactic nucleus. 
    }
    \label{fig:18dyk_manga_data}
\end{figure*}

The MaNGA observations for \dykhost\ were made on Plate 9870-9101 with MaNGA ID 1-199368 (Fig.~\ref{fig:18dyk_manga_data}A). Using the reported coordinates of \dyk, the local environment of \dyk\ is within the central spaxel (31,31), coincident with the galactic nucleus. 

We investigate Baldwin-Phillips-Terlevich \citep[BPT;][]{baldwin_1981_ClassificationParametersEmissionline} and WH$\alpha$ versus $\Nii$/H$\alpha$ \citep[WHAN;][]{cidfernandes_2010_AlternativeDiagnosticDiagrams,cidfernandes_2011_ComprehensiveClassificationGalaxies} diagnostics produced from each spaxel, with a focus on the galactic nucleus (Fig.~\ref{fig:18dyk_manga_data}B~--~\ref{fig:18dyk_manga_data}C). We also explore the stellar velocity dispersion (Fig.~\ref{fig:18dyk_manga_data}D) and D4000 spectral index (Fig.~\ref{fig:18dyk_manga_data}E) measurement across the galaxy.

The per spaxel MaNGA BPT diagram (Fig.~\ref{fig:18dyk_manga_data}B) indicates that \dykhost\ is predominantly a star-forming galaxy with LINER emission diagnostics that serve as potential indications of some AGN activity in its nucleus. The source of these LINER emission-line signatures in \dykhost\ is actively debated in the literature. \citet{frederick_2019_NewClassChanginglook} proposed that \dyk\ was the result of an `AGN turn-on' event where a low intensity AGN-LINER flared into a more active narrow-lined Seyfert-1 type nucleus. In contrast, \citet{huang_2023_AT2018dykRevisitedTidal} favoured a TDE occurring within a LINER galaxy. \citet{masterson_2024_NewPopulationMidinfraredselected} classified \dyk\ as a possible TDE, but did not include it in their `gold' sample of MIR identified TDEs given its potential as an AGN produced contaminant. 

The source of LINER emission can be difficult to identify, with both weak AGN activity and ionization from older stellar populations producing similar ionization signatures. The WHAN diagram was devised to help break this degeneracy and enable more robust classification using two typically strong emission lines, H$\alpha$ and \Nii, with the width of H$\alpha$ and the line ratio of the two being the properties used for classification. Specifically, these diagnostics can be used to delineate LINER emission into two categories: those that are produced by `weak AGN (wAGN)' and those generated by evolved stellar populations, so-called `retired galaxies (RG)'. The WHAN diagram also classifies galaxies as `pure starforming (PSF)', `strong AGN (sAGN)' - analogous to the Seyfert classification on a BPT diagram - and `passive galaxies (PG)'. 
\citet{frederick_2019_NewClassChanginglook} explored the WHAN classification of \dykhost\ as a whole using the SDSS spectrum and determined a somewhat ambiguous RG/wAGN classification. Here we utilise the available MaNGA data to better explore the galaxy's nuclear region. The MaNGA data provide a clear RG classification for the location of \dyk\ and its surroundings, with a SNR per spaxel exceeding 20. Additionally, we have explored the MaNGA AGN Catalog \citep{comerford_2020_Catalog406AGNs}, which includes WISE colour and X-ray and radio diagnostics to identify AGNs within MaNGA observations without using emission line ratios. \dykhost\ is not classified as an AGN by any of these diagnostics and is thus not included in the catalog. Based on the MIR behaviour shown in Section~\ref{subsec:MIR_Evolution} and the discussion above, we conclude that the LINER emission classification for \dykhost\ is the result of an evolved stellar population rather than underlying AGN activity. This conclusion supports the assessment of \citet{huang_2023_AT2018dykRevisitedTidal}.

Initially noted by \citet{arcavi_2014_CONTINUUMHeRICHTIDAL} and expanded upon by \citet{french_2016_TidalDisruptionEvents}, \citet{law-smith_2017_TidalDisruptionEvent}, and \citet{graur_2018_DependenceTidalDisruption}, TDEs are over-represented in quiescent Balmer-strong galaxies (also known as post-starburst or E+A galaxies). \citet{masterson_2024_NewPopulationMidinfraredselected} investigated whether \dykhost\ could be categorised as such a galaxy using the original SDSS spectrum. Whilst the SDSS-Legacy spectrum points to a quiescent central region, there was no indication of the strong Balmer absorption required for a post-starburst classification. We extend this analysis through the MaNGA data to explore whether the local environment of \dyk\ is consistent with this type of stellar population using the spaxel at the location of AT~2018dyk. We find that whilst, like the larger region covered by the SDSS-Legacy spectrum, this central region hosting \dyk is quiescent, with an H$\alpha$ equivalent width of 1.95~$\pm$~0.05~\AA, it again lacks the strong Balmer absorption, with a measured H$\delta_A$ spectral index of 1.28~$\pm$~0.16~\AA, compared to the \citet{french_2016_TidalDisruptionEvents} Balmer-strong threshold of >4~\AA.

We also make use of the MaNGA data to investigate the stellar velocity dispersion, central SMBH mass, and the spatially resolved D4000 spectral index of \dykhost, with these spectroscopically derived host galaxy properties summarised in Table~\ref{tab:spec_host_info}. We find that the central region of \dykhost\ has a stellar velocity dispersion of 122.5~$\pm$~1.9~\kms\ when measured at the spaxel at the coordinates of \dyk. This is consistent with, though higher than the mean, stellar velocity dispersions of the TDE hosts reported by \citet{graur_2018_DependenceTidalDisruption}. When all measured spaxels are considered, we measure a mean value of 85.0~$\pm$~0.2~\kms\ which remains higher than the mean stellar velocity dispersion measured by \citet{graur_2018_DependenceTidalDisruption} of 68.0~$\pm$~1.1~\kms.

To provide an estimate of the central SMBH mass, we use the MaNGA stellar velocity dispersion measurements and the scaling relation from \citet{kormendy_2013_CoevolutionNotSupermassive}. When using the mean stellar velocity measured from all spaxels, a mass of 6.86~$\pm$~0.05~log$_{10}$(\msol) is obtained. A higher mass estimate of 7.56~$\pm$~0.04~log$_{10}$(\msol) is measured when only the central spaxel is considered, corresponding to the higher central velocity dispersion. We adopt the lower mass estimate for our analysis in Section~\ref{subsec:emission_line_behaviour}, as the \citet{kormendy_2013_CoevolutionNotSupermassive} relation was determined using full galaxy, rather than local, spectra.

The D4000 spectral index (a measure of the continuum difference before and after the 4000~\AA\ spectral break) has long been used as a proxy for stellar population ages \citep{poggianti_1997_IndicatorsStarFormation}. Low values of the D4000 index ($\sim$~<~1.6) indicate young stellar populations (which should be closely matched to starforming regions identified through other methods), whilst larger values of D4000 indicate older stellar populations ($\sim$~>~1.6) \citep{loubser_2016_regulationstarformation}. The resolved map of the D4000 index values (Fig.~\ref{fig:18dyk_manga_data}E) matches the properties shown by both the BPT and WHAN diagnostics. The visible spiral arms and outer regions of \dykhost\ have the low D4000 values expected of young stellar populations with D4000 increasing towards the galactic nucleus - well matched by the `retired galaxy' classification of this region in the WHAN diagram.

\subsection{SMBH and Stellar Mass Estimation with \tdemass}
\label{sec:TDEMass}

For comparison to the spectroscopically derived host galaxy SMBH masses, we utilise the \tdemass\ python code \citep{ryu_2020_MeasuringStellarBlack} to estimate both the SMBH mass and the mass of the star undergoing disruption in \dyk. \tdemass\ is an implementation of the slow circularisation and shock driven model for a TDE's optical luminosity, as described by \citet{piran_2015_DISKFORMATIONDISKa}.

We use the peak blackbody luminosity as calculated by \citet{huang_2023_AT2018dykRevisitedTidal} and assume a peak blackbody temperature matching the earliest measurement calculated by \citet{hinkle_2021_SwiftFixNuclear} ($\sim$~24500~$_{-1100}^{+1200}$~K at $\sim$~20~d post peak) as the input parameters for \tdemass. We use both the original \tdemass\ model described by \citet{ryu_2020_MeasuringStellarBlack} and the new model incorporating slow-cooling described by \citet{krolik_2025_FollowMassConcordance} to provide SMBH mass estimates.

\begin{figure*}
    \centering
    \includegraphics[width=\textwidth]{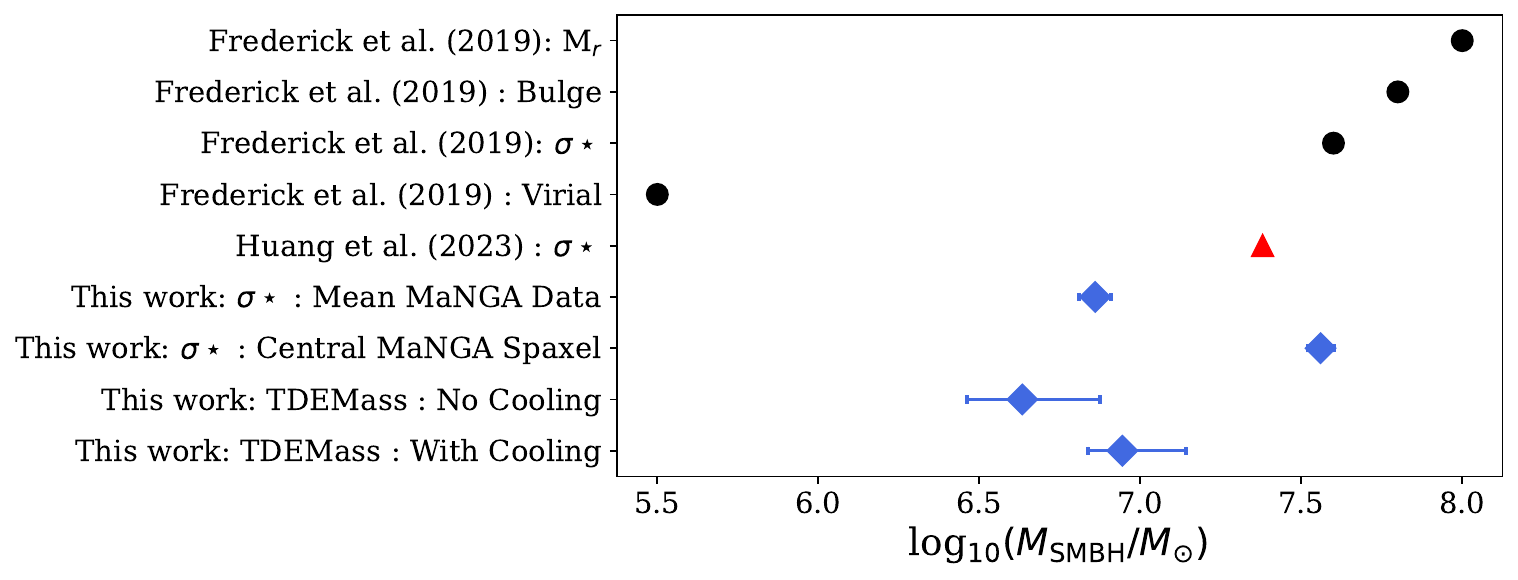}
    \caption{Compiled mass estimates for the SMBH at the centre of \dyk's host galaxy. Data collated from \citet{frederick_2019_NewClassChanginglook}, \citet{huang_2023_AT2018dykRevisitedTidal}, and this work. Note that the error bars reflect measurement uncertainties (where available) only and do not include any statistical uncertainties resulting from intrinsic scatter in the relations used.
    }
    \label{fig:SMBH_Estimates}
\end{figure*}

The model not incorporating slow-cooling finds a SMBH mass of 6.63~$_{+0.17}^{-0.24}$~log$_{10}$(\msol) and the mass of the star undergoing disruption to be 0.4~$_{+0.19}^{-0.18}$~\msol, with the model including cooling measuring the respective parameters to be 6.94~$_{+0.11}^{-0.20}$~log$_{10}$(\msol) and  0.25~$_{+0.15}^{-0.20}$~\msol. Both estimates are on the low end of the range of SMBH black hole estimates for \dyk\ (see Fig.~\ref{fig:SMBH_Estimates}) but are consistent with our estimate measured using the mean stellar velocity dispersion within the MaNGA data.

As described by \citet{ryu_2020_MeasuringStellarBlack}, whilst the dominant factor in the determination of SMBH mass is peak blackbody temperature, for a given temperature, increasing the peak blackbody luminosity will increase the measured SMBH mass. Additionally, the measured values for the stellar mass determined from both models is significantly lower than what is expected of a Sun-like star. The calculation of stellar mass is dominated by peak blackbody luminosity, with the parameters being positively correlated.

From the presence of a significant MIR outburst and CrL emission, it is clear that \dyk\ occurs in a gas-rich environment . Thus, significant absorption of its peak emission could be expected which in turn would lead to a lower measurements of SMBH and stellar mass. As such, direct application of such models should be treated with care as the assumptions used may not be directly applicable to the physical configuration of events such as \dyk.

Alternatively, if the model assumptions do hold, the lower stellar mass (and relatively faint overall luminosity) estimated by \tdemass, could be explained by \dyk\ being produced by a partial disruption of a larger star, as suggested by \citet{huang_2023_AT2018dykRevisitedTidal}.

\section{Discussion}
\label{Sec:Discussion}

\subsection{Optical emission line behaviour}
\label{subsec:emission_line_behaviour}

\begin{figure*}
    \centering
    \includegraphics[width=\textwidth]{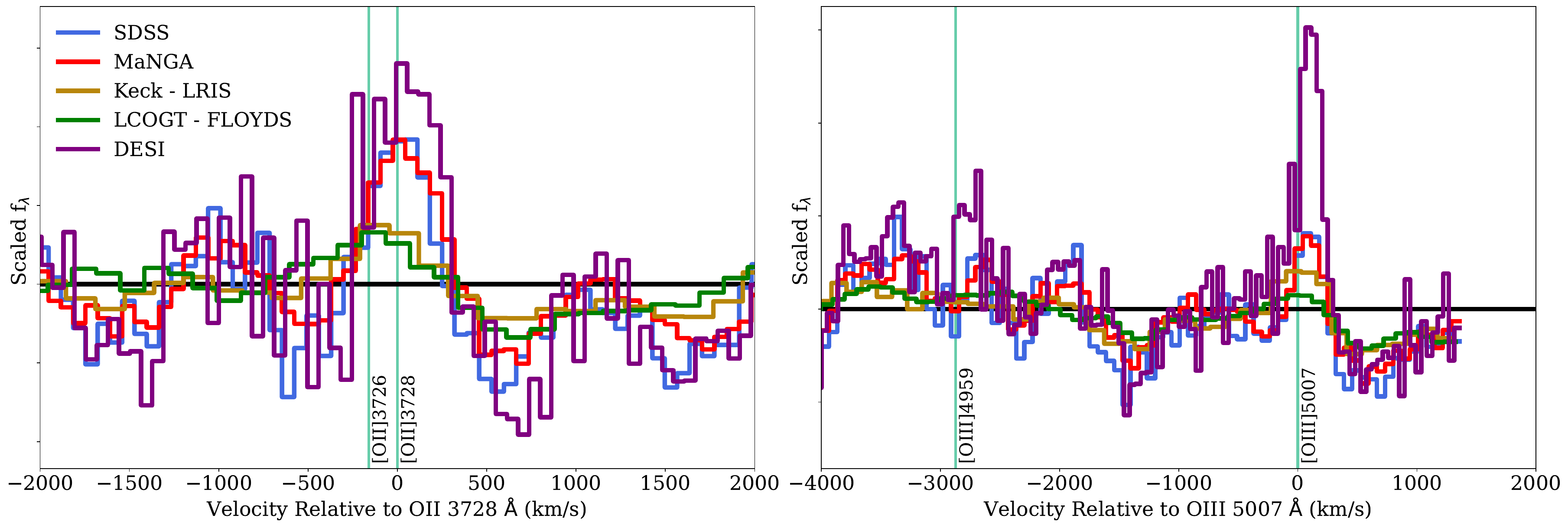}
    \caption{
    \textit{Left:} \Oii~3726+3728~\AA\ doublet line region in velocity space relative to the expected rest position of \Oii~3728~\AA. The relative intensity of the feature drops during outburst before increasing in strength, exceeding the relative intensity of the feature prior to the original outburst.
    \textit{Right:} Line region of the \Oiii~4959 and 5007~\AA\ lines in velocity space relative to the expected position of \Oiii~5007~\AA. The \Oiii~4959~\AA\ line is only observable in the most recent DESI spectrum. A sharp increase in \Oiii~5007~\AA\ emission is clearly observable in the most recent DESI spectrum. Slight redshifts ($\sim$~100~\kms) in the observed line peaks are seen in both line regions. These are explored in detail in Section~\ref{subsec:emission_line_behaviour}.}
    \label{fig:18dyk_oxygen_velocity_comparison}
\end{figure*}

Increasing \Oiii~5007~\AA\ emission has previously been observed in TDE-ECLEs, with this behaviour first identified by \citet{yang_2013_LONGTERMSPECTRALEVOLUTION} in follow-up spectra obtained several years after the \citet{wang_2012_EXTREMECORONALLINE} SDSS discovery spectra. The most notable example of this evolution is seen in the TDE-ECLE SDSS~J1342+0530 \citep{clark_2024_Longtermfollowupobservations}, in which the \Oiii~5007~\AA\ emission line now dominates the spectrum. This increase in \Oiii~5007~\AA\ line strength in SDSS J1342+0530 has also been accompanied by the expected increase in the linked \Oiii~4959~\AA\ emission. The \Oiii~4959~\AA\ emission line is not seen in pre-DESI spectra of \dyk\ as the overall weakness of the feature results in a low SNR. As previously noted, a similar increase in line strength is observed post outburst in the \Oii~3726, 3728~\AA\ doublet with both this and the other oxygen features also being redshifted from the expected 0~\kms\ position by $\sim$~100\kms. We show all three of these oxygen features in velocity space in Fig.~\ref{fig:18dyk_oxygen_velocity_comparison}.

To investigate this emission behaviour further, we conduct line fitting of the three oxygen features showing increased line fluxes post outburst and perform a comparative analysis with SDSS~J1342+0530. Given the small difference in phase between the Keck+LRIS and LCOGT+FLOYDS spectra of \dyk\ and the significantly higher resolution of the Keck+LRIS spectrum, the LCOGT+FLOYDS spectrum is not included in this analysis. For the same reasons, we prefer the DESI spectrum of SDSS J1342+0530 over the one taken with NTT+EFOSC2. Additionally, where possible, we analogously explore the properties of the standard AGN diagnostic lines along with the coronal Fe lines in each spectrum. The emission line fitting makes use of the \textsc{SPECUTILS} \citep{earl_2024_astropyspecutilsv1150} Python package (in turn relying on \textsc{ASTROPY}; \citealt{astropycollaboration_2013_AstropyCommunityPython, astropycollaboration_2018_AstropyProjectBuilding, astropycollaboration_2022_AstropyProjectSustaining}) and includes local region continuum fitting and removal. Given the difficulty in correcting for both stellar continua and the presence of transient emission, we do not attempt to correct the spectra of either object for these effects. Instead, we present these measurements as observed, with a focus on the relative changes in observed features and on their velocity profiles. Specifically for \dyk, the lack of directly observable H$\beta$ emission at phases other than during the early evolution of the TDE outburst, which is dominated by TDE rather than host galaxy flux, prevents the construction of the usual optical diagnostic diagrams on a time varying basis. \citet{frederick_2019_NewClassChanginglook} provide measurements of these corrected line ratios (see, e.g., figures 13, 18--21), work which we do not duplicate here. As our spectral comparisons have shown, the SDSS Legacy, central spaxel MaNGA, and DESI spectra are all very similar. As such, no significant changes in measured line ratios (stellar absorption corrected or otherwise) are expected, with the exception of those involving oxygen, which as previously described, showed significant evolution by the time of the DESI spectrum.

For the \Oii~3726,3728~\AA\ doublet fit, due to the resolution of our spectra, we use a single Gaussian to represent both lines. Additionally, the lines composing the \Sii\ doublet and the pair of \Nii\ lines in the H$\alpha$ complex are tied to have the same width in wavelength space. Where one or more lines are blended, the components are fitted simultaneously. An example of the fitting results is shown in Fig.~\ref{fig:18dyk_example_fit}. The full results of this fitting are given in Appendix~\ref{Appendix:Emission Line Fitting}, split by non-coronal (Table~\ref{tab:Appendix_Tab_Non_CL_Results}) and coronal emission lines (Table~\ref{tab:Appendix_Tab_CL_Results}). A summary of the resulting line ratios is provided in Table~\ref{tab:Appendix_Tab_Line_Ratios}.

\begin{figure}
    \centering
    \includegraphics[width=\columnwidth]{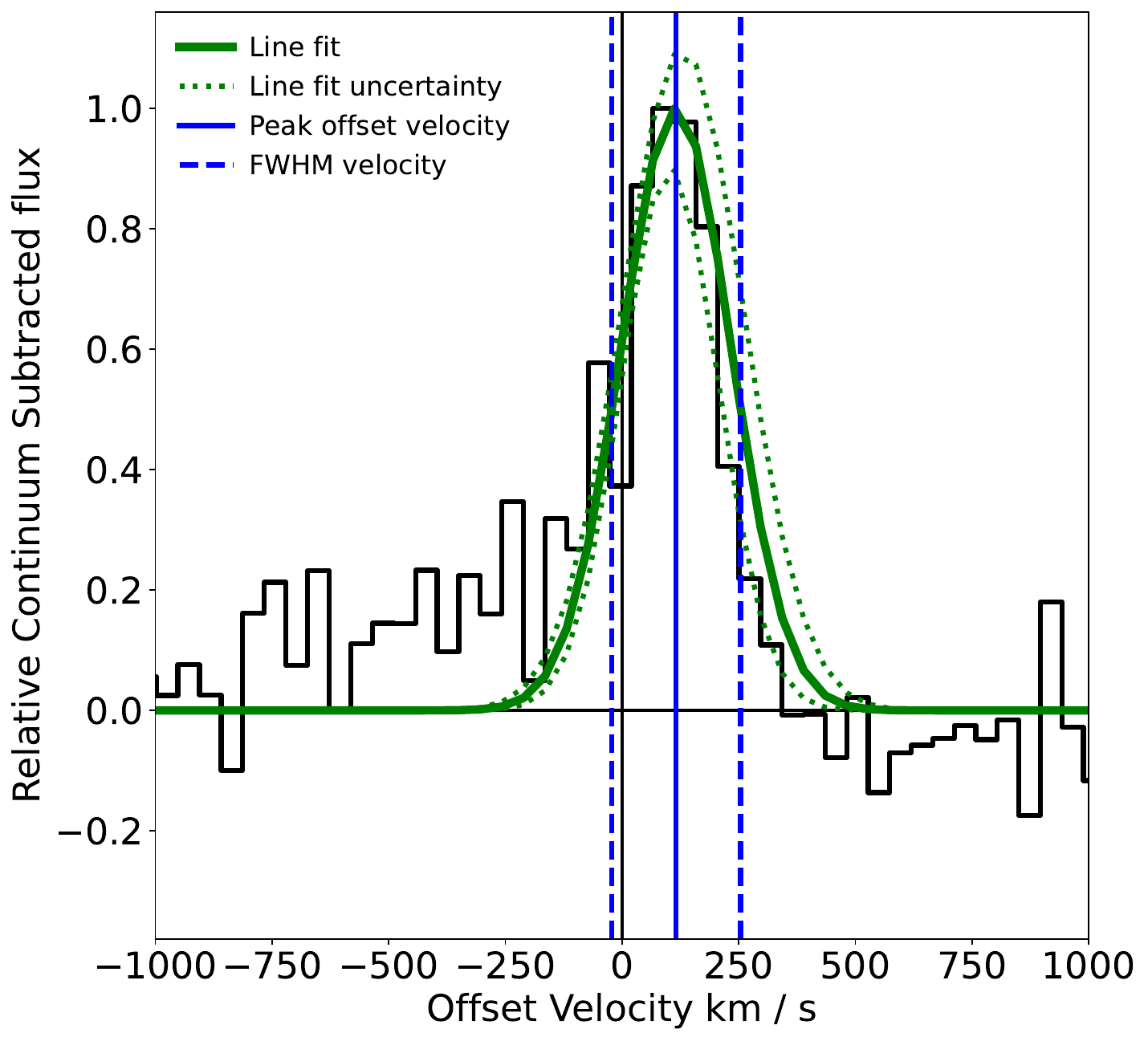}
    \caption{Fit to the \Oiii~5007\AA\ emission line in the DESI spectrum of \dyk.}
    \label{fig:18dyk_example_fit}
\end{figure}

As the latest DESI spectrum now closely matches the pre-outburst SDSS spectrum (with the exception of this increased \Oiii\ emission), and the fitted \Oiii~5007~\AA\ feature is consistent in both peak offset (DESI: 120~$\pm$~10~\kms\ compared to SDSS: 110~$\pm$~20~\kms) and FWHM velocity (DESI: 280~$\pm$~30~\kms\ compared to SDSS: 290~$\pm$~40~\kms) with the pre-outburst SDSS spectrum, we consider the most likely cause of this increased \Oiii\ emission to be the delayed response of more distant material to the transient TDE flux rather than a change in the long-term accretion behaviour of the SMBH within \dykhost. If such a change had occurred, we would expect to also see changes in the other AGN diagnostic lines relative to the SDSS spectrum which is not observed (e.g., \Nii\ and \Sii\ as seen in the bottom panel of Fig.~\ref{fig:18dyk_zoom_spec_evolution}).

Additionally, in the Keck spectrum the observed coronal lines are of comparable strength to, though wider than, the \Oiii~5007~\AA\ emission line. Differences in the FWHM velocities between coronal and non-coronal emission lines were also observed in the CrL-TDE TDE~2022upj and investigated by \citet{newsome_2024_MappingInner01}. 

\begin{figure*}
    \centering
    \includegraphics[width=\textwidth]{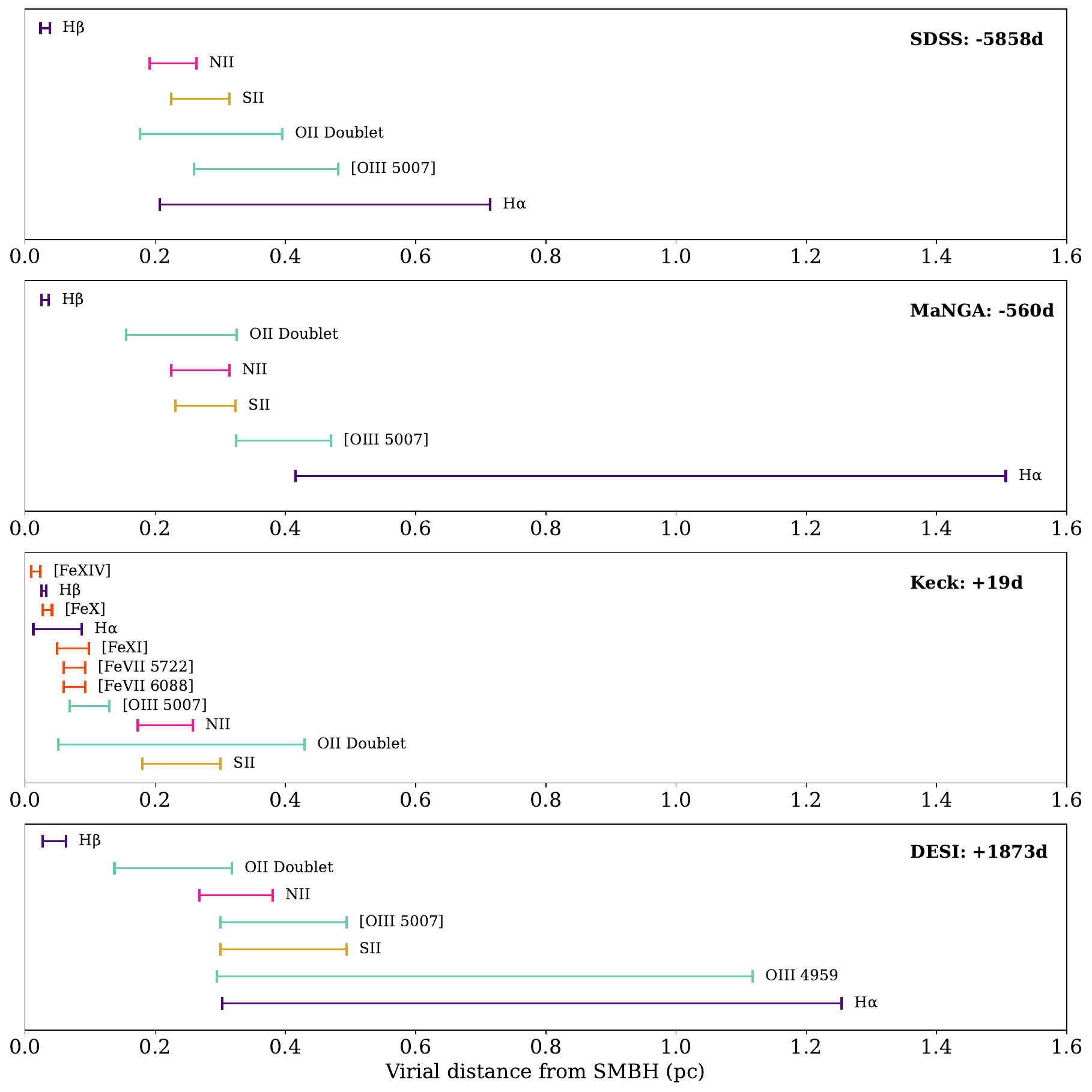}
    \caption{Virial distance estimates for the formation location of the measured narrow emission lines based on their FWHM velocities. Note the scale difference between the lines observed in the Keck spectrum compared the others. This difference is the result of several of the lines in the Keck spectrum being directly affected by the ongoing active TDE phase.}
    \label{fig:line_distances}
\end{figure*}

Following \citet{newsome_2024_MappingInner01}, we use the measured FWHM velocities and our estimate of the SMBH mass determined using the mean MaNGA stellar velocity dispersion data to calculate virial radii around the SMBH, and hence the distance between the SMBH and the emitting material. This calculation was performed for a range of emission lines for each spectroscopic observation of \dyk. Here we again favour the higher resolution Keck spectrum over the similarly timed FLOYDS spectrum. We find that the emission locations of the spectral lines are consistent within the uncertainties between the pre (SDSS and MaNGA) and post (DESI) spectra of \dyk\ (Table~\ref{tab:Appendix_Line_Distances} and Fig.~\ref{fig:line_distances}). The \Oiii~4959~\AA\ line that had developed by the time of the DESI spectrum is also consistent in distance with the other narrow lines. Both the \Oiii~4959~\AA\ and \Oiii~5007~\AA\ lines are located at greater distances from the SMBH than the coronal iron features, consistent with their more delayed line strength increases in response to the TDE outburst. In the Keck spectrum, H$\alpha$ and \Oiii~5007~\AA\ have increased FWHMs with a resulting decrease in the calculated virial distances, which we attribute to the effect of the ongoing TDE. 

As with the measured locations of the coronal lines in TDE~2022upj \citep{newsome_2024_MappingInner01}, we find that the \Fexiv\ emission is located nearest to the black hole, followed by the \Fex\ line. The \Fevii\ lines (the \Fevii~3759~\AA\ line is not included in Fig.~\ref{fig:line_distances} due to its much larger uncertainty but is consistent with the other \Fevii\ lines) and \Fexi~7894~\AA\ are located at the largest distances from the black hole. However, we note that the lower SNR of the \Fexi~7894~\AA\ makes its true distance harder to measure as the FWHM velocity may be underestimated. This layered line location structure is similar to the line structure of TDE~2022upj observed by \citet{newsome_2024_MappingInner01}, indicating similar complexities in the gas structure close to the SMBH. However, the distances determined here for the line formation in \dyk\ are larger than those determined for TDE~2022upj. This discrepancy may be due to the larger mass of the SMBH in \dyk's host galaxy. 

Depending on the mass estimate of TDE~2022upj used, we find the mass ratio between the SMBHs responsible for \dyk\ (when using the value measured from the mean MaNGA stellar velocity dispersion) and TDE~2022upj to be 3.6~--~14.5. The corresponding mean ratio between the emission line distances is 0.9~--~3.5. In both cases, the mean line distance ratio is found to be lower than SMBH mass ratio. One would naively expect a correlation between the mass of the SMBH and the distance to its surrounding ISM; measurements of these ratios in future CrL-TDEs could reveal the exact nature of such a correlation and constrain its underlying physics.

\subsection{MIR photometric behaviour}
\label{subsec:MIR_Photometric_Behaviour}

As described in Section~\ref{subsec:MIR_Evolution}, pre-outburst MIR photometry of AT~2018dyk suggests a galaxy with little underlying variability; a \textit{W1}~-~\textit{W2} colour consistent with no significant AGN activity; and with AllWISE photometry consistent with a spiral galaxy hosting no AGN activity (obscured or otherwise) based on the classification scheme of \citet{wright_2010_WIDEFIELDINFRAREDSURVEY}. During the outburst, \dyk, like other CrL-TDEs, reddened significantly in \textit{W1}~-~\textit{W2} colour, whilst remaining bluer than the \citet{stern_2012_MIDINFRAREDSELECTIONACTIVE} AGN colour cut. ECLE-TDEs and other CrL-TDEs have shown stronger \textit{W1}~-~\textit{W2} reddening during outburst and cross the \citet{stern_2012_MIDINFRAREDSELECTIONACTIVE} colour cut.

This, in combination with the smaller amplitude of the MIR flare of \dyk\ compared to the other coronal line displaying objects in the comparison sample, and the shorter overall duration of its outburst, points to differences in the physical structure of the material responsible for reprocessing the TDE flux into the observed MIR emission. These differences are most likely a combination of reduced covering factor, density, and overall mass of the material surrounding the SMBH. A detailed physical modelling of the ISM around the SMBH may be able to break the degeneracies between these variables, but it is beyond the scope of this work.

When explored collectively, trends are evident in the MIR behaviour of CrL-TDEs (Fig.~\ref{fig:MIR_Delta_Fig}). Firstly, the relationship between $\Delta$\textit{W1} and $\Delta$\textit{W2} is not one-to-one. Instead, the quadratic curve best-fit to the data (as determined by a likelihood-ratio test), reflects the increased strength of the outbursts in the \textit{W2} band compared to \textit{W1} and shown in Equation~\ref{eqn:delta_w1_vs_w2}. 

\begin{equation}
\label{eqn:delta_w1_vs_w2}
     \Delta \mathit{W2} = -0.18(\Delta \mathit{W1})^{2} + 0.7(\Delta \mathit{W1}) - 0.48
\end{equation}

When examined further in $\Delta$(\textit{W1-W2}) vs $\Delta$\textit{W2} colour space, those CrL-TDEs with the brightest MIR outbursts are also seen to have the reddest outbursts. This colour-luminosity relation is best fitted by the linear relation:

\begin{equation}
\label{eqn:delta_w1_minus_w2_vs_w2}
     \Delta(\mathit{W1}-\mathit{W2}) = -0.25(\Delta \mathit{W2}) + 0.19
\end{equation}

Unfortunately, given that only four objects in this parameter space can be used for fitting this relation (the rest are only limits), it is not statistically significant at this time. Whilst further observations will be required to confirm this relation statistically, qualitatively such a relation is likely the result of larger quantities or densities of circumnuclear material near the SMBHs involved, or larger covering fractions reprocessing more TDE emission into the MIR bands (e.g., \citealt{hinkle_2024_MidInfraredEchoesAmbiguous}). We make two further observations on the robustness of this observed relation. The fourth object included in the fit for this relationship (AT~2018gn) was added to the comparison sample during internal review and agreed well with the initial identification of the relationship with the other three included transients (AT~2017gge, \dyk, and AT~2018bcb). Moreover, three additional CrL-TDEs are observed to be close to (or indeed at) their outburst colour peak at the time of most recent MIR observations (TDE~2022fpx, TDE~2022upj, and VT~J1548), though would require additional observations to confirm this. All of these objects have already reached clear peaks in their \textit{W2} light curves (i.e., their $\Delta$\textit{W2} values are fixed); further increases in their $\Delta$(\textit{W1-W2}) values would improve their already close scatter from the determined relation. This is further evidence that MIR observations of this class of objects are essential to fully understand their properties and evolution, with the end of the NEOWISE-R mission opening a significant gap in our observational capabilities.

We also note that all known CrL-TDEs have displayed MIR outbursts in both \textit{W1} and \textit{W2} bands greater than 0.5~mag, consistent with both coronal emission lines and MIR outbursts requiring significant amounts of material in proximity to the SMBH for the reprocessing of TDE emission. Other TDE's (such as TDE~2019azh) show that small MIR outbursts can occur without the development of coronal lines, hinting at a range of environmental configurations and differing thresholds for the occurrence of each feature.

\section{Conclusions}
\label{Sec:Conclusions}

In this work, we have published a new follow-up spectrum of the host galaxy of the nuclear transient \dyk. Using this spectrum, as well as archival data, we have conducted a thorough analysis of the photometric and spectroscopic properties of \dyk\ and its host galaxy. We make the following observations.

We conclude that \dyk\ is the result of a TDE occurring in an environment rich in circumnuclear material, resulting in the reprocessing of TDE flux into both high ionisation Fe coronal lines and a MIR outburst. Spectroscopic analysis of the local region of the host galaxy (\dykhost) in which \dyk\ occurred reveals that the previously reported LINER emission signatures are the result of an evolved stellar population rather than underlying AGN activity.

Additionally, \dykhost - sits within or close to the transitional `green valley', with a centrally concentrated stellar population housing an older quiescent population in its nucleus and some star-formation still ongoing within the outer spiral arms. An analysis of the host properties of \dykhost\ shows that it is consistent with known TDE host populations, though it lies toward the high end of the expected stellar mass range \citep{law-smith_2017_TidalDisruptionEvent, graur_2018_DependenceTidalDisruption, hammerstein_2023_FinalSeasonReimagined}.

Furthermore, the optical evolution of \dyk\ itself is consistent with an outburst produced by a single TDE. The most recent DESI spectrum now closely matches the archival, pre-outburst SDSS Legacy and MaNGA spectra, with the exceptions of significant increases in strength of the \Oiii~5007~\AA\ and \Oii~3728~\AA\ emission features. Similar strengthening of \Oiii~5007~\AA\ emission in particular has been observed in the original TDE-linked ECLE sample by \citet{yang_2013_LONGTERMSPECTRALEVOLUTION} and \citet{clark_2024_Longtermfollowupobservations}. Given the lack of evolution in other AGN diagnostic lines compared to the SDSS spectrum, we attribute this increased emission to more distant material responding to the short-term TDE emission rather than a long-term change of AGN activity state. Emission line fitting also confirms that \dyk\ matches the qualitative definition of an ECLE, with one or more coronal lines being at least 20 per cent the line strength of \Oiii~5007\AA\ at the time of outburst.

In addition to this optical behaviour, \dyk\ displayed MIR evolution consistent with a TDE-linked ECLE: a $\sim$~0.5~mag outburst in both the \textit{W1} and \textit{W2} bands, with an accompanying \textit{W1}-\textit{W2} reddening of $\sim$~0.4~mag. However, this outburst was on a shorter timescale, and had a lower amplitude, than observed in archival TDE linked ECLEs \citep{dou_2016_LONGFADINGMIDINFRARED, clark_2024_Longtermfollowupobservations}. Photometric variability analysis of \dyk\ and the other CrL-TDEs shows lower levels of variability prior to and post outburst than has been seen in either CL-LINERs or AGN-ECLEs, aiding in the confirmation that these classes of transient are distinct from one another in observable properties.

Virial distance estimates based on the FWHM velocities of the narrow emission lines and SMBH masses suggest that the gas structures around the SMBHs responsible for \dyk\ and TDE~2022upj are similar. Whilst the calculated distances to the emitting material are larger in \dyk, the material is relatively closer to the SMBH compared to the mass ratio of the two SMBHs.

Comparisons between \dyk\ and other TDEs displaying MIR outbursts reveal tentative evidence for a colour-luminosity relationship in their MIR evolution. Specifically, those objects with brighter MIR flares show more significant reddening at outburst. This may be a consequence of such objects having larger dust covering fractions or other environmental factors that lead to reprocessing more of the initial UV/optical continuum produced by the TDE. Whilst the number of objects for which these measurements can be made without relying on upper limits is limited (four) and prevents the statistical confirmation of the relation, the range of behaviour already extends over 3~mag in \textit{W2} outburst amplitude, highlighting the diverse range of physical configurations in these systems. MIR observations play a critical role developing our understanding of these objects; the end of the NEOWISE-R mission has created a key gap in our observational capabilities.

In conclusion, we find that \dyk\ was a TDE that went on to excite iron coronal lines in its host galaxy. This object continues to strengthen the link between TDEs and variable ECLEs. It also stresses the necessity for long-term, multi-wavelength follow-up of nuclear transients over a timescale of years in order to properly classify them and study the environments in which they occur.

\section*{Acknowledgements}
\label{Sec:Acknowledgements}

We thank the anonymous reviewer for their helpful comments and feedback.

PC was supported by the Science \& Technology Facilities Council grant ST/V001000/1.

PC and OG were supported by the Science \& Technology Facilities Council [grants ST/S000550/1 and ST/W001225/1].

CG thanks the Science Technology Facilities Council (STFC) for support from the Durham consolidated grant (ST/T000244/1).

This material is based upon work supported by the U.S. Department of Energy (DOE), Office of Science, Office of High-Energy Physics, under Contract No. DE–AC02–05CH11231, and by the National Energy Research Scientific Computing Center, a DOE Office of Science User Facility under the same contract. Additional support for DESI was provided by the U.S. National Science Foundation (NSF), Division of Astronomical Sciences under Contract No. AST-0950945 to the NSF’s National Optical-Infrared Astronomy Research Laboratory; the Science and Technology Facilities Council of the United Kingdom; the Gordon and Betty Moore Foundation; the Heising-Simons Foundation; the French Alternative Energies and Atomic Energy Commission (CEA); the National Council of Humanities, Science and Technology of Mexico (CONAHCYT); the Ministry of Science, Innovation and Universities of Spain (MICIU/AEI/10.13039/501100011033), and by the DESI Member Institutions: \url{https://www.desi.lbl.gov/collaborating-institutions}. Any opinions, findings, and conclusions or recommendations expressed in this material are those of the author(s) and do not necessarily reflect the views of the U. S. National Science Foundation, the U. S. Department of Energy, or any of the listed funding agencies.

The authors are honored to be permitted to conduct scientific research on Iolkam Du’ag (Kitt Peak), a mountain with particular significance to the Tohono O’odham Nation.

Funding for the Sloan Digital Sky 
Survey IV has been provided by the 
Alfred P. Sloan Foundation, the U.S. 
Department of Energy Office of 
Science, and the Participating 
Institutions. 

SDSS-IV acknowledges support and 
resources from the Center for High 
Performance Computing at the 
University of Utah. The SDSS 
website is www.sdss4.org. SDSS-IV is managed by the 
Astrophysical Research Consortium 
for the Participating Institutions 
of the SDSS Collaboration including 
the Brazilian Participation Group, 
the Carnegie Institution for Science, 
Carnegie Mellon University, Center for 
Astrophysics | Harvard \& 
Smithsonian, the Chilean Participation 
Group, the French Participation Group, 
Instituto de Astrof\'isica de 
Canarias, The Johns Hopkins 
University, Kavli Institute for the 
Physics and Mathematics of the 
Universe (IPMU) / University of 
Tokyo, the Korean Participation Group, 
Lawrence Berkeley National Laboratory, 
Leibniz Institut f\"ur Astrophysik 
Potsdam (AIP), Max-Planck-Institut 
f\"ur Astronomie (MPIA Heidelberg), 
Max-Planck-Institut f\"ur 
Astrophysik (MPA Garching), 
Max-Planck-Institut f\"ur 
Extraterrestrische Physik (MPE), 
National Astronomical Observatories of 
China, New Mexico State University, 
New York University, University of 
Notre Dame, Observat\'ario 
Nacional / MCTI, The Ohio State 
University, Pennsylvania State 
University, Shanghai 
Astronomical Observatory, United 
Kingdom Participation Group, 
Universidad Nacional Aut\'onoma 
de M\'exico, University of Arizona, 
University of Colorado Boulder, 
University of Oxford, University of 
Portsmouth, University of Utah, 
University of Virginia, University 
of Washington, University of 
Wisconsin, Vanderbilt University, 
and Yale University.

The Liverpool Telescope is operated on the island of La Palma by Liverpool John Moores University in the Spanish Observatorio del Roque de los Muchachos of the Instituto de Astrofisica de Canarias with financial support from the UK Science and Technology Facilities Council.

\section*{Data Availability}
The data underlying this work are available in the article and in its online supplementary material available through Zenodo \citep{clark_2025_2018dyktidaldisruption}. 



\bibliographystyle{mnras}
\bibliography{MyLibrary.bib} 



\clearpage
\appendix

\section{Sample Information}
\label{Appendix:Sample_Info}

In this appendix, we provide summary information on the properties of all objects used as the comparison sample within this work.

\begin{table*}
\caption{Summary information for the comparison sample of other objects used in this paper. AT~2018dyk is included for convenient reference.}
\label{Tab:Object_Summary_Info}
\begin{tabular}{lcccc}
\hline
Object & RA (J2000) & Dec (J2000) & Redshift ($z$) & Host Name \\ \hline
\textbf{AT~2018dyk} & 15:33:08.0149 & +44:32:08.2039 & 0.037 & SDSS~J153308.01+443208.4 \\
\\
\textbf{CrL-TDEs} \\
AT~2017gge & 16:20:34.9900 & +24:07:26.5000 & 0.067 & SDSS~J162034.99+240726.5 \\
AT~2018gn & 01:46:42.4500 & +32:30:29.3004 & 0.037 & 2MASX J01464244+3230295
\\
AT~2018bcb & 22:43:42.8710 & -16:59:08.4913 & 0.120 & 2MASX~J22434289-1659083 \\
AT~2021dms & 03:21:24.0695 & -11:08:45.7120 & 0.031 & MCG-02-09-033 \\
AT~2021acak & 10:34:47.9900 & +15:29:22.4200 & 0.136 & SDSS~J103447.90+152922.4 \\
TDE~2022fpx & 15:31:03.7420 & +53:24:19.1800 & 0.073 & SDSS~J153104.92+532409.2 \\
TDE~2022upj & 00:23:56.8459 & -14:25:23.2198 & 0.054 & SDSS~J002356.88-142524.0 \\
\\
\textbf{TDE-ECLEs}\\
SDSS~J0748+4712 & 07:48:20.6668 & +47:12:14.2648 & 0.062 & 2MASS~J07482067+4712138 \\
SDSS~J0952+2143 & 09:52:09.5629 & +21:43:13.2979 & 0.079 & 2MASS~J09520955+2143132 \\
SDSS~J1241+4426 & 12:41:34.2561 & +44:26:39.2636 & 0.042 & LEDA~2244532\\
SDSS~J1342+0530 & 13:42:44.4150 & +05:30:56.1451 & 0.037 & 2MASX~J13424441+0530560\\
SDSS~J1350+2916 & 13:50:01.4946 & +29:16:09.6460 & 0.078 & 2MASS~J13500150+2916097\\
\\
\textbf{CrL-TDEs / CrL-AGN} \\
VT~J154843.06+220812.6 & 15:48:43.0662 & +22:08:12.6866 & 0.031 & SDSS~J154843.06+220812.6 \\
\\
\textbf{Multi-epoch CrL-TDEs} \\
AT~2019avd & 08:23:36.7674 & +04:23:02.4598 & 0.028 & SDSS J082338.23+042258.3 \\
TDE~2019qiz & 04:46:37.8800 & -10:13:34.9000 & 0.015 & 2MASX J04463790-1013349 \\
\\
\textbf{AGN-ECLEs} \\
SDSS~J0938+1353 & 09:38:01.6376 & +13:53:17.0423 & 0.101 & SDSS~J093801.63+135317.0\\
SDSS~J1055+5637 & 10:55:26.4177 & +56:37:13.1010 & 0.074 & SDSS~J105526.41+563713.1 \\
\\
\textbf{NonCrL-TDEs} \\
TDE~2019azh & 08:13:16.9450 & +22:38:54.0300 & 0.022 & KUG~0810+227 \\
\\
\textbf{CL-LINERs} \\
iPTF16bco & 15:54:40.2643 & +36:29:51.9540 & 0.237 & SDSS~J155440.25+362952.0 \\
AT~2018gkr & 08:17:26.4190 & +10:12:10.1088 & 0.168 & LEDA~3091244
\\
AT~2018aij & 12:54:03.7882 & +49:14:52.9152 & 0.101 & 2MASX~J12540375+4914533
 \\
AT~2018ivp & 10:40:45.0027 & +26:03:00.0328 & 0.067 & LEDA~1760642
\\
AT~2018lnh & 12:25:50.2978 & +51:08:46.4244 & 0.046 & 2MASS~J12255032+5108463
 \\
ZTF18aasuray & 11:33:55.9457 & +67:01:07.0572 & 0.040 & 2MASX~J11335602+6701073
 \\
\hline
\end{tabular}
\end{table*}

\FloatBarrier
\clearpage
\section{MIR Analysis}

\subsection{Additional Variability Analysis}
\label{Appendix:Additional_MIR_Varaibilty_Analysis}

To better quantify the MIR variability of \dyk\ and the comparison object sample, we explore the standard deviation ($\sigma$) and maximum change in magnitude ($\delta$) in three phases: before outburst (A), during outburst (B), and post-outburst (c). For some objects, one or more of these phases have not been observed and so are omitted from the corresponding tables. The means for each classification of objects are also reported. For both individual WISE filters and in MIR colour, the CrL-TDEs including \dyk, are observed to be less variable than the AGN-ECLEs prior to outburst and display similar colour variability. \dyk\ has also been observed to have returned to pre-outburst levels of variability, with the other CrL-TDEs still in their outburst phases. As expected, during outburst the variability of the CrL-TDEs increases dramatically in both metrics. For comparison processes, \dyk\ is excluded from the calculation of the mean values of the remaining CrL-TDEs.

The $\delta_B$ values (i.e., maximum change in magnitudes during outburst) differ from the $\Delta$ values of overall maximum change as the $\Delta$ calculation includes the quiescent value as a reference point rather than just the internal maximal change in magnitude of the observations comprising the outburst.

\begin{table*}
\caption{Standard deviations of the per-band MIR light curve for each object divided in to three phases. A: Pre-outburst, B: During Outburst and C: Post-outburst. Where an object's per-outburst behaviour has not been observed, `A' phase values are not quoted, likewise for objects not displaying a clear outburst `B' values are not quoted or where an object is still displaying outburst activity at time of the last observation where `C' values are not possible to measure.}
\label{Tab:Varibilty_Sigmas}
\begin{adjustbox}{width=1\textwidth}
\begin{tabular}{lcccccccccc}
\hline
\textbf{Object} & \textbf{Classification} & \textbf{$\sigma$W1\textsubscript{A}} & \textbf{$\sigma$W1\textsubscript{B}} & \textbf{$\sigma$W1\textsubscript{C}} & \textbf{$\sigma$W2\textsubscript{A}} & \textbf{$\sigma$W2\textsubscript{B}} & \textbf{$\sigma$W2\textsubscript{C}} & \textbf{$\sigma$(W1-W2)\textsubscript{A}} & \textbf{$\sigma$(W1-W2)\textsubscript{B}} & \textbf{$\sigma$(W1-W2)\textsubscript{C}} \\ \hline
\\
\textbf{AT~2018dyk} & CrL-TDE & \textbf{0.01} & \textbf{0.14} & \textbf{0.01} & \textbf{0.01} & \textbf{0.27} & \textbf{0.03} & \textbf{0.02} & \textbf{0.14} & \textbf{0.02} \\
 &  &  &  &  &  &  &  &  &  &  \\
AT~2017gge & CrL-TDE & 0.03 & 0.29 & - & 0.05 & 0.29 & - & 0.06 & 0.09 & - \\
AT~2018gn & CrL-TDE & 0.01 & 0.34 & - & 0.02 & 0.48 & - & 0.02 & 0.20 & - \\
AT~2018bcb & CrL-TDE & 0.03 & 0.38 & - & 0.06 & 0.41 & - & 0.06 & 0.09 & - \\
AT~2021dms & CrL-TDE & 0.02 & 0.29 & - & 0.03 & 0.49 & - & 0.04 & 0.23 & - \\
TDE~2022fpx & CrL-TDE & 0.08 & 0.41 & - & 0.10 & 0.61 & - & 0.08 & 0.05 & - \\
TDE~2022upj & CrL-TDE & 0.02 & 0.70 & - & 0.06 & 0.93 & - & 0.05 & 0.25 & - \\
VT~J1548 & CrL-TDE & 0.01 & 0.74 & - & 0.03 & 0.94 & - & 0.03 & 0.24 & - \\
 &  &  &  &  &  &  &  &  &  &  \\
\textbf{Mean} & \textbf{CrL-TDE*} & \textbf{0.03} & \textbf{0.45} & \textbf{-} & \textbf{0.05} & \textbf{0.59} & \textbf{-} & \textbf{0.05} & \textbf{0.15} & \textbf{-} \\ \hline
 &  &  &  &  &  &  &  &  &  &  \\
AT~2019avd & Multi-epoch CrL-TDE / CrL-AGN & 0.01 & 0.28 & - & 0.03 & 0.36 & - & 0.02 & 0.17 & - \\
TDE~2019qiz & Multi-epoch CrL-TDE & 0.02 & 0.24 & - & 0.02 & 0.49 & - & 0.02 & 0.26 & - \\
 &  &  &  &  &  &  &  &  &  &  \\
\textbf{Mean} & \textbf{Multi-epoch CrL-TDE} & \textbf{0.01} & \textbf{-} & \textbf{-} & \textbf{0.02} & \textbf{-} & \textbf{-} & \textbf{0.02} & \textbf{-} & \textbf{-} \\ \hline
 &  &  &  &  &  &  &  &  &  &  \\
SDSS~J0748+4712 & TDE-ECLE & - & 0.21 & - & - & 0.42 & - & - & 0.22 & - \\
SDSS~J0952+2143 & TDE-ECLE & - & 0.15 & - & - & 0.32 & - & - & 0.18 & - \\
SDSS~J1241+4426 & TDE-ECLE & - & 0.08 & - & - & 0.16 & - & - & 0.08 & - \\
SDSS~J1342+0530 & TDE-ECLE & - & 0.19 & - & - & 0.48 & - & - & 0.30 & - \\
SDSS~J1350+2916 & TDE-ECLE & - & 0.25 & - & - & 0.46 & - & - & 0.23 & - \\
 &  &  &  &  &  &  &  &  &  &  \\
\textbf{Mean} & \textbf{TDE-ECLE} & \textbf{-} & \textbf{0.17} & \textbf{-} & \textbf{-} & \textbf{0.37} & \textbf{-} & \textbf{-} & \textbf{0.20} & \textbf{-} \\ \hline
 &  &  &  &  &  &  &  &  &  &  \\
SDSS~J0938+1353 & AGN-ECLE & 0.04 & - & - & 0.03 & - & - & 0.02 & - & - \\
SDSS~J1055+5637 & AGN-ECLE & 0.13 & - & - & 0.14 & - & - & 0.03 & - & - \\
 &  &  &  &  &  &  &  &  &  &  \\
\textbf{Mean} & \textbf{AGN-ECLE} & \textbf{0.08} & \textbf{-} & \textbf{-} & \textbf{0.09} & \textbf{-} & \textbf{-} & \textbf{0.02} & \textbf{-} & \textbf{-} \\ \hline
 &  &  &  &  &  &  &  &  &  &  \\
TDE~2019azh & NonCrL-TDE & 0.02 & - & 0.02 & 0.02 & - & 0.03 & 0.02 & - & 0.02 \\ 
 &  &  &  &  &  &  &  &  &  &  \\ \hline
 &  &  &  &  &  &  &  &  &  &  \\
iPTF16bco ** & CL-LINER & 0.09 & 0.05 & 0.04 & 0.13 & 0.07 & 0.04 & 0.07 & 0.05 & 0.04 \\
AT~2018aij & CL-LINER & 0.03 & 0.11 & 0.03 & 0.06 & 0.12 & 0.04 & 0.06 & 0.04 & 0.00 \\
AT~2018gkr & CL-LINER & 0.30 & - & - & 0.41 & - & - & 0.16 & - & - \\
AT~2018ivp & CL-LINER & - & 0.15 & - & - & 0.17 & - & - & 0.03 & - \\
AT~2018lnh ** & CL-LINER & 0.09 & 0.17 & - & 0.12 & 0.11 &  & 0.06 & 0.09 &  \\
ZTF18aasuray & CL-LINER & - & 0.24 & - & - & 0.34 & - & - & 0.11 & - \\
\\
\textbf{Mean} & \textbf{CL-LINER} & \textbf{0.13} & \textbf{0.15} & \textbf{0.03} & \textbf{0.18} & \textbf{0.16} & \textbf{0.04} & \textbf{0.09} & \textbf{0.07} & \textbf{0.02} \\
\hline
\end{tabular}
\end{adjustbox}
\begin{flushleft}
\textbf{Notes:}\\
* Excluding \dyk\ for comparison purposes. Given its occurrence within an AGN hosting galaxy, AT~2021acak is also excluded from this analysis.\\
** MIR outbursts displayed by these objects and outlined here occur several years after the identification of the `changing look' event.\\
\end{flushleft}
\end{table*}
\raggedbottom

\begin{table*}
\caption{Maximum changes of the per-band MIR light curves for each object divided in to three phases. A: Pre-outburst, B: During Outburst and C: Post-outburst. Where an object's per-outburst behaviour has not been observed, `A' phase values are not quoted, likewise for objects not displaying a clear outburst `B' values are not quoted or where an object is still displaying outburst activity at time of the last observation where `C' values are not possible to measure.}
\label{Tab:Varibilty_Deltas}
\begin{adjustbox}{width=1\textwidth}
\begin{tabular}{lcccccccccc}
\hline
\textbf{Object} & \textbf{Classification} & \textbf{$\delta$W1\textsubscript{A}} & \textbf{$\delta$W1\textsubscript{B}} & \textbf{$\delta$W1\textsubscript{C}} & \textbf{$\delta$W2\textsubscript{A}} & \textbf{$\delta$W2\textsubscript{B}} & \textbf{$\delta$W2\textsubscript{C}} & \textbf{$\delta$(W1-W2)\textsubscript{A}} & \textbf{$\delta$(W1-W2)\textsubscript{B}} & \textbf{$\delta$(W1-W2)\textsubscript{C}} \\ \hline
\\
\textbf{AT~2018dyk} & CrL-TDE & \textbf{0.04} & \textbf{0.38} & \textbf{0.03} & \textbf{0.05} & \textbf{0.63} & \textbf{0.18} & \textbf{0.05} & \textbf{0.35} & \textbf{0.10} \\
 &  &  &  &  &  &  &  &  &  &  \\
AT~2017gge & CrL-TDE & 0.10 & 0.97 & - & 0.15 & 0.88 & - & 0.19 & 0.29 & - \\
AT~2018gn & CrL-TDE & 0.05 & 1.24 & - & 0.05 & 1.81 & - & 0.07 & 0.68 & - \\
AT~2018bcb & CrL-TDE & 0.10 & 1.21 & - & 0.17 & 1.28 & - & 0.18 & 0.27 & - \\
AT~2021dms & CrL-TDE & 0.6 & 0.81 & - & 0.11 & 1.25 & - & 0.12 & 0.60 & - \\
TDE~2022fpx & CrL-TDE & 0.31 & 0.98 & - & 0.43 & 1.41 & - & 0.29 & 0.51 & - \\
TDE~2022upj & CrL-TDE & 0.08 & 1.50 & - & 0.22 & 2.00 & - & 0.17 & 0.55 & - \\
VT~J1548 & CrL-TDE & 0.04 & 2.47 & - & 0.07 & 3.25 & - & 0.08 & 0.90 & - \\
 &  &  &  &  &  &  &  &  &  &  \\
\textbf{Mean} & \textbf{CrL-TDE*} & \textbf{0.11} & \textbf{1.31} & \textbf{-} & \textbf{0.17} & \textbf{1.70} & \textbf{-} & \textbf{0.16} & \textbf{0.54} & \textbf{-} \\ \hline
 &  &  &  &  &  &  &  &  &  &  \\
AT~2019avd & Multi-epoch CrL-TDE / CrL-AGN & 0.04 & 0.98 & - & 0.09 & 1.08 & - & 0.07 & 0.53 & - \\
TDE~2019qiz & Multi-epoch CrL-TDE & 0.06 & 0.70 & - & 0.08 & 1.47 & - & 0.07 & 0.77 & - \\
 &  &  &  &  &  &  &  &  &  &  \\
\textbf{Mean} & \textbf{Multi-epoch CrL-TDE} & \textbf{0.05} & \textbf{-} & \textbf{-} & \textbf{0.08} & \textbf{-} & \textbf{-} & \textbf{0.07} & \textbf{-} & \textbf{-} \\ \hline
 &  &  &  &  &  &  &  &  &  &  \\
SDSS~J0748+4712 & TDE-ECLE & - & 0.77 & - & - & 1.48 & - & - & 0.74 & - \\
SDSS~J0952+2143 & TDE-ECLE & - & 0.61 & - & - & 1.33 & - & - & 0.73 & - \\
SDSS~J1241+4426 & TDE-ECLE & - & 0.30 & - & - & 0.61 & - & - & 0.35 & - \\
SDSS~J1342+0530 & TDE-ECLE & - & 0.67 & - & - & 1.80 & - & - & 1.14 & - \\
SDSS~J1350+2916 & TDE-ECLE & - & 0.91 & - & - & 1.70 & - & - & 0.89 & - \\
 &  &  &  &  &  &  &  &  &  &  \\
\textbf{Mean} & \textbf{TDE-ECLE} & \textbf{-} & \textbf{0.65} & \textbf{-} & \textbf{-} & \textbf{1.38} & \textbf{-} & \textbf{-} & \textbf{0.77} & \textbf{-} \\ \hline
 &  &  &  &  &  &  &  &  &  &  \\
SDSS~J0938+1353 & AGN-ECLE & 0.12 & - & - & 0.12 & - & - & 0.07 & - & - \\
SDSS~J1055+5637 & AGN-ECLE & 0.55 & - & - & 0.60 & - & - & 0.13 & - & - \\
 &  &  &  &  &  &  &  &  &  &  \\
\textbf{Mean} & \textbf{AGN-ECLE} & \textbf{0.33} & \textbf{-} & \textbf{-} & \textbf{0.36} & \textbf{-} & \textbf{-} & \textbf{0.10} & \textbf{-} & \textbf{-} \\ \hline
 &  &  &  &  &  &  &  &  &  &  \\
TDE~2019azh & NonCrL-TDE & 0.07 & - & 0.09 & 0.06 & - & 0.08 & 0.07 & - & 0.09 \\
\\ \hline
\\
iPTF16bco ** & CL-LINER & 0.34 & 0.12 & 0.12 & 0.45 & 0.18 & 0.10 & 0.22 & 0.14 & 0.09 \\
AT~2018aij & CL-LINER & 0.09 & 0.24 & 0.08 & 0.15 & 0.31 & 0.10 & 0.19 & 0.13 & 0.06 \\
AT~2018gkr & CL-LINER & 0.93 & - & - & 1.17 & - & - & 0.55 & - & - \\
AT~2018ivp & CL-LINER & - & 0.54 & - & - & 0.61 & - & - & 0.15 & - \\
AT~2018lnh ** & CL-LINER & 0.33 & 0.41 & - & 0.42 & 0.26 & - & 0.18 & 0.22 & - \\
ZTF18aasuray & CL-LINER & - & 0.71 & - & - & 0.94 & - & - & 0.33 & - \\
 &  &  &  &  &  &  &  &  &  &  \\
\textbf{Mean} & \textbf{CL-LINER} & \textbf{0.42} & \textbf{0.4} & \textbf{0.10} & \textbf{0.55} & \textbf{0.46} & \textbf{0.10} & \textbf{0.28} & \textbf{0.19} & \textbf{0.08} \\ \hline
\end{tabular}
\end{adjustbox}
\begin{flushleft}
\textbf{Notes:}\\
* Excluding \dyk\ for comparison purposes. Given its occurrence within an AGN hosting galaxy, AT~2021acak is also excluded from this analysis.\\
** MIR outbursts displayed by these objects and outlined here occur several years after the identification of the `changing look' event.\\
\end{flushleft}
\end{table*}
\raggedbottom

\FloatBarrier

\subsection{MIR Power-Law Fitting Parameters}
\label{Appendix:MIR_Power_Law_Fits}

\begin{figure*}
    \includegraphics[width=0.95\textwidth]{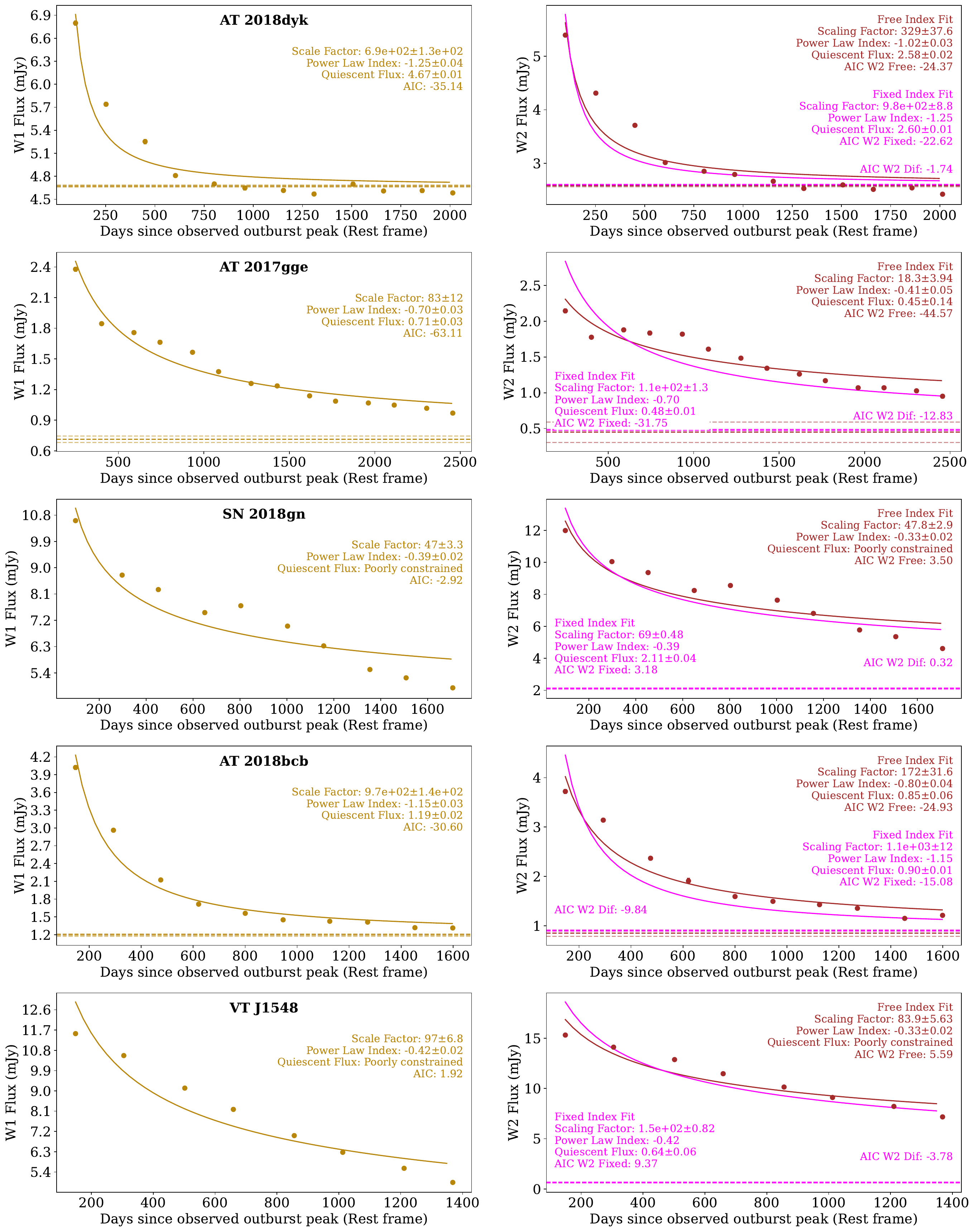}
    \caption{Power-law fits to the \textit{W1} {\it (left)} and \textit{W2} {\it (right)} photometry of the CrL-TDE sample where 5 or more epochs of observation have been obtained post MIR peak. Quiescent-flux values ($C$) are included when constrained by the fitting ($C > 0$ and $\Delta C < 0.15$) and shown by the dashed lines accompanied by the 1$\sigma$ uncertainties. Method follows that of \protect\citet{clark_2024_Longtermfollowupobservations} using the final NEOWISE-R release.}
    \label{fig:CLTDES_MIR_Power_Law_Fit_All_Points}
\end{figure*}

\begin{figure*}
    \includegraphics[width=0.95\textwidth]{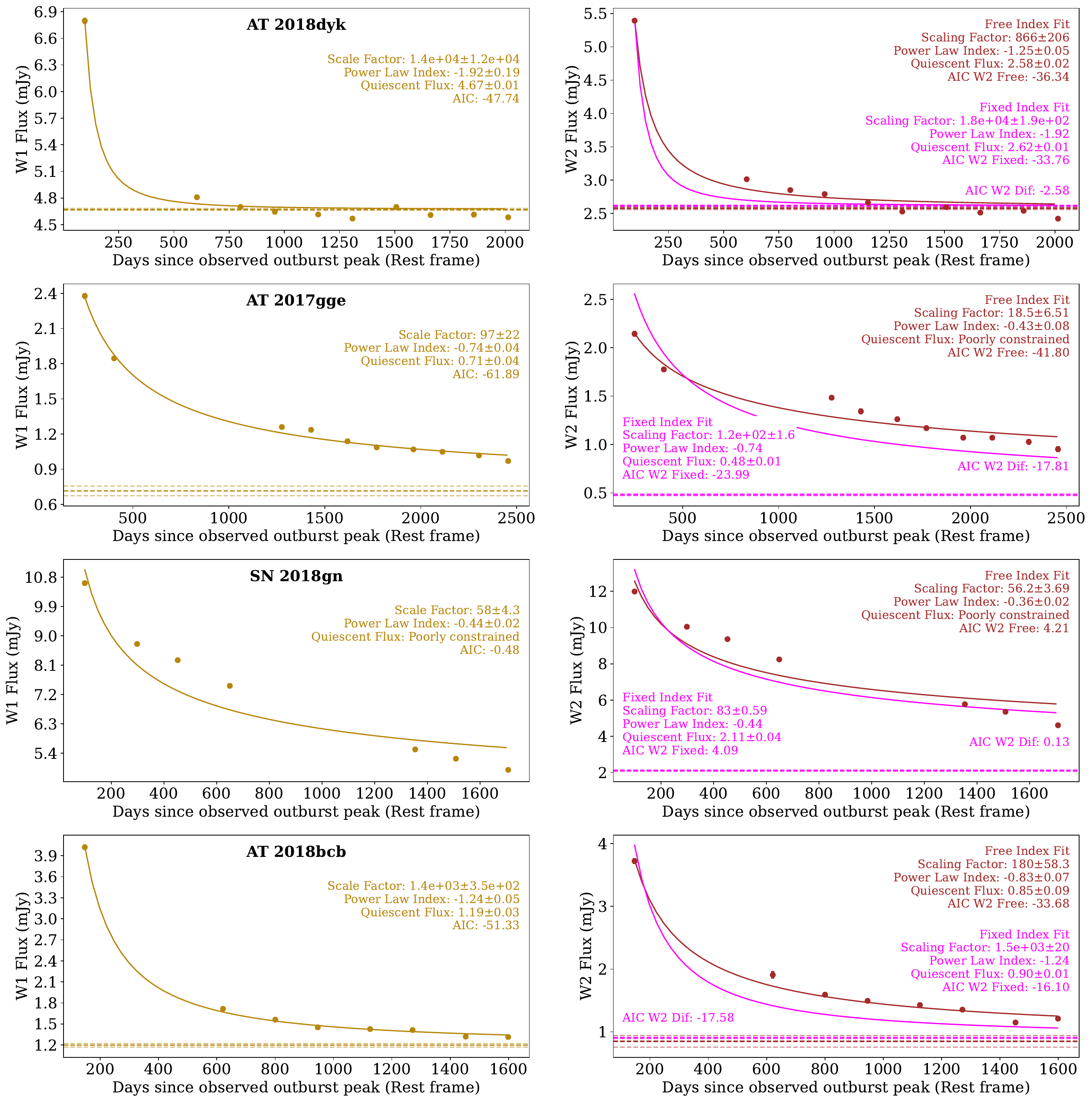}
    \caption{Power-law fits to the \textit{W1} {\it (left)} and \textit{W2} {\it (right)} photometry of the CrL-TDE sample where 5 or more epochs of observation have been obtained post MIR peak and where an early excess is present within the MIR light curve with the observations from this excess removed. Quiescent-flux values ($C$) are included when constrained by the fitting ($C > 0$ and $\Delta C < 0.15$) and shown by the dashed lines accompanied by the 1$\sigma$ uncertainties. Method follows that of \protect\citet{clark_2024_Longtermfollowupobservations} using the final NEOWISE-R release.}
    \label{fig:CLTDES_MIR_Power_Law_Fit_Early_Excess_Excluded}
\end{figure*}

\begin{figure*}
    \includegraphics[width=0.95\textwidth]{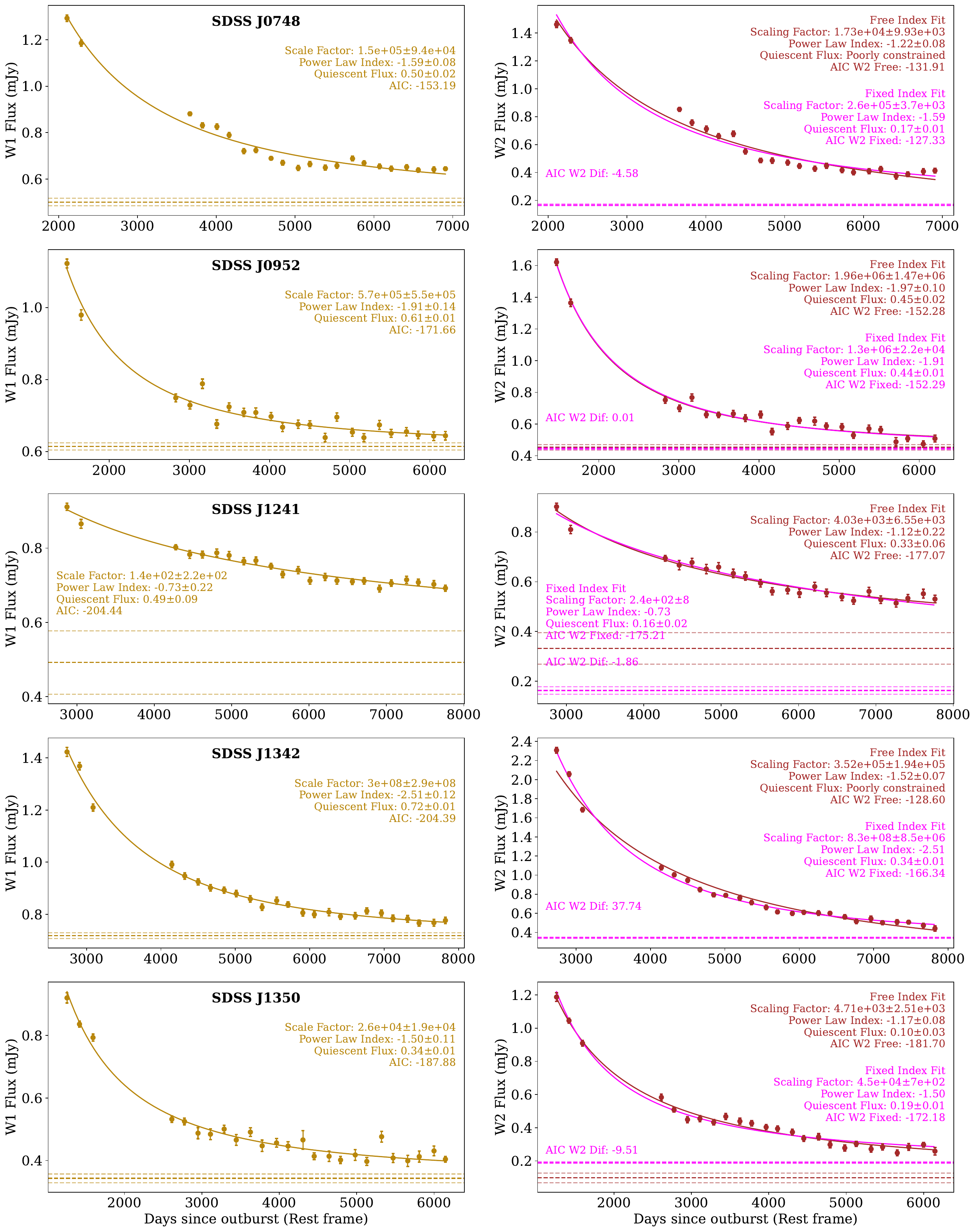}
    \caption{Power-law fits to the \textit{W1} {\it (left)} and \textit{W2} {\it (right)} photometry of the TDE-ECLE sample. Quiescent-flux values ($C$) are included when constrained by the fitting ($C > 0$ and $\Delta C < 0.15$) and shown by the dashed lines accompanied by the 1$\sigma$ uncertainties. Method follows that of \protect\citet{clark_2024_Longtermfollowupobservations} using the final NEOWISE-R release.}
    \label{fig:ECLE_MIR_Power_Law_Fit}
\end{figure*}

Here we present the results of the power-law fits to the MIR data for each of the objects within the comparison sample that have been shown to have variable coronal iron lines and classified as either a TDE-ECLE or a single epoch CrL-TDE. The results are detailed in Table~\ref{tab:MIR_Power_Law_Fits} and presented visually with comparison to the raw data points for the updated fits for the CrL-TDEs considering all data points in Fig.~\ref{fig:CLTDES_MIR_Power_Law_Fit_All_Points} and for the CrL-TDEs where an early excess in the MIR light curves has been excluded in Fig.~\ref{fig:CLTDES_MIR_Power_Law_Fit_Early_Excess_Excluded}. Fits for CrL-TDEs are only shown where there are at least five epochs of observation following peak MIR luminosity. Additionally, the results for the updated fitting using all WISE data available for the TDE-ECLE sample is also given in Fig.~\ref{fig:ECLE_MIR_Power_Law_Fit}, 

\clearpage
\begin{landscape}
\centering 
\begin{table}
\caption{MIR power-law fitting parameters. Model selected reflects a standard power-law decay typical of TDEs with a luminosity floor reflecting the quiescent galaxy luminosity as was favoured in earlier analyses \citep{dou_2016_LONGFADINGMIDINFRARED, clark_2024_Longtermfollowupobservations}. Following \citealt{clark_2024_Longtermfollowupobservations}, fitting in the \textit{W1} and \textit{W2} bands is initially independent (\textit{Free} fits) though due to the poorly constrained nature of some of the \textit{W2} quiescent flux values the results of fitting the \textit{W2} data where the decay index B is fixed to that of the \textit{W1} band are also reported. The results of fitting are compared through an AIC test, with the `Free' fits being preferred in all cases at some level where a statistical preference is displayed with the exception of SDSS~J1342, though as the decay law index is the most important parameter in this analysis, the values included in the main text (Fig.~\ref{fig:18dyk_mir_evolution}) reflect the results of the `Free' fit. As described in the main text, several CrL-TDEs (including AT~2018dyk) display light-curve `bumps' or `shoulders' (deviations from smooth declines) and we report fits to the data both including (All Points) or excluding (Excess Exclusion) these light curve regions from the fitting. The fits to the data themselves are shown in Figures~\ref{fig:CLTDES_MIR_Power_Law_Fit_All_Points}--\ref{fig:ECLE_MIR_Power_Law_Fit}.}
\label{tab:MIR_Power_Law_Fits}
\begin{adjustbox}{width=\columnwidth}
\begin{tabular}{lcccccccccccc}
\hline
\multicolumn{11}{c}{Model : $f(t) = At\textsuperscript{$B$} + C$} &  &  \\ \hline
Object & $A$\textsubscript{W1} & $B$\textsubscript{$W1$} & $C$\textsubscript{$W1$} (mJy) &  & $A$\textsubscript{$W2$ Free} & $B$\textsubscript{$W2$ Free} & $C$\textsubscript{$W2$ Free} (mJy) &  & $A$\textsubscript{$W2$ Fixed} & $C$\textsubscript{$W2$ Fixed} (mJy) & $\Delta$AIC\textsubscript{Free - Fixed}$\dag$ & $\Delta$AIC\textsubscript{Excluded - All}$\square$ \\ \hline
\textbf{AT~2018dyk} &  &  &  &  &  &  &  &  &  &  &  &  \\
All Points & 690.27$\pm$125.00 & -1.25$\pm$0.04 & 4.67$\pm$0.01 &  & 329.29$\pm$37.62 & -1.02$\pm$0.03 & 2.58$\pm$0.02 & & 979.43$\pm$8.85 & 2.60$\pm$0.01 & -1.74 & $\cdots$ \\
Excess Exclusion & 13969.49$\pm$12120.68 & -1.92$\pm$0.19 & 4.67$\pm$0.01 &  & 866.42$\pm$206.39 & -1.25$\pm$0.05 & 2.58$\pm$0.02 & & 18452.22$\pm$187.25 & 2.62$\pm$0.01 & -2.58 & -9.60, -11.97\\
 &  &  &  &  &  &  &  &  &  &  &  &  \\
\textbf{CrL-TDES} &  &  &  &  &  &  &  &  &  &  &  &  \\
AT 2017gge - All Points & 83.31$\pm$11.88 & -0.70$\pm$0.03 & 0.71$\pm$0.03 &  & 18.31$\pm$3.94 & -0.41$\pm$0.05 & 0.45$\pm$0.14 & & 112.70$\pm$1.26 & 0.48$\pm$0.01 & -12.83 & $\cdots$ \\
AT 2017gge - Excess Exclusion & 97.38$\pm$22.22 & -0.74$\pm$0.04 & 0.71$\pm$0.04 &  & 18.55$\pm$6.51 & -0.43$\pm$0.08 & 0.45$\pm$0.20* & & 122.20$\pm$1.57 & 0.48$\pm$0.01 & -17.81 & 1.22, 2.77\\
SN 2018gn - All Points & 46.92$\pm$3.31 & -0.39$\pm$0.02 & 3.36$\pm$0.25* &  & 47.83$\pm$2.90 & -0.33$\pm$0.02 & 2.05$\pm$0.44* & & 68.90$\pm$0.48 & 2.11$\pm$0.04 & 0.32 & $\cdots$ \\
SN 2018gn - Excess Exclusion & 57.55$\pm$4.33 & -0.44$\pm$0.02 & 3.36$\pm$0.21* &  & 56.18$\pm$3.69 & -0.36$\pm$0.02 & 2.05$\pm$0.37* & & 83.19$\pm$0.59 & 2.11$\pm$0.04 & 0.13 & 2.51, 1.01\\
AT 2018bcb - All Points & 968.83$\pm$137.60 & -1.15$\pm$0.03 & 1.19$\pm$0.02 &  & 171.76$\pm$31.61 & -0.80$\pm$0.04 & 0.85$\pm$0.06 & & 1134.85$\pm$12.20 & 0.90$\pm$0.01 & -9.84 & $\cdots$ \\
AT 2018bcb - Excess Exclusion & 1386.85$\pm$345.24 & -1.24$\pm$0.05 & 1.19$\pm$0.03 &  & 179.55$\pm$58.28 & -0.83$\pm$0.07 & 0.85$\pm$0.09 & & 1503.32$\pm$19.59 & 0.90$\pm$0.01 & -17.58 & -20.73, -8.75\\
VT J154843.06+220812.6 & 96.70$\pm$6.77 & -0.42$\pm$0.02 & 1.07$\pm$0.33* &  & 83.91$\pm$5.63 & -0.33$\pm$0.02 & 0.61$\pm$0.74* & & 146.22$\pm$0.82 & 0.64$\pm$0.06 & -3.78 & $\cdots$ \\
 &  &  &  &  &  &  &  &  &  &  &  &  \\
\textbf{TDE-ECLES} &  &  &  &  &  &  &  &  &  &  &  &  \\
SDSS J0748+4712 & 1.54e+05$\pm$9.38e+04 & -1.59$\pm$0.08 & 0.50$\pm$0.02 &  & 1.73e+04$\pm$9.93e+03 & -1.22$\pm$0.08 & 0.00$\pm$0.05 & & 2.63e+05$\pm$3.72e+03 & 0.17$\pm$0.01 & -4.58\\
SDSS J0952+2143 & 5.67e+05$\pm$5.54e+05 & -1.91$\pm$0.14 & 0.61$\pm$0.01 &  & 1.96e+06$\pm$1.47e+06 & -1.97$\pm$0.10 & 0.45$\pm$0.02 & & 1.34e+06$\pm$2.18e+04 & 0.44$\pm$0.01 & 0.01\\
SDSS J1241+4426 & 139$\pm$219 & -0.73$\pm$0.22 & 0.49$\pm$0.09 &  & 4.03e+03$\pm$6.55e+03 & -1.12$\pm$0.22 & 0.33$\pm$0.06 & & 241$\pm$7.96 & 0.16$\pm$0.02 & -1.86\\
SDSS J1342+0530 & 3.03e+08$\pm$2.95e+08 & -2.51$\pm$0.12 & 0.72$\pm$0.01 &  & 3.52e+05$\pm$1.94e+05 & -1.52$\pm$0.07 & 0.00$\pm$0.05 & & 8.26e+08$\pm$8.49e+06 & 0.34$\pm$0.01 & 37.74\\
SDSS J1350+2916 & 2.58e+04$\pm$1.91e+04 & -1.50$\pm$0.11 & 0.34$\pm$0.01 &  & 4.71e+03$\pm$2.51e+03 & -1.17$\pm$0.08 & 0.10$\pm$0.03 & & 4.46e+04$\pm$699 & 0.19$\pm$0.01 & -9.51\\ \hline
\end{tabular}
\end{adjustbox}
\begin{flushleft}
\textbf{Notes:}\\
For the `$W2$ Fixed' parameters, the value of $B$ was set to match that determined by the $W1$ fitting. Fits defined as `Excess Exclusion' have the points post peak showing flux excesses removed prior to the fitting.\\
* Indicates a poorly constrained quiescent flux value, i.e., Where C $\leq
$ 0 or $\Delta$C $\geq
$~0.15\\
$\dag$ In this case, a negative value indicates that the `Free' index fit is more favoured.\\
$\square$ For this comparison, the more favoured `Free' fits are compared and a negative value indicates the fit excluding the early excess points is more favoured. Given as two values: W1 value, W2, value.\\
\end{flushleft}
\end{table}
\end{landscape}
\clearpage

\FloatBarrier

\subsection{MIR outburst peak analysis}

Here in Table~\ref{tab:delta_fitting_parameters}, we include the full fitting parameters and statistical test results for the MIR outburst peak analysis as described in Section~\ref{fig:18dyk_mir_evolution} and Fig.~\ref{fig:MIR_Delta_Fig}.

\begin{table*}
\caption{Fitting parameters obtained in the $\Delta$\ value analysis. Data and fits shown in Fig.~\ref{fig:MIR_Delta_Fig}}
\label{tab:delta_fitting_parameters}
\begin{tabular}{llcrrrr}
\hline
 & Model & Parameter & Value & t-statistic & p-value & $\sigma$ \\ \hline
\textbf{$\Delta$\textit{W2} vs $\Delta$\textit{W1}} & Quadratic $^*$ &  &  &  &  &  \\
 &  & a & -0.18 & -2.26 & 1.09e-01 & 1.60 \\
 &  & b & 0.7 & 3.13 & 5.23e-02 & 1.94 \\
 &  & c & -0.48 & -4.34 & 2.26e-02 & 2.28 \\
 &  &  &  &  &  &  \\
\textbf{$\Delta$(\textit{W1-W2}) vs $\Delta$\textit{W2}} & Linear \textsuperscript{\textsquare} &  &  &  &  &  \\
 &  & m & -0.25 & -5.65 & 2.99e-02 & 2.17 \\
 &  & c & 0.19 & 2.99 & 9.59e-02 & 1.67 \\ \hline
\end{tabular}
\begin{flushleft}
$^*$ Selected through maximum likelihood analysis and AIC value comparison between a fixed constant, a linear model and a quadratic model\\
\textsuperscript{\textsquare} Selected through maximum likelihood analysis and AIC value comparison between a fixed constant and a linear model. A quadratic model was not included in this comparison due to the small number of data points (3) available for inclusion \\
\end{flushleft}
\end{table*}
\raggedbottom

\FloatBarrier
\clearpage
\section{Emission Line Fitting}
\label{Appendix:Emission Line Fitting}

Here we report the results of the emission line fitting conducted for the spectra of AT~2018dyk and the comparison TDE-ECLE SDSS J1342+0530.
Table~\ref{tab:Appendix_Tab_Non_CL_Results} and Table~\ref{tab:Appendix_Tab_CL_Results} summarise the fitting information for the non-coronal line and coronal lines respectively. Similarly, Table~\ref{tab:Appendix_Tab_Line_Ratios} summarises the resulting line ratios. Note that whilst the emission lines here are measured following correction for the local continuum, they have not been specifically corrected for stellar absorption and should thus be used primarily for relative comparisons.

\clearpage
\begin{landscape}
\centering 
\begin{table}
\caption{Emission line fitting results for non-coronal lines. Dots indicate where an emission feature / component was not detected in the given spectrum. Lines are fitted with independent properties, with the exception of the \Nii\ which are tied during fitting to have the same Gaussian standard deviations.}
\label{tab:Appendix_Tab_Non_CL_Results}
\begin{adjustbox}{width=\columnwidth}
\begin{tabular}{lccccccccccccccc}
\hline
 &  &  &  &  &  &  &  &  &  &  &  &  &  &  &  \\
{\ul \textbf{AT 2018dyk}} &  &  &  &  &  &  &  &  &  &  &  &  &  &  &  \\
 & \multicolumn{3}{c}{\textbf{SDSS : -5853 d}} &  & \multicolumn{3}{c}{\textbf{MaNGA: -560 d}} &  & \multicolumn{3}{c}{\textbf{Keck: +23 d}} &  & \multicolumn{3}{c}{\textbf{DESI: +1873 d}} \\
 & EQW & Offset V & FWHM V &  & EQW & Offset V & FWHM V &  & EQW & Offset V & FWHM V &  & EQW & Offset V & FWHM V \\
\textbf{Feature} & (\AA) & (\kms) & (\kms) &  & (\AA) & (\kms) & (\kms) &  & (\AA) & (\kms) & (\kms) &  & (\AA) & (\kms) & (\kms) \\
\Oi~6300~\AA & $\cdots$ & $\cdots$ & $\cdots$ &  & $\cdots$ & $\cdots$ & $\cdots$ &  & $\cdots$ & $\cdots$ & $\cdots$ &  & $\cdots$ & $\cdots$ & $\cdots$ \\
\Oii~Doublet * & -3.6~$\pm$~1.2 & 120~$\pm$~20 & 330~$\pm$~60 &  & -3.8~$\pm$~1.1 & 140~$\pm$~20 & 360~$\pm$~60 &  & -1.0~$\pm$~0.7 & 90~$\pm$~60 & 360~$\pm$~140 &  & -4.9~$\pm$~1.9 & 160~$\pm$~30 & 370~$\pm$~70 \\
\Oiii~4959~\AA & $\cdots$ & $\cdots$ & $\cdots$ &  & $\cdots$ & $\cdots$ & $\cdots$ &  & $\cdots$ & $\cdots$ & $\cdots$ &  & -0.5~$\pm$~0.3 & 140~$\pm$~30 & 210~$\pm$~60 \\
\Oiii~5007~\AA & -1.2~$\pm$~0.3 & 110~$\pm$~20 & 290~$\pm$~40 &  & -1.0~$\pm$~0.1 & 110~$\pm$~10 & 280~$\pm$~20 &  & -2.4~$\pm$~0.6 & 60~$\pm$~30 & 560~$\pm$~80 &  & -2.8~$\pm$~0.6 & 120~$\pm$~10 & 280~$\pm$~30 \\
H$\alpha$ Narrow & -0.5~$\pm$~0.3 & 90~$\pm$~30 & 260~$\pm$~70 &  & -0.4~$\pm$~0.2 & 70~$\pm$~20 & 180~$\pm$~50 &  & -23.9~$\pm$~0.7 & 50~$\pm$~10 & 790~$\pm$~290 &  & -0.5~$\pm$~0.3 & 90~$\pm$~30 & 200~$\pm$~60 \\
H$\alpha$ Broad & $\cdots$ & $\cdots$ & $\cdots$ &  & $\cdots$ & $\cdots$ & $\cdots$ &  & -22.6~$\pm$~1.7 & 80~$\pm$~20 & 2350~$\pm$~1010 &  & $\cdots$ & $\cdots$ & $\cdots$ \\
\Nii\ 6548~\AA & -1.3~$\pm$~0.2 & 50~$\pm$~20 & 370~$\pm$~20 &  & -1.1~$\pm$~0.2 & 70~$\pm$~20 & 340~$\pm$~20 &  & -1.7~$\pm$~0.3 & 30~$\pm$~10 & 380~$\pm$~30 &  & -1.1~$\pm$~0.2 & 50~$\pm$~20 & 310~$\pm$~20 \\
\Nii\ 6584~\AA & -3.2~$\pm$~0.3 & 70~$\pm$~10 & 370~$\pm$~20 &  & -2.9~$\pm$~0.3 & 70~$\pm$~20 & 340~$\pm$~20 &  & -3.1~$\pm$~0.4 & -40~$\pm$~20 & 380~$\pm$~30 &  & -3.5~$\pm$~0.3 & 70~$\pm$~10 & 310~$\pm$~20 \\
\Sii~6717~\AA & -0.9~$\pm$~0.2 & 120~$\pm$~20 & 340~$\pm$~20 &  & -0.8~$\pm$~0.1 & 70~$\pm$~10 & 340~$\pm$~20 &  & -0.5~$\pm$~0.2 & 0~$\pm$~30 & 360~$\pm$~40 &  & -0.9~$\pm$~0.2 & 40~$\pm$~20 & 280~$\pm$~30 \\
\Sii~6731~\AA & -1.0~$\pm$~0.2 & 70~$\pm$~20 & 340~$\pm$~20 &  & -0.8~$\pm$~0.1 & 80~$\pm$~10 & 330~$\pm$~20 &  & -0.7~$\pm$~0.2 & -20~$\pm$~30 & 360~$\pm$~40 &  & -0.9~$\pm$~0.2 & 70~$\pm$~20 & 280~$\pm$~30 \\
H$\beta$ & 2.6~$\pm$~0.1 & 190~$\pm$~40 & 1000~$\pm$~100 &  & 2.9~$\pm$~0.1 & 190~$\pm$~30 & 1000~$\pm$~70 &  & -9.5~$\pm$~0.5 & 30~$\pm$~10 & 1040~$\pm$~30 &  & 1.9~$\pm$~0.1 & 130~$\pm$~70 & 830~$\pm$~160 \\
 &  &  &  &  &  &  &  &  &  &  &  &  &  &  &  \\ \hline
 &  &  &  &  &  &  &  &  &  &  &  &  &  &  &  \\
{\ul \textbf{SDSS J1342+0530}} ** &  &  &  &  &  &  &  &  &  &  &  &  &  &  &  \\
 & \multicolumn{3}{c}{\textbf{SDSS: MJD~52373}} &  & \multicolumn{3}{c}{\textbf{MMT: MJD~55921}} &  & \multicolumn{3}{c}{\textbf{DESI: MJD~59279}} &  &  &  &  \\
 & EQW & Offset V & FWHM V &  & EQW & Offset V & FWHM V &  & EQW & Offset V & FWHM V &  &  &  &  \\
\textbf{Feature} & (\AA) & (\kms) & (\kms) &  & (\AA) & (\kms) & (\kms) &  & (\AA) & (\kms) & (\kms) &  &  &  &  \\
\Oi~6300~\AA & -1.2~$\pm$~0.6 & -70~$\pm$~100 & 820~$\pm$~230 &  & -1.5~$\pm$~0.3 & -80~$\pm$~10 & 260~$\pm$~30 &  & -1.0~$\pm$~0.4 & 100~$\pm$~10 & 150~$\pm$~30 &  &  &  &  \\
\Oii~Doublet * & -9.0~$\pm$~1.5 & 150~$\pm$~10 & 350~$\pm$~30 &  & \multicolumn{3}{c}{Out of wavelength coverage} &  & -10.8~$\pm$~3.7 & 130~$\pm$~30 & 370~$\pm$~70 &  &  &  &  \\
\Oiii~4959~\AA & -0.9~$\pm$~0.3 & 100~$\pm$~20 & 190~$\pm$~40 &  & -5.3~$\pm$~0.5 & -100~$\pm$~10 & 280~$\pm$~20 &  & -12.1~$\pm$~0.9 & 100~$\pm$~10 & 220~$\pm$~10 &  &  &  &  \\
\Oiii~5007~\AA & -4.5~$\pm$~0.4 & 70~$\pm$~10 & 250~$\pm$~10 &  & -14.4~$\pm$~1.4 & -90~$\pm$~10 & 350~$\pm$~20 &  & -38.7~$\pm$~2.1 & 90~$\pm$~10 & 200~$\pm$~10 &  &  &  &  \\
H$\alpha$ Narrow & -13.0~$\pm$~0.5 & 70~$\pm$~10 & 250~$\pm$~10 &  & -13.1~$\pm$~0.7 & -80~$\pm$~10 & 330~$\pm$~10 &  & -17.1~$\pm$~0.6 & 110~$\pm$~10 & 180~$\pm$~10 &  &  &  &  \\
H$\alpha$ Broad & $\cdots$ & $\cdots$ & $\cdots$ &  & $\cdots$ & $\cdots$ & $\cdots$ &  & $\cdots$ & $\cdots$ & $\cdots$ &  &  &  &  \\
\Nii~6548~\AA & -1.8~$\pm$~0.4 & 190~$\pm$~20 & 330~$\pm$~30 &  & -2.0~$\pm$~0.4 & -80~$\pm$~20 & 260~$\pm$~20 &  & -2.4~$\pm$~0.3 & 100~$\pm$~10 & 190~$\pm$~10 &  &  &  &  \\
\Nii~6584~\AA & -3.4~$\pm$~0.5 & 80~$\pm$~10 & 330~$\pm$~30 &  & -5.0~$\pm$~0.6 & -118.39~$\pm$~10 & 260~$\pm$~20 &  & -7.0~$\pm$~0.6 & 70~$\pm$~10 & 190~$\pm$~10 &  &  &  &  \\
\Sii~6717~\AA & -1.4~$\pm$~0.2 & 90~$\pm$~10 & 280~$\pm$~20 &  & -1.5~$\pm$~0.3 & -100.41~$\pm$~10 & 220~$\pm$~20 &  & -1.9~$\pm$~0.4 & 90~$\pm$~10 & 190~$\pm$~20 &  &  &  &  \\
\Sii~6731~\AA & -1.4~$\pm$~0.2 & 100~$\pm$~10 & 280~$\pm$~20 &  & -1.2~$\pm$~0.2 & -65.60~$\pm$~10 & 220~$\pm$~20 &  & -2.0~$\pm$~0.4 & 90~$\pm$~10 & 190~$\pm$~20 &  &  &  &  \\
H$\beta$ & -2.2~$\pm$~0.2 & 40~$\pm$~10 & 210~$\pm$~10 &  & -3.3~$\pm$~0.3 & -90~$\pm$~6 & 300~$\pm$~10 &  & -4.7~$\pm$~0.9 & 110~$\pm$~10 & 190~$\pm$~20 &  &  &  &  \\
 &  &  &  &  &  &  &  &  &  &  &  &  &  &  &  \\ \hline
\end{tabular}
\end{adjustbox}
\begin{flushleft}
\textbf{Notes:}\\
* The `\Oii~Doublet' is a single Gaussian fit to the \Oii~3726~\AA\ and \Oii~3728~\AA\ emission lines which, at the resolution of the available spectra, are too blended to be separated.\\
** The evolutionary phase of this object is not well constrained. As such, we report the MJD of each spectrum rather than phase.
\end{flushleft}
\end{table}
\end{landscape}
\clearpage
\clearpage
\begin{landscape}
\centering 
\begin{table}
\caption{Emission line fitting results for the coronal lines.}
\label{tab:Appendix_Tab_CL_Results}
\begin{tabular}{lccccccccccc}
\hline
 &  &  &  &  &  &  &  &  &  &  &  \\
{\ul \textbf{AT 2018dyk}} * &  &  &  &  &  &  &  &  &  &  &  \\
\textbf{} & \multicolumn{3}{c}{\textbf{Keck: +23 d}} &  & \textbf{} &  &  &  & \textbf{} &  &  \\
 & EQW & Offset V & FWHM V &  &  &  &  &  &  &  &  \\
\textbf{Feature} & (\AA) & (\kms) & (\kms) &  &  &  &  &  &  &  &  \\
\Fevii~3759~\AA & -2.5~$\pm$~0.8 & 200~$\pm$~60 & 840~$\pm$~150 &  &  &  &  &  &  &  &  \\
\Fevii~5160~\AA & $\cdots$ & $\cdots$ & $\cdots$ &  &  &  &  &  &  &  &  \\
\Fevii~5722~\AA & -1.9~$\pm$~0.4 & 120~$\pm$~30 & 640~$\pm$~60 &  &  &  &  &  &  &  &  \\
\Fevii~6088~\AA & -2.1~$\pm$~0.3 & 50~$\pm$~20 & 640~$\pm$~50 &  &  &  &  &  &  &  &  \\
\Fex~6376~\AA & -2.5~$\pm$~0.4 & -50~$\pm$~30 & 950~$\pm$~80 &  &  &  &  &  &  &  &  \\
\Fexi~7894~\AA & -1.0~$\pm$~0.3 & -220~$\pm$~40 & 650~$\pm$~100 &  &  &  &  &  &  &  &  \\
\Fexiv~5304~\AA & -2.6~$\pm$~1.0 & 130~$\pm$~110 & 1370~$\pm$~290 &  &  &  &  &  &  &  &  \\
 &  &  &  &  &  &  &  &  &  &  &  \\ \hline
 &  &  &  &  &  &  &  &  &  &  &  \\
{\ul \textbf{SDSS J1342+0530}} ** &  &  &  &  &  &  &  &  &  &  &  \\
 & \multicolumn{3}{c}{\textbf{SDSS: MJD~52373}} &  & \multicolumn{3}{c}{\textbf{MMT: MJD~55921}} &  & \multicolumn{3}{c}{\textbf{DESI: MJD~59279}} \\
 & EQW & Offset V & FWHM V &  & EQW & Offset V & FWHM V &  & EQW & Offset V & FWHM V \\
\textbf{Feature} & (\AA) & (\kms) & (\kms) &  & (\AA) & (\kms) & (\kms) &  & (\AA) & (\kms) & (\kms) \\
\Fevii~3759~\AA & $\cdots$ & $\cdots$ & $\cdots$ &  & \multicolumn{3}{c}{Out of wavelength coverage} &  & -2.8~$\pm$~1.9 & 200~$\pm$~20 & 150~$\pm$~60 \\
\Fevii~5160~\AA & $\cdots$ & $\cdots$ & $\cdots$ &  & -0.9~$\pm$~0.5 & -70~$\pm$~30 & 250~$\pm$~80 &  & $\cdots$ & $\cdots$ & $\cdots$ \\
\Fevii~5722~\AA & $\cdots$ & $\cdots$ & $\cdots$ &  & -1.7~$\pm$~0.3 & -90~$\pm$~10 & 260~$\pm$~20 &  & -1.8~$\pm$~0.5 & 110~$\pm$~20 & 250~$\pm$~40 \\
\Fevii~6088~\AA & $\cdots$ & $\cdots$ & $\cdots$ &  & -2.2~$\pm$~0.2 & -100~$\pm$~10 & 250~$\pm$~10 &  & -1.2~$\pm$~0.5 & 40~$\pm$~20 & 160~$\pm$~40 \\
\Fex~6376~\AA & -2.3~$\pm$~0.4 & 20~$\pm$~10 & 310~$\pm$~30 &  & -0.8~$\pm$~0.3 & -220~$\pm$~20 & 300~$\pm$~50 &  & $\cdots$ & $\cdots$ & $\cdots$ \\
\Fexi~7894~\AA & -3.2~$\pm$~0.3 & 10~$\pm$~10 & 310~$\pm$~20 &  & \multicolumn{3}{c}{Out of wavelength coverage} &  & $\cdots$ & $\cdots$ & $\cdots$ \\
\Fexiv~5304~\AA & -1.5~$\pm$~0.5 & 40~$\pm$~30 & 340~$\pm$~60 &  & \multicolumn{3}{c}{Out of wavelength coverage} &  & $\cdots$ & $\cdots$ & $\cdots$ \\
 &  &  &  &  &  &  &  &  &  &  &  \\ \hline
 \end{tabular}
\begin{flushleft}
\textbf{Notes:}\\
* Only the Keck spectrum is included here as no coronal lines were observed in any of the other spectra. \\
** The evolutionary phase of this object is not well constrained. As such, we report the MJD of each spectrum rather than phase.
\\
\end{flushleft}
\end{table}
\end{landscape}

\begin{table*}
\caption{Emission line ratios determined through Gaussian fitting.}
\label{tab:Appendix_Tab_Line_Ratios}
\begin{tabular}{lccccccc}
\hline
 &  &  &  &  &  &  &  \\
{\ul \textbf{AT 2018dyk}} &  &  &  &  &  &  &  \\
\textbf{} & \textbf{SDSS : -5853 d} &  & \textbf{MaNGA: -560 d} &  & \textbf{Keck: +19 d} &  & \textbf{DESI: +1873 d} \\
\textbf{Line Ratio} &  &  &  &  &  &  &  \\
log$_{10}$(\Nii\ / H$\alpha)$ & 0.8~$\pm$~0.2 &  & 0.9~$\pm$~0.2 &  & 2.4~$\pm$~0.2 &  & 0.9~$\pm$~0.3 \\
log$_{10}$(\Oiii\ 5007 / H$\beta)$ & $\cdots$ &  & $\cdots$ &  & -0.6~$\pm$~0.1 &  & $\cdots$ \\
log$_{10}$(\Sii\ 6717,6731 / H$\alpha)$ & 0.5~$\pm$~0.2 &  & 0.6~$\pm$~0.2 &  & 2.0~$\pm$~0.3 &  & 0.6~$\pm$~0.3 \\
log$_{10}$(\Oi\ / H$\alpha)$ & 0.3~$\pm$~0.3 &  & 0.5~$\pm$~0.3 &  & 1.9~$\pm$~0.3 &  & 0.3~$\pm$~0.4 \\
 &  &  &  &  &  &  &  \\
\Oiii\ 5007 / \Oiii\ 4959 & $\cdots$ &  & $\cdots$ &  & $\cdots$ &  & 4.6~$\pm$~2.7 \\
\Sii\ 6717 / \Sii\ 6731 & 1.0~$\pm$~0.2 &  & 1.1~$\pm$~0.2 &  & 0.7~$\pm$~0.3 &  & 1.0~$\pm$~0.3 \\
 &  &  &  &  &  &  &  \\
\Fevii\ 3759 / \Oiii\ 5007 & $\cdots$ &  & $\cdots$ &  & 1.1~$\pm$~0.5 &  & $\cdots$ \\
\Fevii\ 5160 / \Oiii\ 5007 & $\cdots$ &  & $\cdots$ &  & $\cdots$ &  & $\cdots$ \\
\Fevii\ 5722 / \Oiii\ 5007 & $\cdots$ &  & $\cdots$ &  & 1.6~$\pm$~0.5 &  & $\cdots$ \\
\Fevii\ 6088 / \Oiii\ 5007 & $\cdots$ &  & $\cdots$ &  & 1.8~$\pm$~0.5 &  & $\cdots$ \\
\Fex\ 6376 / \Oiii\ 5007 & $\cdots$ &  & $\cdots$ &  & 2.1~$\pm$~0.6 &  & $\cdots$ \\
\Fexi\ 7894 / \Oiii\ 5007 & $\cdots$ &  & $\cdots$ &  & 0.8~$\pm$~0.3 &  & $\cdots$ \\
\Fexiv\ 5304 / \Oiii\ 5007 & $\cdots$ &  & $\cdots$ &  & 1.1~$\pm$~0.5 &  & $\cdots$ \\
 &  &  &  &  &  &  &  \\
\Fevii\ 3759 / \Fevii\ 6088 & $\cdots$ &  & $\cdots$ &  & 0.6~$\pm$~0.2 &  & $\cdots$ \\
\Fevii\ 5160 / \Fevii\ 6088 & $\cdots$ &  & $\cdots$ &  & $\cdots$ &  & $\cdots$ \\
\Fevii\ 5722 / \Fevii\ 6088 & $\cdots$ &  & $\cdots$ &  & 0.9~$\pm$~0.2 &  & $\cdots$ \\
\Fex\ 6376 / \Fevii\ 6088 & $\cdots$ &  & $\cdots$ &  & 1.2~$\pm$~0.2 &  & $\cdots$ \\
\Fexi\ 7894 / \Fevii\ 6088 & $\cdots$ &  & $\cdots$ &  & 0.4~$\pm$~0.1 &  & $\cdots$ \\
\Fexiv\ 5304 / \Fevii\ 6088 & $\cdots$ &  & $\cdots$ &  & 0.6~$\pm$~0.3 &  & $\cdots$ \\
 &  &  &  &  &  &  &  \\ \hline
 &  &  &  &  &  &  &  \\
{\ul \textbf{SDSS J1342+0530*}} &  &  &  &  &  &  &  \\
 & \textbf{SDSS: MJD~52373} &  & \textbf{MMT: MJD~55921} &  & \textbf{DESI: MJD~59279} &  & \textbf{} \\
\textbf{Line Ratio} &  &  &  &  &  &  &  \\
log$_{10}$(\Nii\ / H$\alpha)$ & -0.6~$\pm$~0.1 &  & -0.4~$\pm$~0.1 &  & -0.4~$\pm$~0.1 &  &  \\
log$_{10}$(\Oiii\ 5007 / H$\beta)$ & 0.3~$\pm$~0.1 &  & 0.6~$\pm$~0.1 &  & 0.9~$\pm$~0.1 &  &  \\
log$_{10}$(\Sii\ 6717,6731 / H$\alpha)$ & -0.7~$\pm$~0.1 &  & -0.7~$\pm$~0.1 &  & -0.6~$\pm$~0.1 &  &  \\
log$_{10}$(\Oi\ / H$\alpha)$ & -1.0~$\pm$~0.2 &  & -0.9~$\pm$~0.1 &  & -1.2~$\pm$~0.2 &  &  \\
 &  &  &  &  &  &  &  \\
\Oiii\ 5007 / \Oiii\ 4959 & 4.7~$\pm$~1.7 &  & 2.4~$\pm$~0.3 &  & 3.0~$\pm$~0.3 &  &  \\
\Sii\ 6717 / \Sii\ 6731 & 1.0~$\pm$~0.3 &  & 1.2~$\pm$~0.3 &  & 1.0~$\pm$~0.3 &  &  \\
 &  &  &  &  &  &  &  \\
\Fevii\ 3759 / \Oiii\ 5007 & $\cdots$ &  & Out of wavelength coverage &  & 0.1~$\pm$~0.1 &  &  \\
\Fevii\ 5160 / \Oiii\ 5007 & $\cdots$ &  & 0.1~$\pm$~0.1 &  & $\cdots$ &  &  \\
\Fevii\ 5722 / \Oiii\ 5007 & $\cdots$ &  & 0.1~$\pm$~0.1 &  & 0.1~$\pm$~0.1 &  &  \\
\Fevii\ 6088 / \Oiii\ 5007 & $\cdots$ &  & 0.2~$\pm$~0.1 &  & 0.1~$\pm$~0.1 &  &  \\
\Fex\ 6376 / \Oiii\ 5007 & 0.5~$\pm$~0.1 &  & 0.1~$\pm$~0.1 &  & $\cdots$ &  &  \\
\Fexi\ 7894 / \Oiii\ 5007 & 0.5~$\pm$~0.1 &  & 0.2~$\pm$~0.1 &  & $\cdots$ &  &  \\
\Fexiv\ 5304 / \Oiii\ 5007 & 0.4~$\pm$~0.1 &  & Out of wavelength coverage &  & $\cdots$ &  &  \\
 &  &  &  &  &  &  &  \\
\Fevii\ 3759 / \Fevii\ 6088 & $\cdots$ &  & Out of wavelength coverage &  & 1.1~$\pm$~0.9 &  &  \\
\Fevii\ 5160 / \Fevii\ 6088 & $\cdots$ &  & 0.4~$\pm$~0.2 &  & $\cdots$ &  &  \\
\Fevii\ 5722 / \Fevii\ 6088 & $\cdots$ &  & 0.8~$\pm$~0.2 &  & 1.5~$\pm$~0.8 &  &  \\
\Fex\ 6376 / \Fevii\ 6088 & $\cdots$ &  & 0.4~$\pm$~0.1 &  & $\cdots$ &  &  \\
\Fexi\ 7894 / \Fevii\ 6088 & $\cdots$ &  & 1.4~$\pm$~0.2 &  & $\cdots$ &  &  \\
\Fexiv\ 5304 / \Fevii\ 6088 & $\cdots$ &  & Out of wavelength coverage &  & $\cdots$ &  &  \\
\\ \hline
\end{tabular}
\begin{flushleft}
\textbf{Notes}\\
* The evolutionary phase of this object is not well constrained. As such, we report the MJD of each spectrum rather than phase.
\end{flushleft}
\end{table*}
\raggedbottom

\FloatBarrier
\clearpage

\section{Emission Line Virial Distance Estimates}

In this appendix, we summarise the virial distance measurements measured from the FWHM of the narrow emission lines as described in Section~\ref{subsec:emission_line_behaviour}.

\begin{table*}
\caption{Virial distance measurements determined from the FWHM velocities of the narrow emission features.}
\label{tab:Appendix_Line_Distances}
\begin{tabular}{lccccccc}
\hline
\\
{\ul \textbf{AT 2018dyk}} &  &  &  &  &  &  &  \\
 & \textbf{SDSS : -5853 d} &  & \textbf{MaNGA: -560 d} &  & \textbf{Keck: +19 d} &  & \textbf{DESI: +1873 d} \\
 & Virial Distance &  & Virial Distance &  & Virial Distance &  & Virial Distance \\
\textbf{Feature} & (pc) &  & (pc) &  & (pc) &  & (pc) \\
\Fevii\ 5722 & $\cdots$ &  & $\cdots$ &  & 0.08~$\pm$~0.02 &  & $\cdots$ \\
\Fevii\ 6088 & $\cdots$ &  & $\cdots$ &  & 0.08~$\pm$~0.02 &  & $\cdots$ \\
\Fex\ 6376 & $\cdots$ &  & $\cdots$ &  & 0.03~$\pm$~0.01 &  & $\cdots$ \\
\Fexi\ 7894 & $\cdots$ &  & $\cdots$ &  & 0.07~$\pm$~0.02 &  & $\cdots$ \\
\Fexiv\ 5304 & $\cdots$ &  & $\cdots$ &  & 0.02~$\pm$~0.01 &  & $\cdots$ \\
 &  &  &  &  &  &  &  \\
\Oii\ Doublet * & 0.29~$\pm$~0.11 &  & 0.24~$\pm$~0.09 &  & 0.24~$\pm$~0.19 &  & 0.23~$\pm$~0.09 \\
\Oiii\ 4959 \AA & $\cdots$ & & $\cdots$ &  & $\cdots$ &  & 0.71~$\pm$~0.41 \\
\Oiii\ 5007 \AA & 0.37~$\pm$~0.11 &  & 0.40~$\pm$~0.07 &  & 0.10~$\pm$~0.03 &  & 0.40~$\pm$~0.10 \\
H$\alpha$ & 0.46~$\pm$~0.26 &  & 0.97~$\pm$~0.55 &  & 0.05~$\pm$~0.04 &  & 0.78~$\pm$~0.48 \\
H$\beta$ & 0.03~$\pm$~0.01 &  & 0.03~$\pm$~0.01 &  & 0.03~$\pm$~0.01 &  & 0.05~$\pm$~0.02 \\
\Nii\ ** & 0.23~$\pm$~0.04 &  & 0.27~$\pm$~0.05 &  & 0.22~$\pm$~0.04 &  & 0.33~$\pm$~0.06 \\
\Sii\ *** & 0.27~$\pm$~0.05 &  & 0.28~$\pm$~0.05 &  & 0.24~$\pm$~0.06 &  & 0.40~$\pm$~0.10 \\ \hline
\end{tabular}
\begin{flushleft}
\textbf{Notes}\\
* The `\Oii\ Doublet' is a single Gaussian fit to the \Oii~3726~\AA\ and \Oii~3728~\AA\ emission lines which, at the resolution of the available spectra, are too blended to be separated.\\
** `\Nii' here represents both the \Nii\ 6548~\AA\ and \Nii\ 6585~\AA\ emission lines as these were tied to have the same width during the fitting process.\\
*** `\Sii' here represents both the \Sii\ 6717~\AA\ and \Sii\ 6731~\AA\ emission lines as these were tied to have the same width during the fitting process.\\
SMBH mass used for these calculations is 7.27$\times$10$^{6}$~$\pm$~8.64$\times$10$^{5}$~\msol.
\end{flushleft}
\end{table*}


\bsp	
\label{lastpage}
\end{document}